\documentclass[11pt,a4paper]{article}
\usepackage{jheppub}
\usepackage{amsmath}
\usepackage{amsfonts}
\usepackage{graphicx}
\usepackage{cancel}
\usepackage{todonotes}

\newcommand{\meas}[2]{\frac{d^#2 #1}{(2\pi)^#2}}

\newcommand{\Lagr}{\mathcal{L}}

\newcommand{\order}[1]{\mathcal{O}\!\left(#1\right)}

\newcommand{\der}[2]{\frac{\partial#1}{\partial#2}}
\newcommand{\derDelta}[2]{\frac{\delta#1}{\delta#2}}

\newcommand{\qimplies}{\quad\implies\quad}

\newcommand{\non}{\nonumber\\}
\title{{Lessons from $O(N)$ models in one dimension}}
\date{\today}
\author{Daniel Schubring}
\emailAdd{schub071@umn.edu}
\affiliation{Department of Physics, University of Minnesota, Minneapolis, MN 55455, USA}
\abstract{
	Various topics related to the $O(N)$ model in one spacetime dimension (i.e. ordinary quantum mechanics) are considered. The focus is on a pedagogical presentation of quantum field theory methods in a simpler context where many exact results are available, but certain subtleties are discussed which may be of interest to active researchers in higher dimensional field theories as well.
	
	Large $N$ methods are introduced in the context of the zero-dimensional path integral and the connection to Stirling's series is shown. The entire spectrum of the $O(N)$ model, which includes the familiar $l(l+1)$ eigenvalues of the quantum rotor as a special case, is found both diagrammatically through large $N$ methods and by using Ward identities. The large $N$ methods are already exact at $\order{N^{-1}}$ and the $\order{N^{-2}}$ corrections are explicitly shown to vanish. Peculiarities of gauge theories in $d=1$ are discussed in the context of the $CP^{N-1}$ sigma model, and the spectrum of a more general squashed sphere sigma model is found. The precise connection between the $O(N)$ model and the linear sigma model with a $\phi^4$ interaction is discussed. A valid form of the self-consistent screening approximation (SCSA) applicable to $O(N)$ models with a hard constraint is presented. The point is made that at least in $d=1$ the SCSA may do worse than simply truncating the large $N$ expansion to $\order{N^{-1}}$ even for small $N$. In both the supersymmetric and non-supersymmetric versions of the $O(N)$ model, naive equations of motion relating vacuum expectation values are shown to be corrected by regularization-dependent finite corrections arising from contact terms associated to the equation of constraint.}

\makeatletter
\gdef\@fpheader{}
\makeatother

\begin{document}
	
	\maketitle
	
	\section{Introduction}
	
	Quantum field theory (QFT) is a very broad and abstract subject and can be quite difficult for a student to learn. Ordinary quantum mechanics (QM) involving single-particle wavefunctions is rather more concrete, and easier to build intuition on what is calculable and how exactly to do so. But of course ordinary QM may be considered as a special case of QFT where all the fields depend only on a single spacetime parameter. So rather than introducing the major paradigm shift of QFT along with all the additional technical difficulties in higher dimension, such as renormalization, it may be pedagogically prudent to first treat ordinary QM with the same methods that are usually employed for higher dimensional QFT. This has the additional benefit that subtle effects that might be poorly understood or not noticed at all in higher dimensional QFT can be illustrated quite clearly in the one-dimensional case.
	
	Of course, this is hardly a new idea. There have been pedagogical introductions to QFT in one dimension involving the $\phi^4$ and Yukawa interactions \cite{Boozer}. And the notion of first testing new methods of QFT in lower dimension is widespread. An older example of this is a test of the self-consistent screening approximation (SCSA) on linear sigma models in zero \cite{Bray:19740d} and one dimension \cite{Ferrell:1974}, which is a topic that will be discussed later on in these notes. A few selected more recent examples of this practice include studies of resurgence in QM \cite{BogomolnyZJ,Jentschura:2004jg,Dunne:2014bca,Behtash:2015zha}, the consideration of Lefschetz thimbles in zero-dimensional path integrals \cite{Tanizaki:2014tua}, and the consideration of particle production in a nonstationary background field for the $O(N)$ symmetric anharmonic oscillator in QM \cite{Trunin:2021lwg}. Examples where subtle effects in QFT were noticed by considering the one-dimensional case include Bastianelli and van Nieuwenhuizen's detailed study of the one-dimensional path integral of the non-linear sigma model \cite{Bastianelli:1991be}, which is an example rather close to the heart of this paper.
	
	This paper will consider the most familiar special case of the non-linear sigma model, namely the one-dimensional $O(N)$ model, which is a non-linear sigma model with a spherical target space $S^{N-1}$. Two particular values of $N$ which are familiar to every student of ordinary QM (although perhaps not in this language) are the $O(2)$ model, which is the same as a free quantum mechanical particle on a periodic interval, and the $O(3)$ model, which is also known as the \emph{quantum rotor} and can be understood as a quantum mechanical particle constrained to move on a surface a fixed distance from the origin.
	
	The study of the $O(N)$ model is a very old problem regardless of dimension. The notion of the quantum mechanics of a free particle confined to an arbitrary curved manifold was studied already by Podolsky in 1928 \cite{Podolsky1928} who proposed that the Hamiltonian is just the Laplace-Beltrami operator (sometimes just referred to as the Laplacian itself) on the manifold. Solving the quantum mechanics amounts to finding all the eigenvalues and eigenfunctions of the Laplacian, and for a homogeneous manifolds like $S^{N-1}$ this is not too difficult of a problem (see the appendices of \cite{Camporesi:1990wm} and also Appendix \ref{B} here).
	
	This problem is a little more difficult to consider from the operator and path integral points of view. From the operator point of view, the $O(N)$ model involves a constraint in configuration space, and the quantization of constrained systems involves the consideration of Dirac brackets \cite{Dirac:1950pj}. The problem of the path integral for particles on curved manifolds was also considered long ago by DeWitt \cite{DeWitt:1952js}, who in addition to the Laplace-Beltrami operator found that an additional term proportional to the Ricci scalar was necessary. Even fifty years after these original papers issues of the proper quantization of the $O(N)$ model were still being discussed from both points of view \cite{Kleinert:1997em,AbdallaBanerjee}.
	
	In this paper, the one-dimensional $O(N)$ model will be discussed from a different point of view, which is nevertheless common in higher dimensional QFT. We shall take the Euclidean path integral formulation of the $O(N)$ model as given, and study the quantum mechanics through time-ordered correlation functions and Feynman diagrams, which simplify in the limit of large $N$. This perspective is also very old. If the $O(3)$ model path integral is regulated on a lattice it becomes equivalent to the classical Heisenberg model of statistical field theory, and the large $N$ generalization of this was first considered by Stanley in 1968 \cite{Stanley:1968a}. Critical exponents were calculated in the following years \cite{Abe:1972,Ferrell:1972zz}, and its relevance as a toy model for high energy physics was soon noticed (see e.g. \cite{novik} for a review).
	
	But despite its age the large $N$ limit of the $O(N)$ model is still being discussed in the recent literature, particularly for the two-dimensional case, where it is perturbatively renormalizable. In the past few years even the stability of the ground state in large $N$ limit is being debated \cite{Nitta:2017uog,Gorsky:2018lnd,Bolognesi:2019rwq,Pikalov:2020iuh}, and recent studies of renormalons in the $O(N)$ model and closely related models \cite{Ishikawa:2019tnw,Bruckmann:2019mky,Marino:2021six,Schubring:2021hrw,DiPietro:2021yxb} often require careful study of diagrams in the large $N$ expansion.
	
	These notes grew out of considering this two-dimensional case, and checking that all the large $N$ manipulations done in that context were valid. The purpose is primarily pedagogical ---trying to calculate old results which are already known from a wavefunction perspective in new ways which are variants of familiar manipulations in higher dimensional field theories. But along the way we will discover some subtleties which may have been missed previously in the literature.
	
	In the following sections of this introduction we will introduce the notion of considering the spectrum of the quantum rotor from a more quantum field theoretical point of view. One of the main motivations of this paper is to calculate this spectrum solely by considering Feynman diagrams. A detailed summary of the problems considered in this paper is then given in Sec \ref{1.Sec outline}.
	
	\subsection{Spectrum of the quantum rotor}\label{1.Sec spectrum}
To illustrate the change in perspective between quantum mechanics and quantum field theory, consider the quantum rotor, which can be thought of as a single particle of unit mass confined to a sphere of radius $g^{-1}$. This has Hamiltonian
$$H= \frac{g^2}{2}L^2$$
where $L= (L_x,L_y,L_z)$ are angular momentum operators, and this has the familiar spectrum
$$E_l = \frac{g^2}{2}l(l+1),$$
where $l$ is an integer quantifying the total angular momentum of the particle's trajectory on the sphere. Since we are using units with both $\hbar$ and the mass of the particle set to unity, the quantity $g^2$ has dimensions of energy.

Now in the quantum field theory paradigm all of the formulas stay the same but the interpretation dramatically changes. The sphere is no longer thought of as an actual physical space in which a particle lives, it is considered as an abstract space, the \emph{target space} that our field takes values in. The field itself is defined on a one-dimensional spacetime, $d=1$, which has a time coordinate but no spatial extension.

The system no longer describes just one particle. There are many possible particles associated with excitations of the field. The integer $l$ still quantifies something like angular momentum, namely isospin, but as will become clear it can also now be thought of as an integer counting the number of particles. And because space is zero-dimensional, even if these particles repel each other they will form a bound state, simply because they have nowhere else to go.

Let us rewrite this spectrum in a perhaps unfamiliar way,
\begin{align}E_l = l g^2 + \frac{l!}{2!(l-2)!}g^2 \label{1.quantumrotorSpectrum} .\end{align}
The first term represents the non-interacting mass of $l$ particles. Each individual particle has a renormalized mass of $g^2$. The second term is taking into account the two-body interaction between particles. The numerical factor out front is simply the number of distinct two-body combinations, and each pair of particles has a two-body repulsion of energy $g^2$. It is a characteristic of the quantum rotor and all the other related sigma models considered here that all interactions factorize into two-body interactions alone. This is somewhat reminiscent of the factorization of the S-matrix of integrable models like the $O(3)$ model into two-body interactions in $d=2$. \cite{Zamolodchikov:1978xm}

A goal of the first half of this paper is simply to derive this expression for the spectrum $E_l$ using only methods which are also available to us in quantum field theory in higher spacetime dimension $d$. So for instance, to find the two body repulsion we will be drawing Feynman diagrams with two incoming and outgoing lines and finding the location of poles. As is well known, the problem actually simplifies by using large $N$ methods, where instead of considering a target space of $S^2$ we consider a target space of $S^{N-1}$, and treat $1/N$ as a small parameter in which we can do perturbation theory.

For general $N$ the spectrum is
\begin{align}E_l &= l m\left(1-\frac{1}{N}\right)+\frac{l!}{2!(l-2)!}\frac{2m}{N},\qquad m\equiv\frac{Ng^2}{2}.\label{1.ONSpectrum}\end{align}
Here $m$ can be thought of as the \emph{bare mass}. It is a parameter that will appear in our Lagrangian, and it will be corrected by the self-energy of the field. As we can see from this exact expression for the spectrum\footnote{Since from the wavefunction perspective this expression just gives the eigenvalues of the Laplace-Beltrami operator on $S^{N-1}$ it is well known (see e.g. \cite{Kleinert:1997em}). We will review this perspective very briefly in Appendix \ref{B}.}, both the correction to the individual particle mass and the two-body interaction are already exact at the first subleading order in the large $N$ expansion, and we will explicitly see that all the $\order{N^{-2}}$ corrections vanish, although from the diagrammatic perspective this will seem rather mysterious.

A consequence of the exactness of the large $N$ expansion in $d=1$ is that we can extend our results all the way down to $N=2$ where the target space is a circle.\footnote{It can even be extended somewhat formally down to $N=1$ where the target space is two disconnected points, and the Hilbert space is two-dimensional and degenerate in energy.} Indeed, a particle on a circle of radius $g^{-1}$ with circumference parametrized by $x$ has a wavefunction proportional to $\exp(iglx)$, with $l$ an integer, so the spectrum is
$$E_l=\frac{1}{2}\left(gl\right)^2 = l\frac{g^2}{2}+\frac{l!}{2!(l-2)!}g^2,$$
which agrees with $\eqref{1.ONSpectrum}$ for $N=2$.

The exactness of large $N$ expansion might also prompt us to reconsider the \emph{self-consistent screening approximation} (SCSA) which was developed for the closely related linear sigma model, and is widely used in the condensed matter literature. The SCSA agrees with the large $N$ method at $\order{N^{-1}}$ and in some cases does better than simply terminating the large $N$ expansion at that order, especially for low values of $N$ \cite{Ferrell:1974,Bray:19740d}. But clearly if the linear sigma model is in $d=1$ and in a range of parameters close to a non-linear sigma model, even for low $N$ it is better to simply truncate the large $N$ expansion at $\order{N^{-1}}$, since that is already exact for the non-linear sigma model. This matter will be discussed further in Section \ref{6.sec SCSA}.
	
	\subsection{$O(N)$ model path integral}\label{1.secPathInt}
	Rather than starting with an explicit Hamiltonian and Hilbert space, we will begin directly with the path integral formulation in Euclidean space. This will sidestep a lot of the difficulties with quantization of constrained systems which are encountered in other approaches (see e.g. \cite{Kleinert:1997em,AbdallaBanerjee} and references therein). We begin with a real $N$ component field $n_i(\tau)$, satisfying a constraint $\left(n_i\right)^2=g^{-2}$, and the standard $O(N)$ model partition function
	\begin{align*}
		Z=\int \mathcal{D}n \,\exp\left[-\int d\tau\, \frac{1}{2}\left(\dot{n}_i\right)^2\right].
	\end{align*} 
The dot indicates a Euclidean time derivative, the summation convention is implied, and in what follows the index $i$ may be omitted when it is clear by context. Given such a path integral it is straightforward to define correlation functions of the fields. Since our goal is not just to consider the $O(N)$ model as a classical field theory, but to extract information on the quantum mechanical spectrum, in this section we will very briefly review the connection between the Euclidean correlation functions and the underlying quantum mechanics.

As usual, $Z$ is implicitly connected with the Hamiltonian $H$,
$$Z=\lim_{\beta\rightarrow \infty} \text{Tr}\, e^{-\beta H},$$
and correlation functions of local fields $O(\tau)$ may be understood as Euclidean time ordered vacuum expectation values of $\hat{O}$ understood as an operator
$$\langle O_1(\tau_1)O_2(\tau_2)\dots\rangle\equiv \frac{1}{Z}	\int \mathcal{D}n \,n_{i_1}(\tau_1)n_{i_2}(\tau_2)\dots e^{-S[n]}=\langle 0| \mathcal{T}\,\hat{O}_1(\tau_1)\hat{O}_2(\tau_2)\dots |0\rangle,$$
where the time translation of an operator follows the convention
$$\hat{O}(\tau)=e^{+H\tau}\hat{O} e^{-H\tau}.$$
In particular the two-point function is useful for extracting information on the spectrum (indexed by $k$)
$$\langle O(\tau_f)O(\tau_i)\rangle = \sum_k \langle 0|\hat{O}(\tau_f)|k\rangle\langle k|\hat{O}(\tau_i)|0\rangle = \sum_k |\langle k| \hat{O} |0\rangle |^2 e^{-\left(E_k-E_0\right)|\tau_f-\tau_i|}$$
It is convenient to define the energy difference from the vacuum $M_k$ and the normalized amplitude $Z_{k}$,
\begin{align}
	M_k&\equiv E_k-E_0\non
	Z_{k}&\equiv 2M_k|\langle k | \hat{O} |0\rangle |^2,
\end{align}
then the correlation function in momentum space takes the form
\begin{align}
	\langle O(\tau)O(0)\rangle&= \int \frac{dp}{2\pi}\,\sum_k\frac{Z_{k}\, e^{-ip\tau}}{p^2+M_k^2}.
\end{align}
So to find the energy eigenvalue $E_k$ (up to an overall shift by the ground state energy $E_0$) we need only consider the poles of a two-point function of some local operator $\hat{O}$ which has a non-vanishing amplitude $Z_{k}$ with state $|k\rangle$.

At this point the particular characteristics of $O(N)$ model simplify the situation even further. Considering the first nontrivial case, the operator $\hat{n}_i$ will only have non-zero $Z$ factors with states that transform in the fundamental representation of $O(N)$. For a general $O(N)$ symmetric theory we could construct other states that transform in the same way as $\hat{n}_i|0\rangle$ by considering arbitrary functions $f(\hat{n}^2)$ of the scalar $\hat{n}^2$, and then $f(\hat{n}^2)\hat{n}_i|0\rangle$ would represent a distinct linearly independent state. But given the constraint $\hat{n}^2=g^{-2}$, there can only be one linearly independent state with the same transformation properties of $\hat{n}_i|0\rangle$. So it is in fact an eigenvector, and we can disregard the sum over other $k$,
$$\langle n(\tau)n(0)\rangle=\int \frac{dp}{2\pi}\,\frac{Z_1\, e^{-ip\tau}}{p^2+M_1^2}.$$
A similar argument holds for local operators which are products of $l$ components of $n$ all with distinct indices, $$\langle O_l(\tau)O_l(0)\rangle=\int \frac{dp}{2\pi}\,\frac{Z_l\, e^{-ip\tau}}{p^2+M_l^2},\qquad O_l(\tau)\equiv n_{i_1}(\tau)n_{i_2}(\tau)\dots n_{i_l}(\tau).$$
Here of course different choices of the particular values of the indices $i_1, i_2, \dots, i_l$ will lead to distinct operators $O_l$, but it is not indicated explicitly in this notation because by symmetry the values of $M_l$ and $Z_l$ will be the same. Also note that there are other operators involving repeated indices that are in the same irreducible representation of $O(N)$ as the $O_l$ defined above, but for the purposes of finding the spectrum $M_l$, considering $O_l$ alone will suffice.

	\subsection{Outline of the paper}\label{1.Sec outline}
	
	One-dimensional path integrals over general sigma models with curved target spaces have been studied extensively in the past \cite{DeWitt:1952js,Bastianelli:1991be}. What is special about the $O(N)$ model and the other closely related systems considered here is that an approach involving equations of constraint and Lagrange multiplier fields can be usefully applied. This approach is well known, but since it is often encountered in the context of higher dimensional field theories, a student may have some initial doubts as to its mathematical validity. In Sec \ref{2}, this approach is first discussed for $d=0$ path integrals, i.e. ordinary integrals. It is shown that the lowest order evaluation of the integral using the method of Lagrange multiplier fields is just equivalent to Stirling's approximation and higher order corrections in the $1/N$ expansion generate Stirling's series \cite{Stirling}. Feynman diagrams which are a direct analogue to those appearing for genuine field theories are introduced in this context in order to calculate expectation values of polynomials of the components $n_i$.

In Sec \ref{3} we return to genuine path integrals, which even in the $d=1$ case may lead to divergent Feynman diagrams. So along with the calculation of the $\order{N^{-1}}$ correction to the mass $M_1$ the issue of regularization is discussed. By considering two-point correlation functions of the operators $O_l$ the full spectrum is calculated to $\order{N^{-1}}$, and as pointed out above this will already agree with the exact result \eqref{1.ONSpectrum}. In Appendix \ref{A} we go a bit further and consider the $\order{N^{-2}}$ corrections to the spectrum and show that they all vanish. The goal of Sec \ref{3} and Appendix \ref{A} is mostly pedagogical. The exact spectrum \eqref{1.ONSpectrum} can of course be found via wavefunction methods but considering the problem from the perspective of Feynman diagrams can be instructive. There is sometimes something rather satisfying about seeing many distinct Feynman diagrams conspire together to cancel.

In Sec \ref{4} we consider some more general features of $O(N)$ models most of which hold regardless of dimension. The initial motivation is to find ways to simplify the calculation of the $\order{N^{-2}}$ corrections in Appendix \ref{A}, but the discussion will focus on the way that the constraint $n^2=g^{-2}$ and the equations of motion are manifested at the level of diagrams, and will be relevant for the later sections of the paper as well. The most important result, which to the best of my knowledge is new, is that more care may be needed in relating the vacuum expectation value of the Lagrange multiplier field to the `energy density' $\langle\left(\partial n\right)^2\rangle$. As has been discussed previously \cite{novik}, at lowest order of the large $N$ expansion the naive equation relating these expectation values is corrected by a divergent contact term which may be disregarded in a point-splitting regularization. But in Sec \ref{4.secEqM} it is further shown that at $\order{N^{-1}}$ there are corrections due to a more subtle type of contact term coming from the constraint, and these corrections are \emph{finite} and can't be disregarded.

The remaining sections consider further topics and can largely be read independently of each other. In the brief Sec \ref{5} the notion of the Hamiltonian arising directly from the path integral is further considered, and it is shown that despite the many difficulties in finding the Hamiltonian in more general curved spaces \cite{DeWitt:1952js}, and despite the difficulties in constrained quantization \cite{Kleinert:1997em,AbdallaBanerjee}, the naive Hamiltonian can indeed be used in conjunction with Ward identities to find the exact spectrum. This section constitutes a second, independent quantum field theoretic method of finding the spectrum of the quantum rotor, complementing the diagrammatic method in Sec \ref{3}.

In Sec \ref{6} the \emph{linear sigma model} (i.e. the $O(N)$ symmetric $\phi^4$ theory) is considered. As is well known \cite{Bessis:1972sn} and is intuitively plausible, in a certain limit the linear sigma model reduces to the same non-linear sigma model with a hard constraint which we have so far been considering. In this section it is shown in detail how the Hubbard-Stratonovich field of the linear sigma model becomes the Lagrange multiplier field of the non-linear sigma model. Although the diagrams will be largely the same as the non-linear case, it is discussed how the breaking of the constraint is manifested diagrammatically, and how this leads to extra states in the spectrum. The self-consistent screening approximation (SCSA) \cite{BrayRickayzen1972} which is often used for the linear sigma model in a condensed matter context is discussed and compared with the large $N$ approximation, and a consistent way to use the SCSA in the non-linear sigma model limit is presented.

In Sec \ref{7} we consider the $CP^{N-1}$ sigma model in $d=1$, which is closely related to the $O(N)$ model. The guiding theme will be how the difference in geometry between $CP^{N-1}$ and $S^{2N-1}$ is related to gauge symmetry in the path integral. As can be seen from exact wavefunction methods \cite{Camporesi:1990wm}, which are reviewed in Appendix \ref{B}, the energy eigenvalues of a quantum mechanical particle confined to a $CP^{N-1}$ manifold are identical to a subset of the eigenvalues of a particle on $S^{2N-1}$, but those states which are non-gauge invariant disappear from the spectrum. The way in which those states leave the spectrum is most easily seen by considering a particle confined to a \emph{squashed sphere} \cite{SquashedSphereSigma} intermediate between the cases of $S^{2N-1}$ and $CP^{N-1}$. The eigenvalues of a particle on a squashed sphere are found via three independent methods: using a gauge transformation in the path integral, calculating diagrams in the large $N$ expansion, and also via wavefunction methods in Appendix \ref{B}.

Finally, in Sec \ref{8} the supersymmetric (SUSY) $O(N)$ model is considered. It is well known that the Hamiltonian associated to $d=1$ SUSY sigma models is just the Laplacian extended to act on higher differential forms \cite{Witten:1982df}, sometimes referred to as the Laplace-de Rham operator. But this raises a question from the path integral point of view. The Laplace-de Rham operator acting on 0-forms has exactly the same spectrum as the ordinary Laplacian acting on wavefunctions, but it is not at all obvious from the path integral point of view that the bosonic correlation functions are the same, since the fermionic sector is coupled to the bosonic sector.

In Sec \ref{8.sec 0d}, this question is first considered in terms of $d=0$ path integrals, which also have the property that expectation values of polynomials of the bosonic fields $n_i$ are identical to the non-SUSY case. The main virtue of presenting this $d=0$ case first is that it might be a useful pedagogical example with which to introduce the idea of superfields, much like the $d=0$ version of the Wess-Zumino model considered in previous work \cite{MirrorSymmetry}. After this warm-up, the genuine $d=1$ SUSY $O(N)$ model is considered, and it is shown that indeed all supersymmetric corrections to the bosonic spectrum  vanish at $\order{N^{-1}}$, consistent with the argument in terms of the Laplace-de Rham operator above. Finally, as a new result, in Sec \ref{8.sec contact terms} contact term corrections to the equations of motion in the SUSY model are considered. It is shown that much like the non-SUSY case in Sec \ref{4.secEqM} these contact terms lead to finite corrections to the naive relations between various vacuum expectation values in the model.

	\section{Zero-dimensional path integral}\label{2}
	Consider the sphere $S^{N-1}$ embedded in $N$-dimensional Euclidean space with coordinates $n_i$. The radius of the sphere is taken to be $g^{-1}$, so small $g$ corresponds to large radius and vice versa. We can define a $d=0$ `path integral' $Z$ which is just proportional to the volume of the sphere
	\begin{align}
	Z\equiv \int dn_1dn_2\dots dn_N\, \,\delta\left(\frac{1}{2}\left((n_i)^2-\frac{1}{g^2}\right)\right)&=\int d\Omega_N \int dr\,r^{N-1}\,\delta\left(\frac{1}{2}\left(r^2-\frac{1}{g^2}\right)\right)\non&=\frac{2\pi^{N/2}}{\Gamma(N/2)} g^{-(N-1)}\times g.\label{2.Zexact}
	\end{align}
	Here the first factor involving a gamma function is the volume of a unit $(N-1)$-sphere coming from the integration over the angular coordinates $\Omega_{N-1}$, and the second factor is the correct radial factor $r^{N-1}$ evaluated at $g^{-1}$. This is then multiplied by an extra factor of $g$ since the delta function is a function of $r^2$ rather than $r$. But when we consider real path integrals for higher $d$ we will only be interested in correlation functions, so overall factors multiplying the path integral like this won't bother us.
	
	 Note however in passing that the geometry of the sphere is rather special in that the factor produced by integrating out the delta function does not itself depend on $\Omega_{N-1}$ so it will not modify the measure over the target space except for this overall constant factor. This is not true in general so we will not be able to apply the methods reviewed in this paper involving Lagrange multiplier fields to arbitrary geometries. But for the sphere, these methods would work even if we replace the quadratic function of $(n_i)^2$ inside the delta function by any other expression fixing the radius, quartic for instance, although doing so would complicate the large $N$ methods we are about to describe. 
	
	\subsection{Stirling's series}
	
	Now the idea is we want to find $Z$ in another way which we can extend to real path integrals. First rewrite the delta function as a Fourier integral over an auxiliary variable $\lambda$,
	\begin{align}
	Z=\int d^Nn\int \frac{d\lambda}{2\pi}\, \,e^{\frac{1}{2}i\lambda\left(n^2-\frac{1}{g^2}\right) }=e^{\frac{m^2}{2g^2}}\int \frac{d\lambda}{2\pi}\int d^Nn\, \,e^{-\frac{1}{2}\left[m^2 n^2-i\lambda\left(n^2-\frac{1}{g^2}\right)\right] }.\label{2.ActionNL}
	\end{align}
	In the second line we inserted a Gaussian exponential with an arbitrary coefficient $m^2$. Since the delta function fixes $n^2=g^{-2}$ this is compensated by the an overall factor multiplying the path integral. In higher $d$ we will do the same trick but won't keep track of the overall factor.
	
	Next we integrate over the $N$ dimensional Gaussian integral in $n$,
	\begin{align}
	Z=e^{\frac{m^2}{2g^2}}\int \frac{d\lambda}{2\pi}\, \,\left(\frac{2\pi}{m^2-i\lambda}\right)^{N/2}e^{-\frac{i}{2g^2}\lambda }=e^{\frac{m^2}{2g^2}}(2\pi)^{\frac{N-2}{2}}m^{-N}\int d\lambda\,\, e^{-\frac{N}{2}\left[ \ln\left(1-\frac{i\lambda}{m^2}\right)+\frac{i\lambda}{Ng^2}\right]}\label{2.ActionL}
	\end{align}
	Now the idea is that if $N$ is very large, the integral over $\lambda$ will be dominated by values of $\lambda_0$ which are stationary points of the argument of the exponential. In particular we can choose our arbitrary parameter $m$ so as to cause the stationary point to be at $\lambda_0=0$. It is easy to see that this will be true if we set
	\begin{align}
	m^2=Ng^2.\label{2.saddlepoint}
	\end{align}
	In particular this condition $\lambda_0=0$ is equivalent to setting $m^2$ so as to cancel the terms linear in $\lambda$ after expanding the logarithm,
	\begin{align}
	Z=e^{\frac{N}{2}}(2\pi)^\frac{N-2}{2}(Ng^2)^{-\frac{N}{2}}\int d\lambda\,\, e^{-\frac{N}{2}\left[\frac{\lambda^2}{2m^4}+i\frac{\lambda^3}{3m^6}-\frac{\lambda^4}{4m^8}+\mathcal{O}(\lambda^5)\right]}.\label{2.actionExpanded}
	\end{align}
	Now we can treat the terms involving higher powers of $\lambda$ as perturbations on the Gaussian integral in a manner akin to ordinary perturbation theory in QFT. Contracted pairs of $\lambda$ variables will lead to a `propagator' $\frac{2m^4}{N}$, so interaction terms with large powers of $\lambda$ will be suppressed by large inverse powers of $N$ even though each vertex itself comes with a compensating factor of $N$. Note for this expansion to be sensible $m^2$ itself must not change as we vary $N$, which can only true if $g^2$ decreases proportional to $1/N$. This dependence of $g^2$ will need to be true in higher $d$ as well since we will have an analogous expression to \eqref{2.ActionL} in those cases as well, and the $N$ dependence must be just an overall factor multiplying the argument of the exponential.
	
	In any case, if we want to evaluate $Z$ to the lowest order in the $1/N$ expansion, we should just evaluate the Gaussian integral itself ignoring the higher order terms.
	\begin{align}
		Z=e^{\frac{N}{2}}(2\pi)^\frac{N-2}{2}(Ng^2)^{-\frac{N}{2}}\left(4\pi Ng^4\right)^{1/2}\left(1+\mathcal{O}(N^{-1})\right).
		\end{align}
		It is interesting to compare this with the exact expression \eqref{2.Zexact}. The powers of $g$ agree on both sides, and we can solve for the gamma function, defining $\bar{N}\equiv N/2$ for convenience,
		\begin{align}
		\Gamma(\bar{N})=\sqrt{\frac{2\pi}{\bar{N}}}\left(\frac{\bar{N}}{e}\right)^{\bar{N}}\left(1+\mathcal{O}(N^{-1})\right).
		\end{align}
		But this is just Stirling's formula for the Gamma function.
		$$\log \Gamma(\bar{N})=\bar{N}\log(\bar{N})-\bar{N}-\frac{1}{2}\ln \frac{\bar{N}}{2\pi}+\mathcal{O}(N^{-1}).$$
		
\begin{figure}\centering
			\includegraphics[width=0.6\textwidth]{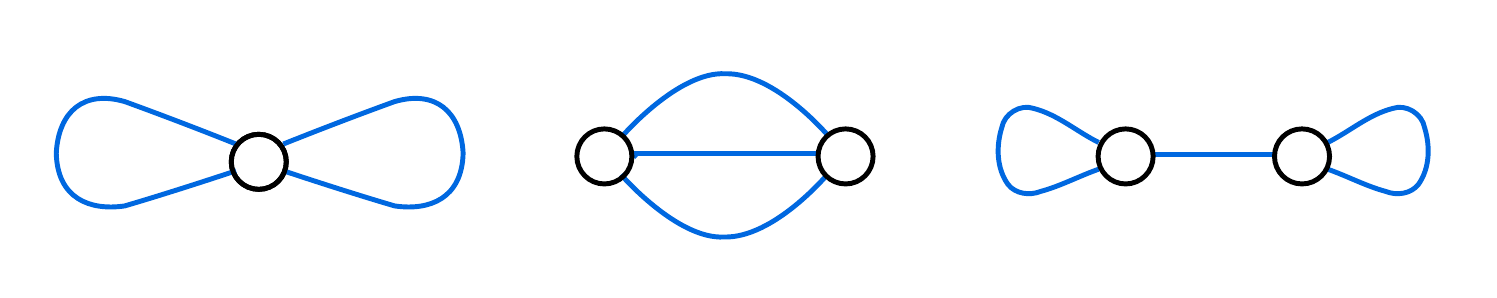}
			\caption{\small Diagrams involved in the $1/N$ correction to $Z$, and thus to Stirling's formula. Each blue line represents a $\lambda$ propagator. The vertices are represented as black circles to remind us that they came from integrating out the $n$ field, which will be represented by black lines later. The order that $\lambda$ lines connect to a vertex will not matter in this $d=0$ case.}\label{2.FigBubble}
		\end{figure}

		Furthermore, the higher order $\mathcal{O}(N^{-1})$ terms in Stirling's formula may be found from perturbation theory using the higher order terms in $\lambda$ in the action \eqref{2.actionExpanded} as vertices and the propagator $2m^4 N^{-1}$ from the quadratic term. In particular, at order $1/N$ there are three distinct `vacuum bubble' diagrams involved in the evaluation of \eqref{2.actionExpanded}, displayed in Fig \ref*{2.FigBubble}. We won't dwell on the details for each diagram in this article, but just for the sake of illustration let's consider the evaluation of these diagrams.

\begin{itemize}
	\item The left diagram of Fig \ref*{2.FigBubble} involves one $\lambda^4$ vertex $\frac{N}{8m^8}$, two $\lambda$ propagators $\frac{2m^4}{N}$, and a combinatorial factor of 3 distinct pairings for a net value of $+\frac{3}{2N}.$
	\item The remaining two diagrams involve two $\lambda^3$ vertices $-i\frac{N}{6m^6}$, three $\lambda$ propagators, and the necessary factor of $1/2!$ from expanding the exponential. The center diagram involves a combinatorial factor of 6, and the right diagram involves a combinatorial factor of 9. So the total value of these two diagrams is
	$-\frac{5}{3N}.$
\end{itemize}
So in total we have a correction of $-\frac{1}{6N}$, and after dividing to the other side of the equation and expressing in terms of the rescaled $N=2\bar{N}$, Stirling's formula is corrected to
		$$\log \Gamma(\bar{N})=\bar{N}\log(\bar{N})-\bar{N}-\frac{1}{2}\ln \frac{\bar{N}}{2\pi}+\frac{1}{12\bar{N}}+\mathcal{O}(N^{-2}).$$
		We could continue on in this manner and we would reproduce what is known as Stirling's series \cite{Stirling}, which is known to all orders. The point is that this chain of manipulations starting from an $n$ field and delta function and ending up with an auxiliary field $\lambda$ which can have complex saddle points and an unusual propagator may seem rather far-fetched at first, but it really is producing a well defined asymptotic expansion.
		
	\subsection{The `propagator' of $n$}\label{2.sec propagator}

	The kinds of calculations that will be more useful in higher dimension are calculations of correlation functions and expectation values.
	
\begin{figure}
			\centering
		\includegraphics[width=0.3\textwidth]{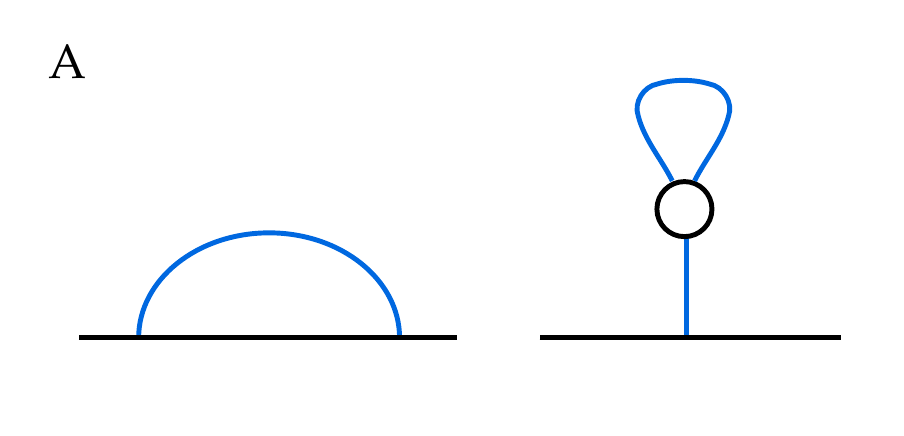}
			\includegraphics[width=0.3\textwidth]{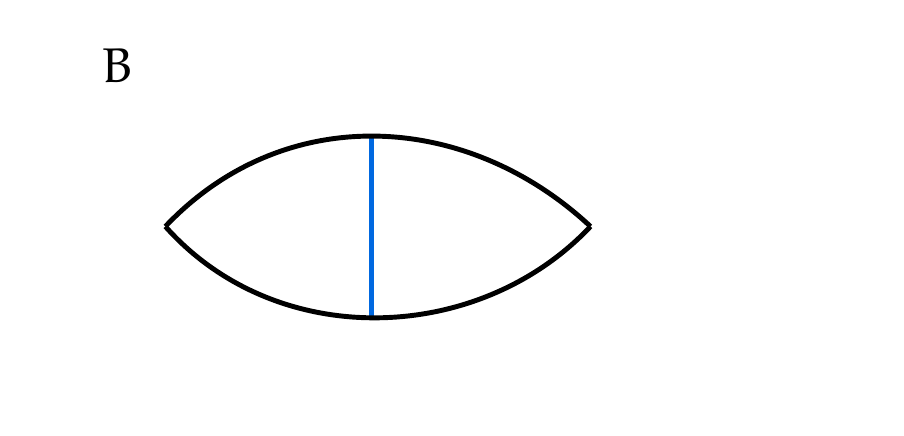}
		\caption{\small Diagrams involved in the $\order{N^{-1}}$ corrections to the $n$ propagator on the left and the two-body $O_2$ propagator on the right.}\label{2.Fig n and n2}
	\end{figure}

	The two-point $n$ correlation function is defined as
	$$\langle n_i n_j \rangle = \frac{1}{Z}\int d\lambda d^N n  \, n_i n_j \, e^{-\frac{1}{2}\left[m^2 n^2-i\lambda\left(n^2-\frac{1}{g^2}\right)\right] },$$ 
and we will not be as picky about keeping track of overall factors in the definition of $Z$ anymore. At lowest order in the large $N$ expansion, this correlation function is just the bare propagator $m^{-2}\delta_{ij}$. Indeed, this should be true to all orders, since the combination of constraint, symmetry, and saddle point condition \eqref{2.saddlepoint} leads to $(n_i)^2=\frac{1}{Ng^2}=m^{-2}$. Let's show explicitly that the possible corrections due to the diagrams in Fig \ref*{2.Fig n and n2}.A vanish.
	
	After integrating out $n$, the propagator becomes
	\begin{align}
	\langle n_i n_j\rangle = \frac{1}{Z}\int d\lambda \frac{\delta^{ij}}{m^2 -i\lambda}\,e^{-\frac{N}{2}\left[\frac{\lambda^2}{2m^4}+i\frac{\lambda^3}{3m^6}-\frac{\lambda^4}{4m^8}+\mathcal{O}(\lambda^5)\right]}.
	\end{align}
	This can be expanded as a geometric series in $i\lambda/m^2$, so every $\lambda$ line insertion in the $n$ propagator in Fig \ref*{2.Fig n and n2}.A comes with a factor of $i$. This could also be seen directly from the $\frac{1}{2}i\lambda n^2$ term in the action before we integrate out $n$, and in practice in higher $d$ it is usually more convenient to work with the action involving both $\lambda$ and $n$, integrating out vertices as needed.
	
	In any case, accounting for the factors of $i$, and the extra combinatorial factor of $3$ for the tadpole diagram, we indeed see that the arc diagram produces $-2m^2/N$ and cancels with the tadpole diagram contribution of $+2m^2/N$. This cancelation will be explained from a more general (but still diagrammatic) perspective in Sec \ref{4.sec tadpole}.
	
	\subsection{Two-body propagator}\label{2. Sec two-body}
	Now let's calculate something a little more non-trivial, the correlation function $\left\langle \left(O_2\right)^2\right\rangle\equiv\langle n_i n_j \,\,n_i n_j\rangle$, with $i\neq j$ and no sum implied. This could be calculated by integrating over $S^{N-1}$ using a trick similar to the well-known trick of calculating the volume of $S^{N-1}$ using gamma functions and Gaussian integrals, and in Sec \ref{5} amplitudes like this will be calculated from recursion relations. The exact result is
	\begin{align}
\langle n_i n_j \,\,n_i n_j\rangle=\frac{1}{(Ng^2)^2}\left(1+\frac{2}{N}\right)^{-1}.
	\end{align}
	It is actually quite easy to calculate this to order $\order{N^{-1}}$, using our large $N$ perturbation theory. At lowest order it is just two $n$ propagators giving the overall factor of $m^{-4}$, and at subleading order, it is given by Diagram B of Fig \ref*{2.Fig n and n2}, with four $n$ propagators, one $\lambda$ propagator, and no combinatorial factors, for a total result of $m^{-4}\left(1-\frac{2}{N}\right)$. At next order it starts to get more difficult, the relevant diagrams are essentially the same as the $d=1$ case in Fig \ref{A.fig n2} which has Fig \ref{4.fig lambda ON2} as a sub-diagram. However in $d=0$ with no momentum integration it is still very manageable\footnote{Some additional simplifications in $d=0$ are the vanishing of the $n$ propagator corrections, so diagrams A and B in Fig \ref{A.fig n2} vanish, and also the order that $\lambda$ lines connect to a $n$ line is irrelevant so diagrams F and G take the same value. }, and the correct result of $+\frac{4}{N^2m^4}$ is found.
	\section{Spectrum of the $d=1$ $O(N)$ model}\label{3}
	Now we will consider a true path integral in $d=1$. The analogue of \eqref{2.ActionNL} in $d=1$ involves a kinetic term,
	\begin{align}
	Z=\int \mathcal{D}n\mathcal{D}\lambda\,e^{-\frac{1}{2}\int d\tau\left[\dot{n}^2+m^2 n^2-i\lambda\left(n^2-\frac{1}{g^2}\right)\right] }.\label{3.ActionNL}
	\end{align}
	
	In principle this proceeds in the same way as the $d=0$ case. We integrate out the quadratic $n$ action leading to a $-\frac{N}{2}\log$ term that can be expanded to produce the $\lambda$ propagator and all higher order vertices. In practice it is more convenient to only explicitly integrate out $n$ for the terms linear and quadratic in $\lambda$ in order to find both the saddle point condition on the parameter $m$ and the $\lambda$ propagator, respectively. For any vertices involving higher powers of $\lambda$, or external $n$ propagators coming from correlation functions, we instead make use of the $\lambda n^2$ vertex and bare $n$ propagator in the original action.
	
	Since the terms linear in $\lambda$ in the original action involve the factor $(n^2-\frac{1}{g^2})$, we immediately see that upon integrating out $n$ for the linear terms to cancel we must have the saddle point condition
	\begin{align}
	N\int\frac{d^dk}{(2\pi)^d}\frac{1}{k^2+m^2}=\frac{1}{g^2}\qimplies m=\frac{Ng^2}{2}\quad (d=1).\label{3.saddle point}
	\end{align} 
	On the left hand side the integral is written as an arbitrary function of $d$ to emphasize that this same condition holds in any dimension. For $d= 2$ the integral is logarithmically divergent, and enforcing this condition with fixed $m$ immediately implies the renormalization group flow of $g$ with cutoff scale. But in $d=1$ the integral converges to $\frac{1}{2m}$ and we have the equation for $m$ on the right hand side.
	
\begin{figure}\centering	\includegraphics[width=0.7\textwidth]{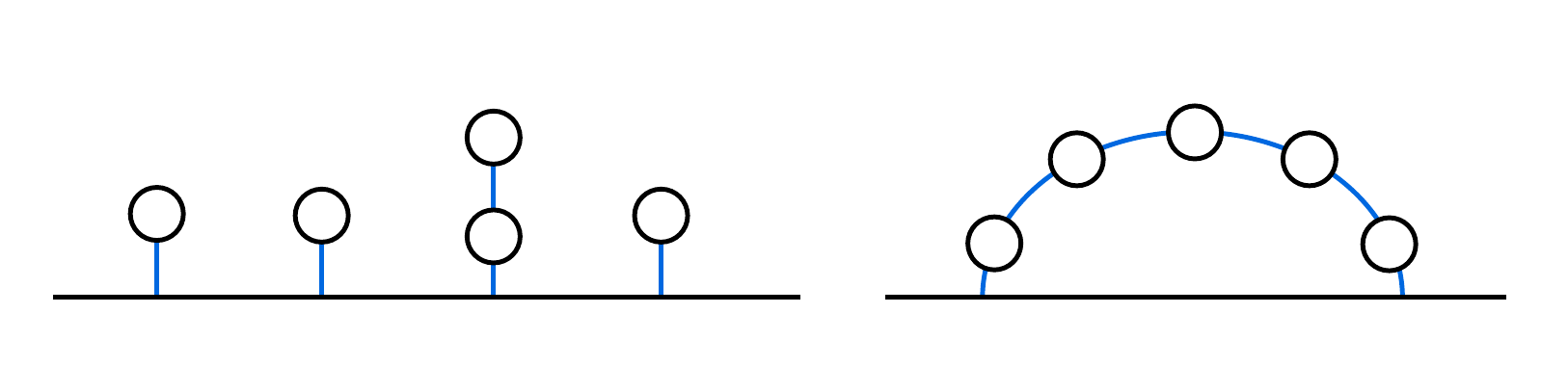}
			\caption{\small By simply counting the factors of $N$ involved in $\lambda$ propagators and $n$ loops, the left diagram would appear to be a $\order{N^{0}}$ correction to the propagator. Likewise the right diagram would appear to be a $\order{N^{-1}}$ correction. In fact since these simple $n$ bubbles and tadpoles have already been taken into account in defining the bare 1- and 2-point functions of $\lambda$, diagrams such as this must be excluded.} \label{3.Fig Excluded diagrams}
		\end{figure}

	To find the propagator of $\lambda$ we integrate out two $\lambda n^2$ vertices, leading to the term
	$$-\frac{1}{2}\int \frac{dp}{2\pi}\,\lambda(-p)\left[\frac{N}{2}J(p^2)\right]\lambda(p)$$
	where $J(p^2)$ is the integral
	\begin{align}
J(p^2)&\equiv\int\frac{d^dk}{(2\pi)^d}\frac{1}{(k^2+m^2)\left((p-k)^2+m^2\right)}\quad=\frac{1}{m\left(p^2+4m^2\right)}.\label{3.J def}
	\end{align}
	This is originally defined for arbitrary dimension since many formulas will hold for any dimension when expressed in terms of $J(p^2)$. The $\lambda$ propagator is
	\begin{align}
	\frac{2}{N J(p^2)}=\frac{2m}{N}\left(p^2+4m^2\right). \label{3. lambda prop}
	\end{align}
	Now that the 1-point and 2-point $\lambda$ vertices have been taken into account to cancel the $\lambda g^{-2}$ term and to define the propagator, they should not appear in diagrams involving both $\lambda$ and $n$ lines, see Fig \ref*{3.Fig Excluded diagrams}.
		\subsection{Regularization and correction to the $n$ propagator}
		Now let's consider the $\order{N^{-1}}$ corrections to the propagator as in Diagram A of Fig \ref{2.Fig n and n2}. Using the $\lambda$ propagator above, the arc diagram on the left gives\footnote{To be clear, $\langle n(-p)n(p)\rangle$ is used to indicate the Fourier transform $\int d\tau \langle n_i(\tau) n_i(0)\rangle e^{ip\tau}$ with no summation over $i$.}
		$$\langle n(-p)n(p)\rangle_{\text{arc}}=-\frac{1}{\left(p^2+m^2\right)^2}\frac{2m}{N}\int\frac{dk}{2\pi}\frac{(p-k)^2+4m^2}{k^2+m^2}.$$
		This is linearly divergent so some form of regularization is necessary. We will use the following simple cutoff scheme:
		\begin{align}
		\langle n(-p)n(p)\rangle_{\text{arc}}&=	-\frac{1}{\left(p^2+m^2\right)^2}\frac{2m}{N}\int \frac{dk}{2\pi}\left(1+\frac{\bcancel{-2pk }+p^2+ 3m^2}{k^2+m^2}\right)\non
		&= 	-\frac{1}{\left(p^2+m^2\right)^2}\frac{2m}{N}\left(\Lambda+\frac{p^2+3m^2}{2m}\right).\label{3.propArc}
		\end{align}
		First the power-law divergence is subtracted off and labeled as $\Lambda$
		\begin{align}
		\Lambda \equiv \int \frac{dk}{2\pi}1.\label{3.regularization}
		\end{align}
		Then the logarithmically divergent part linear in $k$ in the numerator is taken to vanish due to symmetric integration, and only a finite part remains.
		
		This may seem a rather naive prescription, but in fact it is closely related to the rules of dimensional regularization. The main practical difference from a dimensional regularization treatment of the integral is that in this case the power law divergences are kept track of through $\Lambda$ instead of being taken to vanish. This is useful since all power law divergences either cancel between diagrams, or have a real interpretation as contact terms, as we will discuss later.
		
		There may however be certain multi-loop integrals that will appear to give different results under different orders of integration in this scheme, and in that case we can fall back on ordinary dimensional regularization to make sense of them. These problematic integrals will actually be related to problems with naive use of equations of motion in calculating expectation values, and much more will be said about them in Section \ref{4.secEqM}.
		
		Returning to the issue at hand of calculating the correction to the $n$ propagator, we now consider the tadpole diagram in Fig \ref{2.Fig n and n2}.
		\begin{align}
			\langle n(-p)n(p)\rangle_{\text{tadpole}}&=\frac{1}{\left(p^2+m^2\right)^2}\frac{2m}{N}\times 4m^3\int\frac{dp}{2\pi}\frac{dk}{2\pi}\frac{(p-k)^2+4m^2}{(k^2+m^2)(p^2+m^2)^2}\non
			&=\frac{1}{\left(p^2+m^2\right)^2}\frac{2m}{N}\left(\Lambda+2m\right).\label{3.propTad}
		\end{align}
		So adding both diagrams, the $\Lambda$ divergences indeed cancel, and the total result including the zeroth order propagator is
		\begin{align}\langle n(-p)n(p)\rangle=\frac{1}{p^2+m^2}\left(1-\frac{1}{N}\frac{p^2-m^2}{p^2+m^2}\right)=\frac{1-N^{-1}}{p^2+m^2(1-N^{-1})^2}+\order{N^{-2}},\label{3.n propagator}\end{align}
		where the second line follows from the standard trick of taking the self-energy correction to repeat as geometric series to bring it to the denominator.
		
		So the renormalized mass to this order is $M_1=m(1-N^{-1})$, which is already the exact answer in \eqref{1.ONSpectrum}. The correction to the amplitude, $Z_1=1-N^{-1}$, is also what is necessary to ensure that
		$$\langle n(0)\cdot n(0)\rangle =N\int \frac{dp}{2\pi}\frac{1-N^{-1}}{p^2+m^2(1-N^{-1})^2}=\frac{N}{2m}=\frac{1}{g^2},$$
		which is consistent with the constraint.
		
		\subsection{Full spectrum to $\order{N^{-1}}$}\label{3.sec spectrum}
		As reviewed in section \ref{1.secPathInt}, the $j$-body masses $M_j$ in the spectrum may be found from the poles of the 2-point function of the operators
		$$O_j\equiv n_{i_1}n_{i_2}\dots n_{i_j},$$
		where all indices are required to be distinct.
		
		Using a superscript to indicate the order in the large $N$ expansion, the $\langle O_j(-p)O_j(p) \rangle^{(0)}$ diagram is just a set of $j$ bare propagators, all with mass $m$, beginning and ending on the same point. It is useful to generalize this slightly and allow the lines to have distinct masses $m_i$. Then the convolution integral of all the propagators evaluates to
		\begin{align}
			\frac{\sum_i m_i}{2^{j-1}\prod_i m_i}\frac{1}{p^2+\left(\sum_i m_i\right)^2}.\qquad\text{(propagator for $j$ non-interacting masses)}.\label{3.nonInt prop}
		\end{align}
	This is easy to show either by considering the problem in real space, where it simply involves multiplying exponentials, or by using a simple induction argument in momentum space.
	
	So in particular considering  $\langle O_j(-p)O_j(p) \rangle^{(0)}$ where $m_i=m$ for all $i$, we see that the zeroth order mass is naturally just $M_j^{(0)}=jm$. The next order correction $\langle O_j(-p)O_j(p) \rangle^{(1)}$ involves the correction of all the single particle propagators, and also a possible 2-body interaction as in Diagram B of Fig \ref{2.Fig n and n2}. At $\order{N^{-2}}$ 3-body interactions and many more types of 2-body interaction diagrams will come into play, but at $\order{N^{-1}}$ this single 2-body interaction diagram is all that can be written.

	So first let us calculate the correction to the 2-body propagator $\langle O_2(-p)O_2(p) \rangle$, which already has all these features. The correction arising from the single particle propagators involves correcting the mass $m$ to $M_1$ and multiplying by two corrected amplitudes $Z_1^2=(M_1/m)^2$ (see \eqref{3.n propagator}), and the correction arising from the 2-body interaction in Fig \ref{2.Fig n and n2} is straightforward given the Feynman rules developed so far. The result is,
	\begin{align}
				\langle O_2(-p)O_2(p) \rangle = \frac{1}{m}\frac{1-\frac{2}{N}}{p^2+4m^2}.\label{3.o2 prop}
	\end{align}
Note the mass $M_2=2m$ agrees with the exact spectrum \eqref{1.ONSpectrum}, and the integral $$\int \frac{dp}{2\pi}\langle O_2(-p)O_2(p)\rangle = \frac{1}{\left(2m\right)^2}\left( 1-\frac{2}{N}\right)$$
agrees with the calculation of the expectation value $\langle n_in_j \,n_i n_j\rangle$ in the $d=0$ path integral in Sec \ref{2. Sec two-body}.

Now the result for the full spectrum follows from combining this 2-body result, the formula for convoluting propagators of non-interacting masses \eqref{3.nonInt prop}, and some combinatorics. Since heavy use will be made of the convolution formula, some simplifying notation will be helpful. The correlation functions will be abbreviated $$g_j(p)\equiv \langle O_j(-p)O_j(p)\rangle,$$
and the convolution will be denoted by an asterisk $\ast$. For instance,
$$\left(g_1 \ast g_2\right)(p) = \frac{M_1+M_2}{2 M_1 M_2}\frac{Z_1 Z_2 }{p^2+\left(M_1+M_2\right)^2}.$$

Then the $j$-body correlation function may be expressed as
\begin{align}
	g_j= g_1^{\ast j}+\frac{j(j-1)}{2}\left( g_1^{\ast \left(j-2\right)}\ast g_2 - g_1^{\ast j}\right) +\order{N^{-2}}.\label{3. gj convolution}
\end{align}
The first term involves only correcting single lines, and the second term involves choosing two lines to interact as a $g_2$ propagator and subtracting out the non-interacting case to avoid overcounting. Using \eqref{3.nonInt prop} (perhaps with uncorrected propagators wherever appropriate to this order) and simplifying, this becomes
\begin{align} g_j(p)=&\frac{jm}{2^{j-1}m^j}\frac{1-\frac{1}{N}\left(j^2-2j+2\right)}{p^2+\left(jm\right)^2\left(1+\frac{j-2}{N}\right)^2}+\order{N^{-2}}.\label{3. gj 1st order}
\end{align}
So we have the spectrum
\begin{align}
	M_j = jm + \frac{m}{N}j(j-2)+\order{N^{-2}}\label{3.spectrum}
\end{align}
which agrees with the known result \eqref{1.ONSpectrum}. Also the amplitude will agree to this order with the exact result \eqref{5.Aj} found later in Sec. \ref{5}.

Recall the discussion in Sec \ref{1.Sec spectrum} on the appearance of only 2-body interactions in the spectrum, which evokes the situation for integrable theories in $d=2$ \cite{Zamolodchikov:1978xm}. From the perspective of this section, the lack of any higher $n$-body interactions is simply due to the fact that only 2-body interaction diagrams can be constructed at this order in the large $N$ expansion, and the fact that the spectrum is exact at this order. But of course the latter point isn't obvious at all in the diagrammatic perspective!

To show that $\order{N^{-2}}$ corrections to the spectrum indeed do vanish from this perspective, it will be useful to consider general features of the diagrams in the $O(N)$ model, most of which hold for any dimension. This will be the focus of the first half of Sec \ref{4} below. Even after showing that the corrections vanish to the next order in Appendix \ref{A}, this falls short of a diagrammatic proof that it vanishes to all orders. A somewhat more substantial step along these lines is given in Sec \ref{5} in which a recursion relation for the spectrum is found without making any large $N$ approximations, but still only using time-ordered correlation functions, in keeping with the paradigm of this paper.

\section{Generalities in the $O(N)$ model}\label{4}

	\begin{figure}
	\centering
		\includegraphics[width=0.8\textwidth]{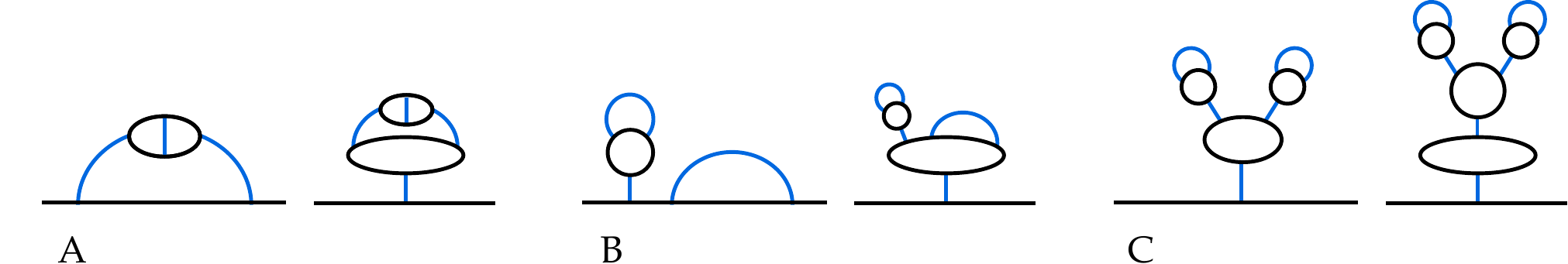}\caption{A few $\order{N^{-2}}$ corrections to the $n$ propagator. The right diagram in C is not a valid diagram, see the discussion surrounding Fig. \ref{3.Fig Excluded diagrams}.}\label{4.fig tadpole cancellation}
\end{figure}
An initial motivation for considering the diagrams in the $O(N)$ model a bit more carefully is to see if we can simplify the calculation of the $\order{N^{-2}}$ corrections, or even better, to try to find a diagrammatic proof that the spectrum at $\order{N^{-1}}$ is exact. It turns out that a purely diagrammatic proof is not so easy to find, but as a consolation prize we will get a better understanding of how the constraint and equations of motion in the $O(N)$ model are manifested in terms of diagrams.

\subsection{Tadpoles, non-tadpoles, and their relation to the constraint} \label{4.sec tadpole}
Let us first review what we know non-perturbatively about the two-point function $\langle n(-p)n(p)\rangle$. We know that due to the constraint its integral over momentum is fixed to be $1/(Ng^2)$. Furthermore we know that it only has a single pole, due to both the constraint and symmetry considerations (recall the discussion in Sec \ref{1.secPathInt}). So its most general possible form is just
$$\langle n(-p)n(p)\rangle=\frac{M_1}{m}\frac{1}{p^2+M_1^2}.$$
We are looking for some clever way to use this in order to argue that $M_1$ is exact at $\order{N^{-1}}$. Consider the self-energy $\Pi$, defined as usual by $\langle n(-p)n(p)\rangle=\left(p^2+m^2+\Pi(p^2)\right)^{-1}$,
\begin{align}
	\Pi(p^2)=-\frac{M_1^{(1)}}{m}\left(p^2-m^2\right)+\left[\left(\frac{M_1^{(1)}}{m}\right)^2-\frac{M_1^{(2)}}{m}\right]p^2+\frac{M_1^{(2)}}{m}m^2+\order{N^{-3}}\label{4.self energy}
\end{align}
The diagrams which lead to this self-energy may be divided into two classes: \emph{tadpoles}, which are insertions of a single $\lambda$ field into the propagator, and (for lack of a better word) \emph{non-tadpoles} which are any other correction. Only non-tadpoles may lead to the $p^2$ dependent terms in the self-energy. So it is clear that we do not need to calculate all $\order{N^{-2}}$ corrections to the propagator in order to determine whether $M^{(2)}_1=0$. It suffices to consider only non-tadpole one-particle irreducible diagrams such as the left diagram in Fig. \ref*{4.fig tadpole cancellation}.A.

Moreover, if we do know the value of the non-tadpole corrections then it is easy to calculate the tadpole corrections as well. The basic idea is illustrated in Fig \ref*{4.fig tadpole cancellation}, and was noticed also already in \cite{Bray:19740d}. Each non-tadpole diagram has a unique corresponding tadpole diagram. This is true even for one-particle reducible diagrams such as Fig. \ref*{4.fig tadpole cancellation}.B. The same trick can not be used to construct a new tadpole diagram from a previous tadpole diagram as shown in Fig. \ref*{4.fig tadpole cancellation}.C, since this produces an $n$ bubble, and these were already accounted for in formulating the bare $\lambda$ propagator.

If a non-tadpole contribution to the self-energy evaluates to the general form $Ap^2+B$, then the corresponding tadpole diagram will evaluate to $-Am^2-B$, leading to the sum $A\left(p^2-m^2\right)$. The cancelation of the constant part $B$ of a non-tadpole diagram with its corresponding tadpole diagram is a general feature of $O(N)$ models which holds in any dimension $d$, and is just due to the form of the $\lambda$ propagator in terms of $J(p)$ (defined in \eqref{3.J def} and \eqref{3. lambda prop}). For instance in $d=2$ this mechanism leads to the cancelation of power-law divergences between the two $\order{N^{-1}}$ corrections to the propagator (shown in Diagram A of Fig \ref{2.Fig n and n2}). And as we saw in Sec \ref{2.sec propagator}, in $d=0$ it leads to the cancelation of all corrections to $\langle n_i n_i\rangle$ since there is no $p$ dependent part of non-tadpole diagrams in $d=0$.\footnote{We see already in $d=0$ that this cancelation between tadpoles and non-tadpoles has something to do with the constraint $n^2=g^{-2}$, as we will elaborate further below.}

Given the coefficient of $p^2$ in brackets in Eq \eqref{4.self energy}, the self-energy due to $\order{N^{-2}}$ non-tadpole one-particle irreducible diagrams and their corresponding tadpoles must be
$$\left[\left(\frac{M_1^{(1)}}{m}\right)^2-\frac{M_1^{(2)}}{m}\right]\left(p^2-m^2\right).$$
Then if this is to be consistent with the form of \eqref{4.fig tadpole cancellation}, the tadpoles corresponding to \emph{reducible} diagrams as in Fig \ref*{4.fig tadpole cancellation}.B must lead to self-energy $+\left(M_1^{(1)}\right)^2$. And indeed, a direct calculation using the $\order{N^{-1}}$ self-energy confirms this. One might at first hope to use this same idea at $\order{N^{-3}}$ to find a contradiction if $M_1^{(2)}\neq 0$, but in fact the calculation ends up being consistent regardless of the value of $M_1^{(2)}$.

To better see what is happening here, consider a general tadpole to all orders, which is equivalent to considering the vacuum expectation value (VEV) $\langle \lambda\rangle$.
	\begin{figure}[t]
	\centering
	\includegraphics[width=0.8\textwidth]{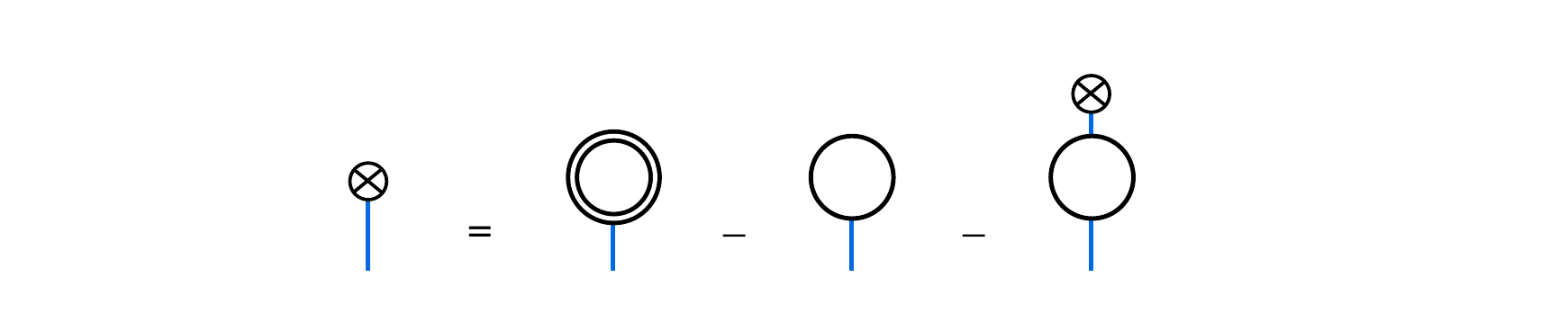}\caption{An equation for $\langle \lambda\rangle$, indicated as a $\lambda$ line ending in a crossed circle representing general interactions. A general interaction can be written in terms of the exact $n$ propagator, indicated by a double black line.}\label{4.fig constraint}
\end{figure}

In Fig \ref*{4.fig constraint} $\langle \lambda\rangle$ is written in terms of a loop of the exact propagator with the bare propagator and tadpole corrections subtracted out, since these terms would lead to the disallowed diagrams of Fig \ref{3.Fig Excluded diagrams}. Writing this as an equation, and keeping the dimension $d$ general, we find,
\begin{align}
	\langle \lambda\rangle &= \frac{i}{J(0)}\left[\left(\int \meas{k}{d}\langle n(-k)n(k)\rangle\right)-\frac{1}{Ng^2}-i\int \meas{k}{d}\frac{1}{\left(k^2+m^2\right)^2}\langle \lambda\rangle\right]\non&\qquad\qimplies \int \meas{k}{d}\langle n(-k)n(k)\rangle =\frac{1}{Ng^2}.
\end{align}
In other words, the fact that every tadpole diagram can be written in terms of some non-tadpole correction of the propagator is equivalent to enforcing the constraint $\lim_{x\rightarrow 0}\langle n(x)\cdot n(0)\rangle= g^{-2}$ in the exact correlation function. This is true even in QFT for higher $d$, where there are divergences so the integral over $k$ or limit as $x\rightarrow 0$ must be understood with some form of regularization present.

When the $\lambda$ propagator is such that the equation represented by Fig \ref{4.fig constraint} is only approximate, which will be the case in the linear sigma model in Sec \ref{6}, then the constraint on the two-point function will also only be approximate, as we will show in Eq \eqref{6. relaxed constraint}. For our purposes here, the constraint on the two-point function is equivalent to choosing the $Z_1$ factor to be $Z_1=M_1/m$, which led to the particular form of the self-energy in \eqref{4.self energy}. So from this perspective it is not surprising that the form of the self-energy together with the relation between tadpoles and non-tadpoles can not be used to restrict the pole $M_1$ further.
\subsection{The $\lambda$ propagator and contact terms from constraints}\label{4.sec lambda and contact terms}

The special simplifications that occurred in the last section which led to the diagrammatic enforcement of the constraint had to do with the bare $\lambda$ propagator \eqref{3. lambda prop} being proportional to the inverse of $J$, which is essentially a single bubble of bare $n$ propagators. It would be useful if this could be extended so that the exact $\lambda$ propagator which includes higher order large $N$ corrections is also equal to the inverse of a corrected $n$ bubble. In fact this will turn out to be true, with some minor caveats. Note that the diagrams at $\order{N^{-2}}$ become much more complicated and numerous. But for instance the diagrams of Fig \ref{4.fig tadpole cancellation}.A can just be understood as the $\order{N^{-1}}$ diagrams in Fig \ref{2.Fig n and n2} with a corrected $\lambda$ propagator, so it is useful to understand this propagator better.

\begin{figure}[t]
	\centering
		\includegraphics[width=0.7\textwidth]{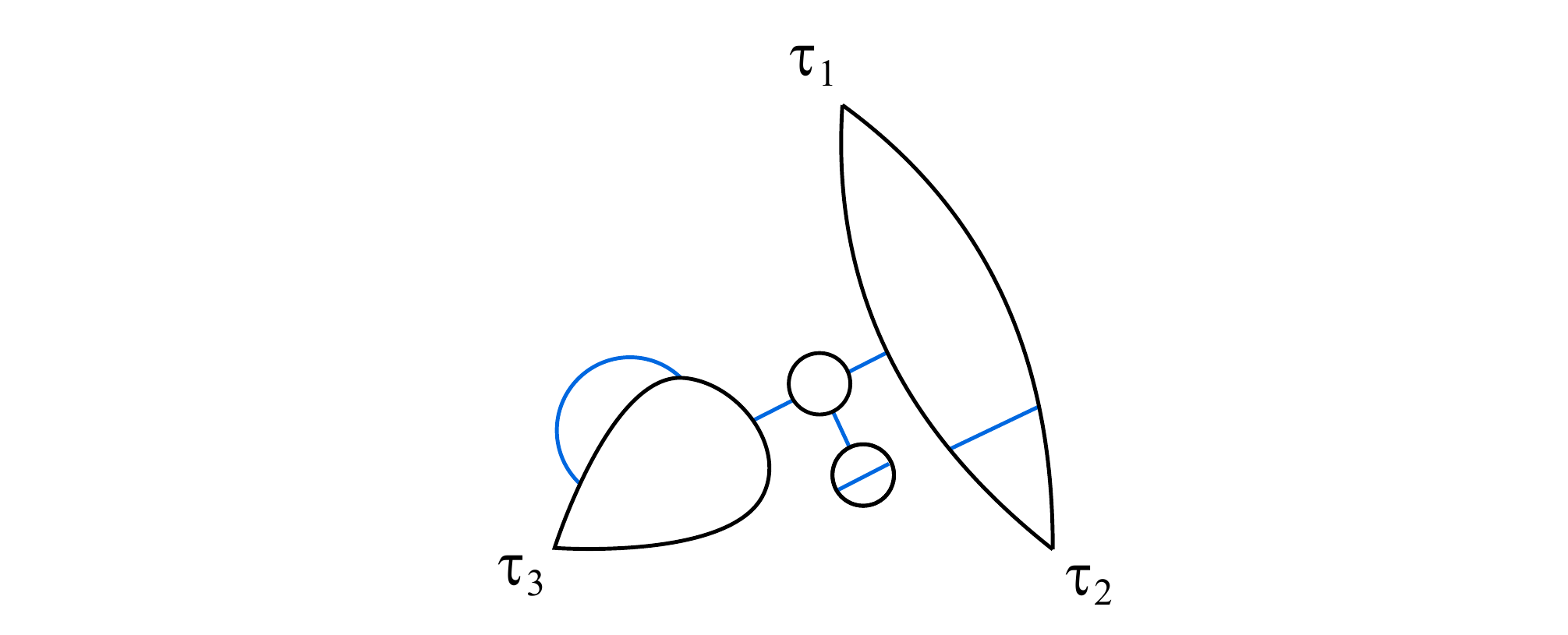}\caption{A rather arbitrary diagram which connects the factors in the correlation function $\langle n^2(\tau_1)n^2(\tau_2)n^2(\tau_3)\rangle$. Somehow all connected diagrams must conspire to cancel.}\label{4.fig connecting n2 factors}
\end{figure}

We may also consider what might at first seem to be an entirely unrelated question. A correlation function such as $\langle n^2(\tau_1)n^2(\tau_2)n^2(\tau_3)\rangle$ must factor into a disconnected product of expectation values $\langle n^2\rangle^3$ since by the constraint the operators $n^2$ are just proportional to a constant. But diagrams connecting the different $n^2$ factors can be drawn, for instance see Fig \ref*{4.fig connecting n2 factors}. Somehow the sum of all the diagrams which connect the factors of $n^2$ must vanish, and it would be interesting to understand better how this happens.

\begin{figure}[t]
	\centering
	\includegraphics[width=0.8\textwidth]{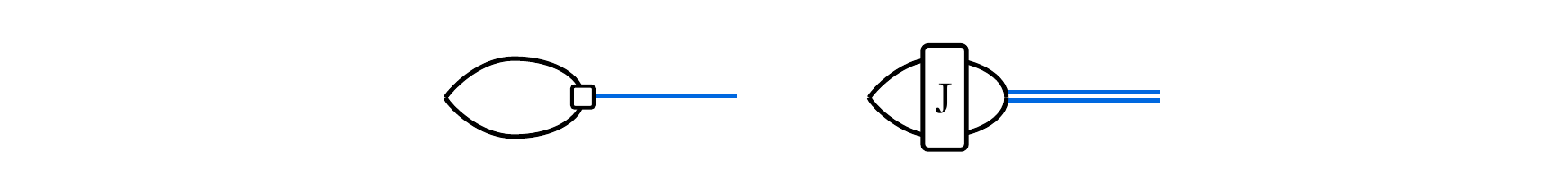}\caption{The connected correlation function $\langle n^2(\tau)\lambda(0)\rangle_\text{connected}$. On the left the boxed $n^2\lambda$ vertex represents the sum of arbitrary connected sub-diagrams. This can be rewritten as the form on the right, where the box labeled $J$ represents the sum of arbitrary sub-diagrams which can't be disconnected by cutting a single $\lambda$ line. The double blue line represents the full corrected $\lambda$ propagator.}\label{4.fig n2lambda connected}
\end{figure}

Both problems mentioned above are elucidated by considering the amplitude $\langle n^2(\tau)\lambda(0)\rangle$. This will indeed have a disconnected piece $\langle n^2\rangle\langle \lambda\rangle$, but as above, there will also be connected diagrams which we can draw, shown in general in Fig \ref*{4.fig n2lambda connected}.

Despite the constraint, the connected diagrams in this case \emph{do not vanish}. This can easily be seen by considering the lowest order interaction vertex, which is just given by the uncorrected term $\frac{i}{2}n^2\lambda$ in the Lagrangian. The bubble of two $n$ propagators leads to $NJ(p^2)$, the $\lambda$ propagator gives $\frac{2}{NJ(p^2)}$, and the vertex (along with a combinatorial factor) gives $i$, for a total of
$$\langle n^2(-p)\lambda(p)\rangle^{(0)}_\text{connected}= 2i. $$

In fact this is the exact result. All higher order corrections vanish. That is because this correlation function is only non-zero due to a contact term associated to the constraint as an equation of motion.

Consider the standard derivation of contact terms from the path integral. The path integral of a total variation in $\lambda$ vanishes, as is true for ordinary integrals,
\begin{align}
	0&=\frac{1}{Z}\int \mathcal{D}{n}\mathcal{D}{\lambda}\derDelta{}{\lambda(y)}\left[\lambda(x)e^{-S[n,\lambda]}\right]=\delta(x-y)+\frac{i}{2}\langle n^2(y)\lambda(x)\rangle-\frac{i}{2g^2}\langle \lambda\rangle,\nonumber
\end{align}
The last term is subtracting the disconnected piece $\langle n^2\rangle\langle \lambda\rangle$, so moving to Fourier space,
\begin{align}
	 \langle n^2(-p)\lambda(p)\rangle_\text{connected}=2i.\label{4.lambda n^2 contact term}
\end{align}
This now allows us to find the exact $\lambda$ propagator in terms of a suitable generalization of the bubble function $J(p^2)$. Considering Fig \ref{4.fig n2lambda connected}, a general connected interaction vertex on the left may be rewritten so that all 1-particle reducible $\lambda$ lines are grouped in the corrected $\lambda$ propagator, and the two $n$ lines may involve arbitrary interactions as long as they are 1-particle irreducible. This corrected $n$ bubble will be denoted as $NJ_\text{1PI}(p^2)$, and then the exactness of the contact term implies
\begin{align}
	\langle \lambda(-p)\lambda(p)\rangle = \frac{2}{NJ_\text{1PI}(p^2)}.
\end{align}
Nothing about this argument depended on $d=1$, so this is true in higher dimensional $O(N)$ models as well. It is useful to separate the exact $J_\text{1PI}$ into a lowest order term and corrections $\delta J$
$$J_\text{1PI}(p^2)\equiv J(p^2)+\delta J(p^2).$$
We see that $\frac{N}{2}\delta J$ just represents the self-energy of the $\lambda$ field,
$$\langle \lambda(-p)\lambda(p)\rangle = \frac{2}{NJ(p^2)}\left(1-\frac{N}{2}\delta J(p^2)\frac{2}{NJ(p^2)}+\left(\frac{N}{2}\delta J(p^2)\frac{2}{NJ(p^2)}\right)^2-\dots\right).$$
This is illustrated in Fig \ref*{4.fig lambda ON2} which shows the $\order{N^{-2}}$ corrections to the $\lambda$ propagator.

\begin{figure}[t]
	\centering
	\includegraphics[width=0.8\textwidth]{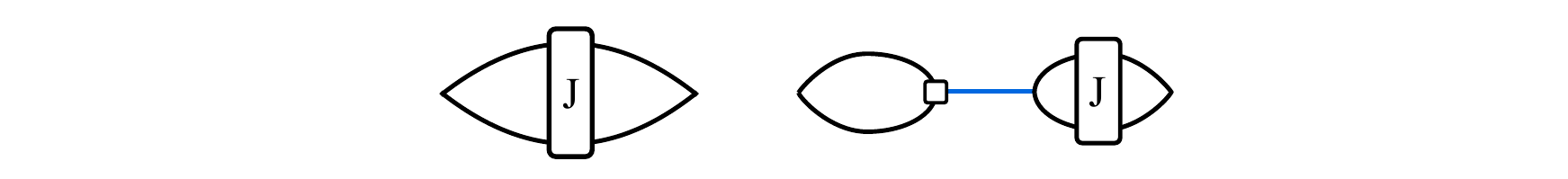}\caption{General diagrams involved in $\langle n^2(-p)n^2(p)\rangle_\text{connected}$. The notation is as in Fig \ref{4.fig n2lambda connected}.}\label{4.fig n2n2}
\end{figure}

Now we return to the second question presented at the beginning of this subsection. How do correlation functions like $\langle n^2(\tau_1)n^2(\tau_2)\dots\rangle$ factor? We will first show this for the two-point case $\langle n^2(-p)n^2(p)\rangle$ which is considered in Fig \ref{4.fig n2n2}. The connected diagrams can be divided into two cases: those which are 1PI, and those which can be written as a 1PI sub-diagram on the right (by convention) connected to a $n^2\lambda$ sub-diagram on the left. Since the $n^2\lambda$ sub-diagram evaluates to $2i$, and the interaction vertex connecting to the 1PI sub-diagram is $i/2$, the second class of diagrams exactly cancels the first. So indeed the connected part of $\langle n^2(-p)n^2(p)\rangle$ vanishes.

The argument for higher order correlation functions like the three-point diagram in Fig \ref{4.fig connecting n2 factors} is similar. Any connected $l$-point diagram can be organized into $m$ factors of $n^2$ connecting directly to an $l$-point 1PI sub-diagram, and the remaining $l-m$ factors of $n^2$ will connect to the 1PI sub-diagram indirectly via $n^2\lambda$ sub-diagrams. For instance, in the particular case of Fig \ref{4.fig connecting n2 factors}, the $\tau_1$ and $\tau_2$ vertices lie on a 3-point 1PI sub-diagram, and the $\tau_3$ vertex connects to this indirectly via a reducible $\lambda$ line. If we summed over all diagrams where $\tau_3$ connects reducibly to the same 1PI sub-diagram, we would just get an overall minus sign times that 1PI sub-diagram\footnote{In fact since the particular $n^2\lambda$ sub-diagram involving $\tau_3$ in Fig \ref{4.fig connecting n2 factors} is $\order{N^{-2}}$ it will cancel with all the other possible choices of $\order{N^{-2}}$ sub-diagrams, since the contact term is $\order{N^{0}}$.}, just as in the two-point case considered above. Returning to the general case, if we consider summing over all possible connected diagrams with the same 1PI sub-diagram, the overall combinatorial factor associated to diagrams with $m$ factors connected directly to the 1PI sub-diagram is
$$(-1)^{l-m}\frac{l!}{m!(l-m)!},$$
and the sum over $m$ vanishes since this is just the binomial expansion of $(1-1)^l$.

\subsection{Equations of motion and non-trivial contact terms}\label{4.secEqM}
So far in this section we have been primarily considering how the constraint $n^2=g^{-2}$ is manifested in terms of diagrams. The constraint is just the equation of motion associated to the $\lambda$ field. As we have just discussed, correlation functions involving both $n^2$ and $\lambda$ fields as in \eqref{4.lambda n^2 contact term} lead to contact terms which are associated to diagrams where the $\lambda$ field connects to the $n^2$ factor. Now we will consider instead the equation of motion associated to $n$ field itself, and we will once again see that diagrams where $\lambda$ connects to a $n^2$ factor that we would naively set equal to $g^{-2}$ are important.

The equation of motion associated to $n$ is
\begin{align}
	(-\partial_\tau^2 +m^2)n =i\lambda n.\label{4.eqMot}
\end{align}
A manipulation which is often done in $O(N)$ models in higher dimensions is to contract this equation of motion with another factor of $n$ at the same spacetime point, and use the constraint, which leads to
\begin{align}
	n\cdot \left(-\partial_\tau^2+m^2\right) n \cong g^{-2}i\lambda,\label{4.naive}
\end{align}
where the $\cong$ sign is used to indicate a naive equality which ignores contact terms.

This relation is quite useful as it relates the $\lambda$ field directly to an operator constructed in terms of the $n$ field alone, and despite the qualification ``naive," it does work in some sense.\footnote{See for instance \cite{Schubring:2021hrw} where this and similar expressions discussed in the supersymmetric $O(N)$ model are used to demonstrate cancelations of renormalon ambiguities in the operator product expansion.} The operator in terms of $n$ is also a rather physical quantity, it is related to the energy operator $H=-\frac{1}{2}\dot{n}^2$ found from classical action considerations (see Sec \ref{5}). The relation \eqref{4.naive} may be rearranged in terms of $H$ as
	$$	H \cong \frac{1}{2g^2}\left(m^2-i\lambda\right).$$
This relation might cause us some concern because it implies $\langle H\rangle$ is non-zero, but the ground state of the Laplace-Beltrami operator which is the presumably the Hamiltonian of this same system from the wavefunction perspective has zero energy. Of course since we are not keeping track of overall factors multiplying the partition function, we may only hope to find energies up to an overall constant. Moreover, there has been much discussion on the correct form of the $d=1$ path integral corresponding to a sigma model in curved space. There may be corrections to the naive Hamiltonian by constant multiples of the Ricci scalar or even non-covariant terms which depend on choice of regularization \cite{DeWitt:1952js,Bastianelli:1991be} (see also \cite{AbdallaBanerjee} and the references therein).
	
	In the case of $S^{N-1}$ such additional terms would only lead to an overall constant added to the Hamiltonian, but combined with the subtleties encountered below they may cause some doubt as to whether $H$ is really the proper Hamiltonian. This matter will be considered below in Sec \ref{5}.
	 The goal of the present section is to demonstrate the limitations of taking \eqref{4.naive} as an exact equality even if we make no identification of the Hamiltonian. Since in $d=1$ we have an exact expression for the correlation function of $n$, and can easily calculate the vacuum expectation value of $\lambda$ on the right hand side, it is easy to see how this relation fails. 

The exact correlation function is $\langle n(\tau)\cdot n(0)\rangle = g^{-2}e^{-M_1|\tau|}$, which implies
\begin{align*}
\langle \partial^2_\tau n(\tau)\cdot n(0)\rangle = g^{-2}M_1^2e^{-M_1|\tau|}-2g^{-2}M_1\delta(\tau),
\end{align*}
where the delta function is arising from the cusp due to the absolute value in the argument of the exponential. Now taking the limit $\tau\rightarrow 0$, and using $M_1=m(1-N^{-1})$, the saddle point condition $m=Ng^2/2$, and $\delta(0)=\Lambda$ (which is consistent with our regularization \eqref{3.regularization}),
\begin{align}
	\langle n \cdot(-\partial^2_\tau +m^2)n\rangle = N\Lambda + \left(-\Lambda+m\right)-\frac{m}{2N}.\label{4.nddn}
\end{align}
On the other hand, we have already calculated $i\langle \lambda\rangle$ to $\order{N^{-1}}$ as the tadpole correction to the propagator in \eqref{3.propTad}. So the right hand side of \eqref{4.naive} is
\begin{align}g^{-2}i\langle \lambda\rangle = g^{-2}\left(\frac{2m}{N}\left(\Lambda+2m\right)+\order{N^{-2}}\right)=(\Lambda+2m)+\order{N^{-1}}.\label{4.lambda vev}\end{align}
Even if we ignore the divergent terms proportional to $\Lambda$ and only consider the finite part, this is in no way equal to \eqref{4.nddn}, so the naive expression \eqref{4.naive} simply does not work. We will later calculate $\langle \lambda\rangle$ to next order and see that the terms at $\order{N^{-1}}$ also do not agree.

\begin{figure}[t]
	\centering
	\includegraphics[width=0.8\textwidth]{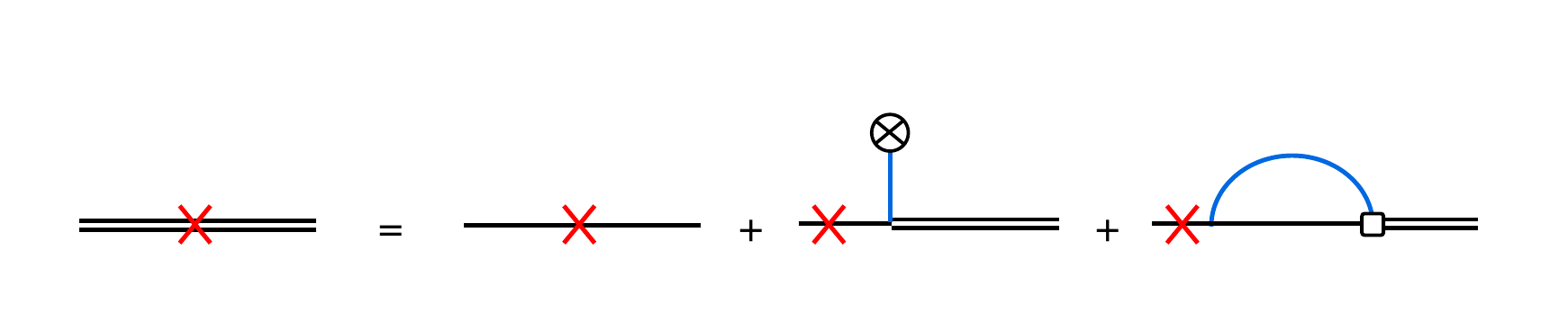}\caption{The Schwinger-Dyson equation. A red cross indicates the operator $-\partial^2+m^2$, which cancels out a bare propagator. The first term on the right hand side leads to the `trivial' contact term, the middle term leads to the naive relation \eqref{4.naive} and the last term leads to the `non-trivial' contact term.}\label{4.fig schwinger dyson}
\end{figure}

To see what is missing let us consider the equations of motion for the correlation function in diagrammatic form (i.e. the Schwinger-Dyson equations). In Fig.\ref{4.fig schwinger dyson} the diagrams associated to $\langle \left(-\partial^2+m^2\right)n(\tau)\cdot n(0)\rangle$ are shown. The differential operator $-\partial^2+m^2$ in each diagram cancels one bare propagator, which is indicated by a crossed out $n$ line. The lowest order diagram is a single propagator which is crossed out. The sum over $N$ components gives an overall factor of $N$, and the integration over all momentum when we set $\tau\rightarrow 0$ gives a factor of $\Lambda$. So this lowest order diagram just gives the term $N\Lambda$ in \eqref{4.nddn}, and we may refer to this as the \emph{trivial contact term}. If the trivial contact term is ignored, which would be the case in a point-splitting regularization where the two factors of $n$ are not taken at exactly the same point, then the naive relation does work at the very lowest order $\order{N^1}$. The divergence from this trivial contact term has been discussed in detail before in the $d=2$ case \cite{novik}.

\begin{figure}[t]
	\centering
	\includegraphics[width=0.8\textwidth]{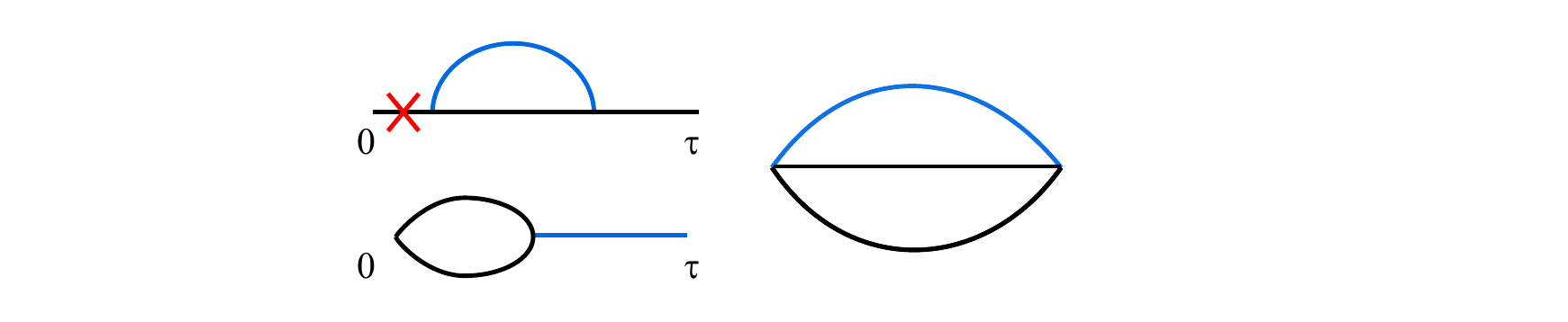}\caption{The non-trivial contact term. The top-left is the term arising from the Schwinger-Dyson equation. The bottom-left is the connected correlation function $\langle n^2(0)\lambda(\tau)\rangle$. As $\tau\rightarrow 0$ both diagrams formally go to the right two-loop diagram, but the evaluation depends on order of integration.}\label{4.fig nontrivial contact}
\end{figure}

All higher order diagrams involve an insertion of $i\lambda n$ next to the crossed out leg, and this is how the equation of motion \eqref{4.eqMot} is seen in diagrammatic form. There are diagrams where this $\lambda$ contracts into a tadpole, and these are what lead to the term $g^{-2}i\lambda$ in the naive expression \eqref{4.naive} when we set $\tau\rightarrow 0$. But besides these tadpole diagrams there are also clearly diagrams where the $\lambda$ field reconnects with the $n$ propagator. When we set $\tau\rightarrow 0$ these lead to terms we may refer to as \emph{non-trivial contact terms}. The lowest order diagram for a non-trivial contact term is shown in top left of Fig. \ref{4.fig nontrivial contact}.

This lowest order diagram is of course directly related to the arc diagram in the $n$ propagator at $\order{N^{-1}}$ which we have already calculated in \eqref{3.propArc}. Using that result, we evaluate the diagram to be
\begin{align}-\int\frac{dp}{2\pi}\frac{N}{p^2+m^2}\left(\frac{2m}{N}\int\frac{dk}{2\pi}\frac{k^2+4m^2}{(k-p)^2+m^2}\right)=-2\Lambda - m.\label{4.contact}
	\end{align}
So combined with the above result for $g^{-2}i\langle\lambda\rangle$ and the trivial contact term $N\Lambda$, this indeed produces the correct value for $\langle n \cdot(-\partial^2_\tau +m^2)n\rangle$ \eqref{4.nddn}, both for finite and divergent terms.

The reason we refer to these terms as contact terms is because they have the same origin as the contact terms associated to the constraint discussed in the previous section. The misstep in the derivation of the naive identity \eqref{4.naive} was setting $n^2=g^{-2}$ even though it was multiplying $\lambda$ at the same point. And formally at least, the two loop diagram in Fig. \ref{4.fig nontrivial contact} is also the limit of $\langle \lambda(\tau)\,n(0)\cdot n(0)\rangle$ as $\tau\rightarrow 0$, which is precisely the contact term we calculated earlier \eqref{4.lambda n^2 contact term}.
$$-\int\frac{dk}{2\pi}\frac{2m}{N}\left(k^2+4m^2\right)\left(\int\frac{dp}{2\pi}\frac{N}{\left((k-p)^2+m^2\right)\left(p^2+m^2\right)}\right)=-2\Lambda.$$

But we see that there is a subtlety in that the result depends on the order of integration. The non-trivial contact term \eqref{4.contact} is not just a delta function, there is a finite part as well, and this finite part is necessary to produce the correct VEV \eqref{4.nddn}. There is a bit of a puzzle as to how the two-loop diagram on the right of Fig \ref{4.fig nontrivial contact} can possibly depend on order of integration, since the manipulations in our regularization scheme are formally very close to dimensional regularization. If we integrate over $k$ first in \eqref{4.contact} using proper dimensional regularization, gamma functions and all,  there is no power-law divergence, but the finite result is indeed $-m$. If we integrate over $p$ first, the inner integral is convergent and proportional to $\left(k^2+4m^2\right)^{-1}$, so the outer $k$ integral is over a constant, which evaluates to zero by the rules of dimensional regularization.

A resolution of this apparent violation of the freedom to change order of integration in dimensional regularization is that it is invalid to take $d\rightarrow 1$ in the convergent inner $p$ integral before evaluating the outer $k$ integral. If the limit is not taken until the end it can be shown that both orders of integration produce the $-m$ result \cite{StackExchangeAnswer}. Still this is a somewhat formal resolution to the problem. After all the $k^2+4m^2$ factor originally came from $J(k^2)^{-1}$ in the $\lambda$ propagator, and as has been shown throughout this section the perfect cancelation of the two factors for any dimension $d$ is related to the equation of constraint. It may be better to simply consider the two-loop diagram in Fig \ref{4.fig nontrivial contact} as something which really doesn't have a well-defined value without some additional prescription in terms of a limit of a more well-defined diagram.

So the VEV \eqref{4.nddn} gets a finite correction from a two-loop diagram that depends on order of integration, and which in an alternate order can be understood as an ordinary contact term associated to the equation of constraint. This situation may seem rather strange but the same scenario will play out in the supersymmetric $O(N)$ model in Sec \ref{8.sec contact terms} as well, for two additional types of two-loop diagrams.

\subsection {Comparison with lattice regularization}\label{4.sec lattice}

To further consider the relation \eqref{4.naive} between $(\dot{n})^2$ and $\lambda$, it may be enlightening to consider some exact results which do not depend on the large $N$ expansion. If the $d=1$ $O(N)$ model is regularized on a lattice with spacing $a$ it becomes exactly solvable by directly integrating the partition function.

 The action becomes
 \begin{align}
S=\frac{a}{2}\sum_r \left[\frac{1}{a^2}(n_{r+1}-n_r)^2-i\lambda_i\left(n_r^2 - g^{-2}\right)\right].\label{4.lattice action}
 \end{align}
The correlation function for this system is found for instance in \cite{Mussardo:2020rxh},
$$\langle n_{r+s} \cdot n_r\rangle= \frac{1}{g^2}\left(\frac{I_{\frac{N}{2}}(\frac{1}{ag^2})}{I_\frac{N-2}{2}(\frac{1}{ag^2})}\right)^s,$$
where $I_N(x)$ is a modified Bessel function of the first kind, which has the asymptotic expansion as $x\rightarrow \infty$,
$$I_N(x)=\frac{e^x}{\sqrt{2\pi x}}\left[1-\frac{4N^2-1}{8x}+\frac{(4N^2-1)(4N^2-9)}{2(8x)^2}+\mathcal{O}(x^{-3})\right].$$
So expanding to order $a^2$,
\begin{align}
	g^2\langle n_{r+1}\cdot n_r\rangle=1-a\frac{g^2}{2}(N-1)+a^2\frac{g^4}{8}(N^2-4N+3)+\mathcal{O}(a^3g^{6})\label{4.corr function lattice}
\end{align}
In the continuum the correlation function satisfies, $\lim_{\tau\rightarrow 0^+}g^2\langle \partial_\tau{n}(\tau)\cdot n(0)\rangle= -M_1$. And indeed the exact lattice correlation function also satisfies,
$$\lim_{a\rightarrow 0}g^2\left\langle\left(\frac{n_{r+1}-n_r}{a}\right)\cdot n_r\right\rangle=-\frac{g^2}{2}\left(N-1\right),$$
which is the exact value $-M_1=-m(1-N^{-1})$.

Now consider the equations of motion given the lattice action \ref{4.lattice action},
\begin{align}
	-\langle n_r\cdot \partial^2 n_r \rangle_\text{lattice}&=\frac{N}{a}+i\langle \lambda_r n^2_r\rangle_\text{lattice},\qquad \partial^2 n_r\equiv \frac{n_{r+1}-2n_r+n_{r-1}}{a^2}.\label{4.eqMot lattice}
\end{align}
This is the analogue of the Schwinger-Dyson equation in Fig \ref{4.fig schwinger dyson}, where the first term is the trivial contact term, and the expectation value $\langle \lambda n^2\rangle$ can be decomposed as $\langle \lambda\rangle g^{-2}+\langle \lambda n^2\rangle_\text{connected}$, which are the tadpole term and the non-trivial contact term, respectively.

$\langle n_r\cdot \partial^2 n_r\rangle_\text{lattice}$ may be evaluated directly using \eqref{4.corr function lattice}, leading to 
\begin{align}\label{4.nddn lattice}
		-\langle n_r\cdot \partial^2 n_r \rangle_\text{lattice}=-\frac{1}{g^2}m^2+(N-1)a^{-1}+2m-\frac{3}{2}\frac{m}{N}.
\end{align}
This agrees with the continuum case \eqref{4.nddn} at leading order, and to all orders in the divergence $\delta(0)=\Lambda=a^{-1}$, but the finite part disagrees at sub-leading order. This is because in the lattice regularization the finite corrections to the VEV due to the non-trivial contact term do not appear.

In lattice regularization, the expectation value of the connected part of $\langle \lambda_r n_r^2\rangle$ is just an ordinary integral, and there is no problem with integrating by parts even though $\lambda_r$ and $n^2_r$ are at the same point.
\begin{align*}
\left\langle \lambda_r\left(n_r^2-\frac{1}{g^2}\right)\right\rangle_\text{lattice}=\frac{1}{Z}\int \dots d\lambda_r \, \lambda_r \,\left(-\frac{2i}{a}\right)\der{}{\lambda_r}e^{-S[n,\lambda]}=\frac{2i}{a},
\end{align*}
this is equivalent to the $2i\delta(0)$ from the ordinary contact term associated to the equation of constraint. The lack of any subtleties here are due to the fact that naturally in lattice regularization there can be no difference in the two `limits' represented in Fig \ref{4.fig nontrivial contact},
$$\lim_{s\rightarrow r}\left\langle \lambda_r n_s^2\right\rangle_\text{lattice}=\lim_{s\rightarrow r}\left\langle \lambda_r n_r\cdot n_s\right\rangle_\text{lattice}.$$

Finally, note that although the VEVs $\left\langle n\cdot \partial^2 n\right\rangle$ disagree at sub-leading order in the two regularizations, the VEV $\langle \lambda\rangle$ is exactly the same. This may be calculated from the contact term $2i/a$ above, and the two equations \eqref{4.eqMot lattice} and \eqref{4.nddn lattice},
\begin{align}
	g^{-2}i\langle \lambda\rangle_\text{lattice}=a^{-1}+2m -\frac{3}{2}\frac{m}{N}.\label{4.lambda vev lattice}
\end{align}
This agrees with the continuum value to the order calculated so far \eqref{4.lambda vev}. The next order term will also be shown to agree in Appendix \ref{A}.

So to summarize, the two-point correlation function is known exactly in both the continuum and lattice case. In both regularizations the VEV $\left\langle n\cdot \partial^2 n\right\rangle$ may be calculated exactly, leading to equations \eqref{4.nddn} and \eqref{4.nddn lattice} respectively. In both cases the VEVs are decomposed into a `trivial' contact term due the $n$ equation of motion at lowest order, a term proportional to $\langle\lambda\rangle$ which appears in the commonly seen relation \eqref{4.naive}, and a `non-trivial' contact term due to the equation of constraint at sub-leading order. Both regularizations agree in the first two terms, but differ in the non-trivial contact term. In the lattice case, this contact term looks like an ordinary delta function, but in the continuum case this involves additional finite corrections which are associated with a non-trivial limit as the component operators in $n(\tau)\cdot \partial^2 n(0)$ approach each other.

\section{Spectrum from Ward identities}\label{5}
After the previous discussion we might be wary of trusting the naive form of the Hamiltonian $H=-\frac{1}{2}\dot{n}^2$ to have the known eigenvalues \eqref{1.ONSpectrum} associated with the Laplace-Beltrami operator, since not only is the vacuum energy non-zero, it depends on regularization in a rather subtle way. But in fact in this section we will show that this is indeed a valid energy operator and gives us an alternate method to derive the spectrum without relying on any large $N$ approximations, but still within the framework of standard quantum field theoretic methods.

The Ward identity associated with the energy operator is derived in a standard way. Consider a time ordered correlation function involving $n(\tau_0)$ and arbitrary insertions of operators at other times, denoted by elipsis,
$$\langle n(\tau_0) \dots\rangle =\frac{1}{Z}\int \mathcal{D}n' \, n'(\tau_0)\dots e^{-\frac{1}{2}\int d\tau \left(\dot{n}'\right)^2}$$
We may change variables in the path integral to a field $n$ which is time translated with respect to $n'$,
$$n'(\tau)= n\left(\tau+\epsilon(\tau)\right)=n(\tau)+\epsilon(\tau)\dot{n}(\tau)+\order{\epsilon^2}.$$
Expanding the expression for the correlation function to first order in $\epsilon$ we find
\begin{align*}
	-\frac{1}{2}\der{}{\tau}\langle \dot{n}^2(\tau)n(\tau_0) \dots\rangle=\delta(\tau-\tau_0)\langle \dot{n}(\tau_0) \dots \rangle
\end{align*}
where in principle there should also be contact terms associated to the fields in elipsis, but we will integrate $\tau$ over a small interval around $\tau_0$ so these will not contribute:
\begin{align*}
\langle [H(\tau_0),n(\tau_0)] \dots\rangle=\langle \dot{n}(\tau_0) \dots \rangle,\qquad H(\tau)\equiv -\frac{1}{2}\dot{n}(\tau)^2.
\end{align*}
Here the operator $H$ has the proper commutation relations for a Hamiltonian since fields as a function of Euclidean time are defined as $n(\tau)=e^{H\tau}n e^{-H\tau}$. Strictly speaking, $H(\tau_0)$ in the commutator must be understood  as a limit from times above and below $\tau_0$ in order to get the proper operator ordering. Throughout this section we will not necessarily bother to indicate the times of the fields explicitly unless it affects the details of the calculation.

Now the goal of this section is to find the eigenvalues of $H$ directly from the combinatorics of time ordered correlation functions, without using large $N$ methods. The end result will be a recursion relation for the energy eigenvalues which agrees with the known spectrum. This will demonstrate that there are no issues with using the `naive' energy operator in the path integral here, despite all the subtleties involved in the canonical approach in constrained spaces \cite{Kleinert:1997em,AbdallaBanerjee}, and despite the subtleties involved in the calculation of $\langle H\rangle$ in the last section.

To begin, consider the following equal time expectation values of distinct components of $n$
\begin{align}
A_j&\equiv \langle n_1^2n_2^2\dots n_{j-1}^2 n_j^2\rangle\\
B_j&\equiv \langle n_1^2n_2^2\dots n_{j-2}^2 n_{j-1}^4\rangle.
\end{align}
These expressions are useful because we will want to consider the expectation value of $H$ with respect to an eigenvector $\hat{O}_j|0\rangle$ of energy $E_j$ (recall $O_j\equiv n_1 n_2\dots n_j$ with all indices distinct)
\begin{align}
\left\langle O_j [H, O_j]\right\rangle = \left(E_j-E_0\right)A_j\equiv M_j A_j.\label{5. Commutator}
\end{align}
The commutator ensures that the results do not depend on an overall constant shift $E_0$ in $H$.

There are many ways to calculate the values $A_j$. $A_2$ was already discussed in the context of the $d=0$ path integral in Sec \ref{2. Sec two-body}. Here they will be calculated by using the constraint $n_j^2 = g^{-2}-\sum_{k\neq j}n_k^2$, which implies
 $$A_j=\frac{1}{g^2} A_{j-1}-(j-1)B_j-(N-j)A_j.$$
 Next note that $B_j=3A_j$ which is easily seen in the path integral approach via Wick contractions. Thus,
 \begin{align}
 	A_j = \frac{g^{-2}}{N+2(j-1)}A_{j-1}.\label{5.Aj}
 \end{align}
which determines all $A_j$ through the initial condition $A_0=\langle1\rangle=1$.

Now first consider the subtracted term in the commutator in \eqref{5. Commutator}, which is not so different from calculation of $\langle H\rangle$ in the previous section
\begin{align}
	-\langle O_j^2 H\rangle=-\langle n_1^2\dots n_j^2\rangle\langle H\rangle=-\frac{1}{2g^2}A_jM_1^2.\label{5. term1}
\end{align}
Here we are considering a point-splitting regularization so the power law divergent terms in $\langle H\rangle$ arising from the cusp of the correlation function are ignored\footnote{From the perspective of last section, these divergences arise from the delta functions from both types of contact terms as well as the divergence in the VEV $\langle \lambda\rangle$. The finite part of the non-trivial contact term still implicitly appears here.}

The first term in the commutator $\left\langle O_j [H,O_j]\right\rangle$ may be written
\begin{align*}
\left\langle O_j H O_j\right\rangle=-\frac{1}{2}\sum_k\lim_{\tau_f,\tau_i\rightarrow 0}\partial_{\tau_f}\partial_{\tau_i}\left\langle O_j n_k(\tau_f)n_k(\tau_i)O_j\right\rangle.
\end{align*}
This is somewhat simpler for the terms $k>j$ since if the identity operator in the form of a sum over states is inserted between the two factors of $n_k$, only states with energy $E_{j+1}$ survive,
 \begin{align}
 -\frac{1}{2}\sum_{k>j}^N\lim_{\tau_f,\tau_i\rightarrow 0}\partial_{\tau_f}\partial_{\tau_i}\left\langle O_j n_k(\tau_f)n_k(\tau_i)O_j\right\rangle&= -\frac{1}{2}\sum_{k>j}^N\lim_{\tau_f,\tau_i\rightarrow 0}\partial_{\tau_f}\partial_{\tau_i}\left\langle O_{j+1}^2\right\rangle e^{-\left(M_{j+1}-M_{j}\right)\left(\tau_f-\tau_i\right)}\non
 &=\frac{1}{2}(N-j)A_{j+1}\left(M_{j+1}-M_{j}\right)^2.\label{5. term2}
 \end{align}
For $k\leq j$, there may be states with either energy $E_{j+1}$ or $E_{j-1}$ which survive in the middle identity operator. The state $O_j n_k|0\rangle$ for $k\leq j$ may be decomposed into $O_{j-1}|0\rangle$ (with norm $\sqrt{A_{j-1}}$) and some unit eigenvector $|E_{j+1}\rangle$, and the coefficients are not difficult to find in terms of the quantities $A$ and $B$,
$$O_j n_k|0\rangle =\frac{A_j}{A_{j-1}}O_{j-1}|0\rangle+\sqrt{B_{j+1}-\frac{A_j^2}{A_{j-1}}}|E_{j+1}\rangle.$$

So then the remaining terms in the Hamiltonian are
\begin{align}
	-\frac{1}{2}\sum_{k=1}^j\lim_{\tau_f,\tau_i\rightarrow 0}\partial_{\tau_f}\partial_{\tau_i}\left\langle O_j n_k(\tau_f)n_k(\tau_i)O_j\right\rangle&= \frac{j}{2}\left[\left(\frac{A_j^2}{A_{j-1}}\right)\left(M_j-M_{j-1}\right)^2+\left(3A_{j+1}-\frac{A_j^2}{A_{j-1}}\right)\left(M_{j+1}-M_{j}\right)^2\right].\label{5. term3}
\end{align}
Then combining all the terms \eqref{5. term1}\eqref{5. term2}\eqref{5. term3} into the relation $\langle O_j [H,O_j]\rangle = M_j A_j$, and using the recursion relation for $A_j$ \eqref{5.Aj}, we find a recursion relation for the spectrum,
\begin{align}
	M_j
	&=\frac{N}{4m} \left[(M_{j+1}-M_j)^2-M_1^2-\frac{(M_{j+1}-M_{j})^2-(M_j-M_{j-1})^2}{N+2(j-1)}\right].
\end{align}
Indeed the exact spectrum $M_j = jm+j(j-2)\frac{m}{N}$ satisfies this relation, so $H$ found from the path integral really is acting as a genuine energy operator despite the subtleties in calculating the vacuum expectation value. Conversely, if we take the validity of the $H$ as given, this gives a recursion relation which determines the entire spectrum once any given value $M_j/m$ is fixed as a function of $N$. For instance, $M_2$ may be solved in terms of $M_1$,
$$M_2=M_1\left(1+\sqrt{1+\frac{4m}{M_1(N-1)}}\right).$$ In an alternate history where these large $N$ and Ward identity methods were developed without knowledge of the wavefunction approach and the exact spectrum, the fact that $M_1=m(1-N^{-1})$ leads to an expression for $M_2$ without radicals (and indeed the entire spectrum to the calculated order does), would be taken as strong circumstantial evidence that  the spectrum at $\order{N^{-1}}$ is exact, but still fall short of a proof. Of course we could go further along in this Ward identity direction and look at the operators associated to generators of the $O(N)$ symmetry group. If we play around with correlation functions of these generators and polynomials of the field like the $O_l$ operators we should be able to reconstruct the well known ladder operator method for finding the spectrum. The line between standard QM methods and standard QFT methods is not so clear in this case precisely because Ward identities involve a notion of commutation relations.

\section{Linear sigma model and SCSA}\label{6}
Consider the linear sigma model, which involves an $N$-component unconstrained real scalar field $n_i$ with an ordinary quadratic kinetic term and quartic interaction term like
$$\kappa \left(n^2-\frac{1}{g^2}\right)^2.$$
Here the coefficient $\kappa$ is some parameter with dimensions of mass cubed in $d=1$. It is intuitively obvious that as the dimensionless parameter $\kappa g^{-6}$ becomes very large the constraint $n^2=g^{-2}$ is enforced and the model reduces to the $O(N)$ model which we have been discussing. This was first shown more precisely by Bessis and Zinn-Justin \cite{Bessis:1972sn}. Here we will show this a different way by putting the linear sigma model in a form in which the connection to the large $N$ limit of the $O(N)$ model is even more obvious. Then it is discussed how the various diagrammatic identities associated to the constraint in Sec \ref{4} are violated for a linear sigma model with finite $\kappa$. Finally, the SCSA \cite{BrayRickayzen1972,Bray:19740d} is discussed and it is put in a form which is valid in the hard constraint limit where $\kappa\rightarrow \infty$ and which differs somewhat from previous appearances in the literature on spin glasses \cite{AvoidedCritical1999,PrincipiKatsnelson2016}.

		    \subsection{$\lambda$ as a Hubbard-Stratonovich field}
		    Rewriting the coefficient $\kappa$ of the quartic term above in terms of a parameter $\mu$ with dimensions of mass, the partition function of the linear sigma model is
\begin{align}
	Z=\int \mathcal{D}n\, e^{-\frac{1}{2}\int d^2 x \Lagr},\qquad\Lagr= (\partial n)^2 + \frac{g^2\mu^2}{2}\left(n^2-\frac{1}{g^2}\right)^2.
\end{align}
$\mu$ is just the `bare mass' that appears upon expanding the quartic term\footnote{Any overall constants in the action will be ignored in equalities here.}, 
	$$\Lagr= (\partial n)^2 -\mu^2 n^2+ \frac{g^2\mu^2}{2}n^4.$$
	Now let's introduce a Hubbard-Stratonovich field $\lambda$ in order to cancel the the quartic term, $\mu_0$ will be an arbitrary parameter,
\begin{align}
	\Lagr	& =(\partial n)^2 -\mu^2 n^2+ \frac{g^2\mu^2}{2}n^4+\frac{1}{2\mu^2 g^2}\left(\lambda +i\mu_0^2- i g^2 \mu^2 n^2\right)^2 \non
	&= (\partial n)^2 +m^2 n^2+ \frac{1}{2\mu^2 g^2}\lambda^2 - i\lambda\left( n^2-\frac{\mu^2_0}{\mu^2 g^2} \right),\qquad m^2\equiv \mu_0^2-\mu^2.\label{6.lagr}
\end{align}
Except for the quadratic term for $\lambda$, which vanishes as $\mu\rightarrow \infty$, this looks suspiciously like the Lagrangian for the $O(N)$ model \eqref{3.ActionNL}. $\mu_0$ may be chosen so that $\langle \lambda\rangle = 0$ at the saddle point, which is equivalent to requiring the linear term in $\lambda$ to vanish when $n$ is integrated out. So the saddle point condition is
\begin{align}
	\frac{N}{2m}=\frac{\mu^2_0}{\mu^2 g^2} \qimplies \frac{m^3}{\mu^2}+m-\frac{Ng^2}{2}=0.\label{6.saddle point}
\end{align}
This is a cubic equation for $m$, but clearly the root which stays finite for $\mu/m\rightarrow \infty$ is
\begin{align}
m=\frac{Ng^2}{2} + \order{\frac{\left(Ng^2\right)^3}{\mu^2}}.
\end{align}
So indeed the saddle point mass $m$ converges to that of the $O(N)$ model.

So the large $N$ linear sigma model perturbation theory is diagrammatically exactly the same as the $O(N)$ model, but the bare mass $m$ is shifted from the $O(N)$ value by an $\mathcal{O}(m^3\mu^{-2})$ term, and the $\lambda$ propagator has correction for finite $\mu$ compared to the $O(N)$ case \eqref{3. lambda prop},
\begin{align}
\left[\frac{1}{2\mu^2 g^2}+\frac{N}{2}J(p^2)\right]^{-1}\qquad \text{($\lambda$ propagator).}\label{6. lambda prop}
\end{align}

This is actually a crucial difference, because it means that a $\lambda$ propagator will no longer exactly cancel the integral arising from a bubble of two $n$ particles, which was the source of all the identities associated to the constraint in Sec \ref{4}. This implies that the states $O_l|0\rangle$ considered previously are no longer eigenstates and involve a whole tower of radial excitations as well.

In particular, the scalar excitations may be found through the poles of the correlation function $\langle n^2(\tau) n^2(0)\rangle$. In Sec \ref{4.sec lambda and contact terms} it was shown that the connected part of this correlation function vanishes in the $O(N)$ model so there are no scalar states except the vacuum. The key point in that argument was that the connected correlation function $\langle n^2(\tau) \lambda(0)\rangle$ is exactly $2i$ to all orders, but that is no longer true given the form of the lowest order $\lambda$ propagator above.

Consider again the diagrams in Fig \ref{4.fig n2lambda connected} for  $\langle n^2(\tau) n^2(0)\rangle$ now understood as being for the linear sigma model with finite $\mu$. The sum of the lowest order diagrams where the boxes are trivial just leads to
	$$\frac{2N/m}{p^2+4m^2+\left(\frac{Ng^2}{2m}\right)2\mu^2}.$$
This implies that to lowest order in the large $N$ expansion the mass of the first scalar excitation $M_{\text{scalar}}$ is
\begin{align}
	M_{\text{scalar}}&=\sqrt{2}\mu\sqrt{\frac{Ng^2}{2m}+2\frac{m^2}{\mu^2}}+\order{N^{-1}}\non
	&=\sqrt{2}\mu + \mathcal{O}\left(\frac{\left(Ng^2\right)^2}{\mu}\right)+\order{N^{-1}}.
\end{align}	

This agrees with the naive mass found by expanding the Lagrangian as is usual for spontaneous symmetry breaking (SSB). One component of the $n$ field is chosen to equal $g^{-1}+\sigma$ so that the constraint $n^2=g^{-2}$ is satisfied up to a field $\sigma$ parametrizing radial fluctuations, and the remaining components become massless Goldstone modes $\phi$,
$$\Lagr_{\text{SSB}} = (\partial \phi)^2 + (\partial \sigma)^2 + 2\mu^2 \sigma ^2 + \Lagr_{\text{interaction}} .$$
So to lowest order in $m/\mu$ and large $N$, the mass $M_{\text{scalar}}$ found above agrees with the bare $\sigma$ mass in the Lagrangian. But the full expression for $M_{\text{scalar}}$ above includes fluctuations to all orders in the small parameter $m/\mu$, and it would be straightforward to consider higher order large $N$ diagrams of the form Fig \ref{4.fig n2n2} in order to find the $\order{N^{-1}}$ correction to $M_\text{scalar}$.\footnote{The next order correction should also contain a pole corresponding to the second radial excitation.}

\subsection{Self-consistent screening approximation}\label{6.sec SCSA}
Some of the earliest works on the large $N$ limit of the linear sigma model were not explicitly written in this form involving a Hubbard-Stratonovich field $\lambda$ and a saddle-point value of $m$. Instead the lowest order of the large $N$ limit was thought of as the Hartree aproximation, and the $\lambda$ propagators in the $\order{N^{-1}}$ diagrams were written as a infinite chain of $n$ field bubbles connected by the $\left (n^2\right)^2$ vertex. The restriction to these bubble chain diagrams was referred to as the \emph{screening approximation} \cite{Ferrell:1972zz}. Although the connection to the large $N$ limit was immediately realized, this perspective naturally led to an approach that focused on self-consistent equations between Green's functions \cite{BrayRickayzen1972,Bray:19740d} referred to as the self-consistent screening approximation (SCSA).

As will be shown below, the SCSA effectively contains all $\order{N^{-1}}$ corrections and partially contains corrections for all higher orders in $1/N$ as well, so it is often understood as being more accurate than any finite order of the large $N$ expansion for small values of $N$. This was explicitly tested in the $d=0$ case \cite{Bray:19740d}, and indeed there are many successful applications of the SCSA which are found consistent with other methods.\footnote{For instance consider a very recent application to elastic membranes \cite{MauriKatsnelson2020}.} But we wish to make the simple point that considering that the spectrum of the $O(N)$ model in $d=1$ is \emph{exact} to $\order{N^{-1}}$, partially including higher order corrections can only do worse than simply truncating the large $N$ expansion, even for arbitrarily small values of $N$. This statement is specific to $d=1$ and large values of $\mu/m$, and also to the specific problem of calculating poles of correlation functions, but it may be worth reconsidering the applicability in higher dimension as well.

\begin{figure}[t]
	\centering
	\includegraphics[width=0.8\textwidth]{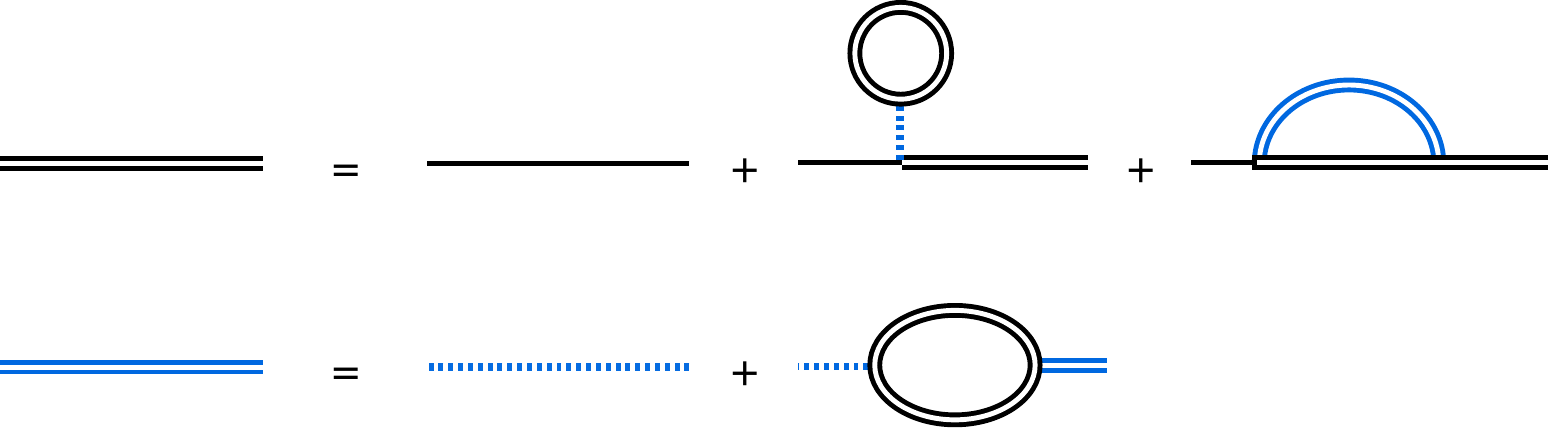}\caption{The self-consistent equations for $g(p)$ and $v(p)$ represented by double black and blue lines, respectively. The dotted blue line here refers to the bare propagator $2\mu^2 g^2$ for $\lambda$.}\label{6.fig SCSA}
\end{figure}

As a step in this direction, the SCSA will be reviewed and formulated in a way in which it is valid for the $O(N)$ model with an exact constraint as well. The SCSA is based on self-consistent equations for the single particle propagator $g(p)\equiv\langle n(-p)n(p)\rangle$ and the $\lambda$ propagator $v(p)\equiv \langle \lambda(-p)\lambda(p)\rangle$, which are expressed in terms of a self-energy $\Pi(p)$ and a bubble function $J(p)$,
\begin{align}
g(p)=\frac{1}{p^2+\mu_0^2-\mu^2+\Pi(p)},\qquad v(p)= \frac{1}{\left(2\mu^2g^2\right)^{-1}+\frac{N}{2}J(p)}.\label{6. Propagators}
\end{align}
Although similar notation is used for similar quantities elsewhere in this paper, throughout this section the functions $g, v, \Pi, J$ should be understood as being functions within the SCSA, which is expressed diagrammatically in Fig \ref{6.fig SCSA}:
\begin{align}
\Pi(p)&=\mu^2 g^2 \left(N\int\frac{dk}{2\pi}g(k)-\frac{\mu_0^2}{\mu^2 g^2}\right)+\int \frac{dk}{2\pi}v(k)g(k+p)\label{6. Pi eq}\\
J(p)&= \int\frac{dk}{2\pi}g(k)g(k+p).\label{6. J eq}
\end{align}
These equations are somewhat modified from what usually appears in the literature since the arbitrary parameter $\mu_0$ is included. This $\mu_0$ need not actually be set to the saddle point value \eqref{6.saddle point}. The lowest order of SCSA, which is referred to as the Hartree approximation in this context, will automatically produce the correct lowest order pole $m^2$,
\begin{align*}
	m^2\equiv \mu_0^2-\mu^2 +\Pi^{(0)}=\bcancel{\mu_0^2}-\mu^2 +\mu^2 g^2\left(N\int\frac{dk}{2\pi}\frac{1}{k^2+ m^2}-\bcancel{\frac{\mu_0^2}{\mu^2 g^2}}\right).
\end{align*}
Solving for $m^2$ just leads back to the saddle point equation \eqref{6.saddle point}. So we may as well just choose $\mu_0$ so that $\Pi^{(0)}=0$ and $m^2=\mu_0^2-\mu^2$ in the first place, in which case \eqref{6. Pi eq} may be written as
\begin{align*}
	\Pi(p)=\mu^2 Ng^2 \int\frac{dk}{2\pi}\left(g(k)-g^{(0)}(k)\right)+\int \frac{dk}{2\pi}v(k)g(k+p).
\end{align*}
As $\mu\rightarrow\infty$ it may seem like the first term could diverge, but recall from Sec \ref{4.sec tadpole} that in the $O(N)$ model limit the constraint $n^2=1/g^2$ implies that the integral of $g(k)-g^{(0)}(k)$ vanishes. So in fact the first term may have a finite limit as $\mu\rightarrow\infty$. The argument associated to Fig \ref{4.fig constraint} may be repeated in this case if the single blue lines are again understood as first order $\lambda$ propagators $v^{(1)}(p)$ \eqref{6. lambda prop} that take into account all lowest order bubble corrections. The result is
\begin{align}
\int\frac{dk}{2\pi}\left(g(k)-g^{(0)}(k)\right)=\frac{\langle \lambda\rangle}{\mu^2Ng^2 i}.\label{6. relaxed constraint}
\end{align}
Substituting this back into \eqref{6. Pi eq} leads to a form of the SCSA which is applicable even in the $\mu\rightarrow \infty$ $O(N)$ model limit.
\begin{align}
\Pi(p)=-i\langle \lambda\rangle+\int \frac{dk}{2\pi}v(k)g(k+p). \label{6.Pi eq lambda}
\end{align}

Of course, this equation could be inferred directly from the form of the self-energy in Fig \ref{6.fig SCSA} without considering integrals of $g$ as an intermediary. But the integral form is more useful for the case with finite $\mu$ since it involves one less parameter. Here, much like $g, v, \Pi,$ and $J$, $\langle \lambda\rangle $ should be understood as a quantity which only includes corrections implied by the SCSA self-consistent equations, and not the exact value. It should be adjusted to whatever value is necessary to ensure that the integral of $g(k)-g^{(0)}(k)$ vanishes. As discussed in Sec \ref{4.sec tadpole}, this is equivalent to including all tadpoles which correspond to non-tadpoles which are kept in the SCSA (including the non-1PI corrections). In higher dimensional $O(N)$ models these tadpoles cancel the power law divergences arising from the arc diagrams. In an earlier version of a SCSA for $O(N)$ models \cite{PrincipiKatsnelson2016} such a $\langle \lambda\rangle$ term was not considered, and strong dependence on the cutoff was noted, although it is possible this is due to unrelated effects.

It should be clear now from this discussion and the form of \eqref{6.Pi eq lambda} that the SCSA does include both of the $\order{N^{-1}}$ corrections to $g(p)$ shown in Diagram A of Fig \ref{2.Fig n and n2}. But at $\order{N^{-2}}$ already many diagrams are excluded. As noted in Sec \ref{4.sec lambda and contact terms} the exact self-energy of $\lambda$ propagator should involve the function $J_{\text{1PI}} $, which is qualitatively similar to the definition of $J$ in the SCSA, but for instance the exact $\order{N^{-2}}$ $\lambda$ propagator shown in Fig \ref{4.fig lambda ON2} contains two extra diagrams on the right which are missed in the SCSA approach. Considering the non-tadpole corrections to the self-energy of the $n$ propagator in Fig \ref{A.fig n}, only Diagram A is fully captured by the SCSA approach. As Diagram B involves the corrected $\lambda$ propagator it is also partially included in the SCSA. A similar story holds for the tadpole corrections due to the correspondence with non-tadpole diagrams, although the extra $\order{N^{-2}}$ tadpoles corresponding to two $\order{N^{-1}}$ corrections are also included.

As the order of the large $N$ expansion increases the diagrams which are included in the SCSA further thins out. Of course one of the purported advantages of the SCSA is that diagrams need not be considered at all. The functions $g$ and $v$ are solved for by iteratively using the SCSA equations, and now as discussed above we also introduce the parameter $\langle \lambda\rangle$ which is adjusted at each step to ensure $g$ obeys the constraint. If this procedure converges then $g$ is effectively including at least some diagrams to all orders of the large $N$ expansion, and for this reason the SCSA has the potential to perform better than a truncated large $N$ expansion for small values of $N$. But once again we point out that this need not always be true, and at least for the $O(N)$ model in $d=1$ truncating the expansion at the first sub-leading order is exact and performs better than the seemingly more sophisticated SCSA method.

\section{Gauge theories and $CP^{\bar{N}-1}$}\label{7}
Besides the $O(N)$ model which we have been discussing so far, there is another useful way to form a large $N$ extension of the $O(3)$ model with target space $S^2$. Odd dimensional spheres $S^{2\bar{N}-1}$ may straightforwardly be written in terms of $\bar{N}$ constrained complex fields $z_a$ rather than the real fields $n_i$ (e.g. $z_a = \frac{1}{\sqrt{2}}\left(n_{2a}+i n_{2a+1}\right)$). If we project the sphere $S^{2\bar{N}-1}$ in this notation onto the space formed by identifying complex unit vectors which only differ by an overall phase $z\sim e^{i\phi}z$, then by definition we recover complex projective space $CP^{\bar{N}-1}$. In the special case of $\bar{N}=2$ this map from $S^{3}\rightarrow CP^1$ is just the Hopf fibration, and $CP^1$ is just another formulation of the sphere $S^2$.

Unlike the $O(N)$ model for $N>3$, the $CP^{\bar{N}-1}$ model has instantons in $d=2$ and a Kahler target space which allows for extended supersymmetry. For this and many other reasons it was first introduced as a toy model in the high energy literature \cite{Eichenherr:1978qa}, but it also appears as an effective field theory in condensed matter physics (see e.g. \cite{Read:1990zza,IrkhinEtAl1996}). The Lagrangian takes the form
\begin{align}
	\Lagr= \partial z^\dagger \cdot \partial z + 2g^2\left(z^\dagger \cdot \partial z\right)^2, \qquad z^\dagger z =\frac{1}{2g^2}.
\end{align}
The additional term compared to the $O(N)$ model just serves to cancel any derivatives of $z$ in the phase direction, so that the model is gauge invariant under the transformation $z(\tau)\rightarrow e^{\phi(\tau)}z(\tau)$.

If $z_a$ is written in terms of explicit coordinates $\varphi^\beta$ on $CP^{\bar{N}-1}$ then this Lagrangian just reduces to the usual sigma model form $\frac{1}{2}g_{\alpha\beta}\dot{\varphi}^\alpha\dot{\varphi}^\beta$ with $g$ the proper Fubini-Study metric, so in $d=1$ the associated Hamiltonian of the system should indeed be the Laplace-Beltrami operator on $CP^{\bar{N}-1}$.

But the problem of finding the spectrum of this operator has an interesting result, which is reviewed in Appendix \ref{B}. Roughly speaking, the spectrum of the $CP^{\bar{N}-1}$ sigma model is exactly the same as the $O(2\bar{N})$ model it is related to, but states which are not invariant under the $U(1)$ gauge symmetry are filtered out. It is our goal in this section to see how this happens from the perspective of the path integral and large $N$ diagrams. To consider the problem in a more concrete form, the special case of $CP^1$ is first discussed.
\subsection{$CP^1$ model}
To introduce the $CP^{1}$ model, let us rewrite the $O(3)$ model in a different form. Define the two-component complex fields $z_a$ in terms of the three Pauli matrices $\sigma_i$ and the $O(3)$ field $n_i$,
\begin{align}
	n_i \equiv \frac{g}{4}z^\dagger \sigma_i z.
\end{align}
As usual, the indices on $z_a$ and the Pauli matrices $\sigma_{i,ab}$ will not be indicated unless necessary. The overall phase of $z$ is not fixed by this definition and can even be chosen to vary with time. The transformed field $z'(\tau)= e^{i\phi(\tau)}z(\tau)$ leads to the exact same $n_i(\tau)$ regardless of $\phi(\tau)$. This will mean that the theory described in terms of $z$ will have gauge invariance.

There is a constraint on $z$ arising from the constraint on $n$.
\begin{align}
n^2 = \frac{g^2}{16}\left(z^\dagger z\right)^2\qimplies z^\dagger z = \frac{4}{g^2}.\label{7.cp1 constraint}
\end{align} 
The Lagrangian may be expressed in terms of $z$ as well, by making use of the constraint and its derivative,
\begin{align}
	\frac{1}{2}\partial n\cdot \partial n = \frac{1}{2}\left[\partial z^\dagger \cdot \partial z + \frac{g^2}{4}\left(z^\dagger \cdot \partial z\right)^2\right].\label{7. action}
\end{align}
The quartic term may be eliminated by introducing the Hubbard-Stratonovich field $A$,
\begin{align}
	\Lagr&= \frac{1}{2}\left[\partial z^\dagger \cdot \partial z + \frac{g^2}{4}\left(z^\dagger \cdot z\right)^2+\frac{4}{g^2}\left(A+i\frac{g^2}{4}z^\dagger\cdot \partial z\right)^2\right]\non
	&=\frac{1}{2}\left(\partial +iA\right)z^\dagger \cdot \left(\partial - iA\right) z.
\end{align}
Now this looks quite a bit like a gauge theory, and indeed if introducing the quadratic term for $A$ is not to spoil the original gauge invariance of the theory for $z$, then $A$ must transform as
\begin{align}
	z\rightarrow e^{i\phi}z,\qquad A\rightarrow A+\partial \phi.
\end{align}
Gauge theories in $d=1$ are very simple. Since $A$ has only one component it can always be set to zero by a gauge transformation. Simply choose
$$\phi(\tau)=-\int_{\tau_0}^\tau d\sigma A(\sigma).$$
Then the Lagrangian becomes just $\frac{1}{2}\partial z^{\dagger}\cdot \partial z$. And since $z$ has four real components, this is seemingly identical to the $O(4)$ model, except that the normalization of the constraint is slightly different. But this was originally the $O(3)$ model, which has a different spectrum and correlation functions. How is this resolved?

Recall that the general spectrum with constraint $n^2=g^{-2}$ is $M_j = \frac{g^2}{2}j\left(j+N-2\right)$. So for the $O(4)$ model with the constraint \eqref{7.cp1 constraint}, the spectrum is
$$M_j = \frac{g^2}{8}j\left(j+2\right).$$
This means that the even levels $j=2l$ become
$$M_{2l}=\frac{g^2}{2}l\left(l+1\right),$$
which is exactly the spectrum of the $O(3)$ model. This makes sense since the $O(3)$ $n$ fields are a linear combination of two components of $z$ fields. But there are only 3 independent single particle states in the $O(3)$ model, whereas there are 9 independent two particle states in the $O(4)$ model. So not only are the odd energy levels of $O(4)$ entirely absent in $O(3)$, the multiplicity of the even levels is drastically reduced too.

In some sense what is happening is that the inclusion of the gauge field $A$ is filtering out everything but the gauge invariant sector of the theory, but it is leaving that sector completely unchanged. This may be understood from various points of view. In particular, how this occurs diagrammatically due to the additional $A$ fields in large $N$ diagrams is rather obscure, and that will be our main focus.

\subsection{Large $\bar{N}$ perturbation theory}
\begin{figure}
	\centering
	\includegraphics[width=0.8\textwidth]{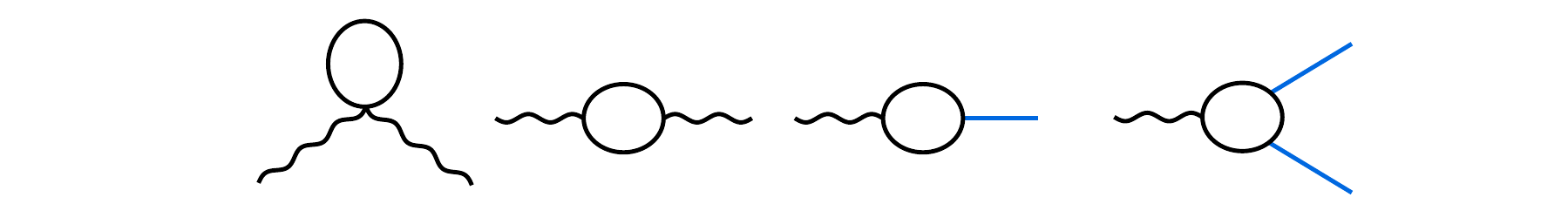}\caption{Some effective vertices involving the $A$ field, which is represented by a wavy black line. The $z$ field which is integrated out of the action is represented by an ordinary black line, much as the $n$ field. When relevant, an arrow will indicate the direction from $z$ to $z^\dagger$. Of the possible 3-point interactions, only the $A\lambda^2$ interaction on the right survives.}\label{7.fig 2point vanish}
\end{figure}

Once we extend to higher $\bar{N}$, $CP^{\bar{N}-1}$ will no longer be equivalent to a sphere, so rather than the constraint $\eqref{7.cp1 constraint}$ it will make more sense to choose a normalization which agrees with an ordinary ungauged $O(2\bar{N})$ model,
\begin{align}
\Lagr=\left(\partial +iA\right)z^\dagger \cdot \left(\partial - iA\right) z + m^2 z^\dagger \cdot z-i\lambda\left(z^\dagger \cdot z - \frac{1}{2g^2}\right).
\end{align}
This leads to the saddle point condition
\begin{align}
	m = \bar{N}g^2 = \frac{Ng^2}{2},
\end{align}
so this is consistent with our earlier conventions for the $O(N)$ model. Upon integrating out the $z$ field the $\lambda$ propagator ends up being the same as before \eqref{3. lambda prop} as well. But now there are additional possible vertices involving $A$, as shown in Fig \ref*{7.fig 2point vanish}. But in fact all the possible two-point $A$ diagrams vanish. In $d=2$ the 2-point diagram involving the $A^2 |z|^2$ vertex would combine with the diagram involving two $iA(z^\dagger \partial z -\partial z^\dagger  z)$ vertices to form the usual gauge invariant propagator in terms of the antisymmetric field strength tensor. But in $d=1$ there can be no non-vanishing field strength tensor and the diagrams simply cancel. The possible mixing term between $A$ and $\sigma$ also cancels due to symmetric integration. Higher order vertices such as an $A\lambda^2$ vertex seem possible (The $A^3$ vertex and $A^2\lambda$ vertex cancel), but without a bare quadratic term to form propagators it is difficult to see how to make use of this.

It is rather unclear at this point how the inclusion of the $A$ field in the large $N$ theory serves to remove states such as the fundamental representation $z_a|0\rangle$ from the spectrum, since it is not clear how to calculate diagrams with additional $A$ lines. So let us consider a seemingly ad hoc solution and put in a quadratic term for $A$ by hand,
\begin{align}
\Lagr_{\epsilon}=\frac{\epsilon}{2g^2}A^2.
\end{align}
$\epsilon$ is some dimensionless regulating parameter that will be set to zero at the end of the calculation. The factor of $g^2$ is correct on dimensional grounds and has the proper $N$ dependence so that $\Lagr_{\epsilon}$ is of the same order as the other terms in the Lagrangian upon integrating out $n$.

\begin{figure}
	\centering
	\includegraphics[width=0.8\textwidth]{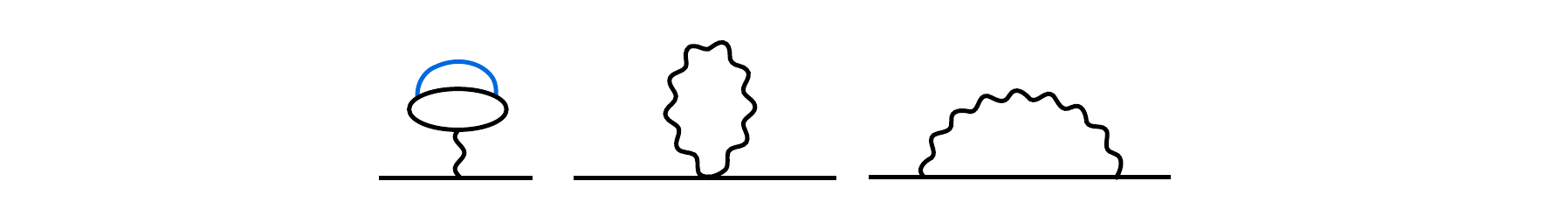}\caption{$\order{N^{-1}}$ corrections to $\langle z^\dagger(-p)z(p)\rangle$ involving the $A$ field. Diagrams involving effective $A^2\lambda$ and $A^3$ vertices which were shown to cancel in general are not included.}\label{7.fig n prop}
\end{figure}

Now consider the correction to the mass $M_1$ of a single particle (i.e. the fundamental representation). The additional $\order{N^{-1}}$ corrections to the self-energy are shown in Fig \ref*{7.fig n prop}. The tadpole-like diagram on the left ends up vanishing. Given that the bare propagator for $A$ is just a constant $g^2\epsilon^{-1}$, the correction due to the $A^2n^\dagger n$ vertex leads to an overall divergence, but this is canceled by the $A$ arc diagram. In total, the correction to the self-energy is
$$\Pi_\epsilon(p^2) = -\frac{g^2 }{2m\epsilon}\left(p^2-m^2\right).$$
Note that considering the discussion in Sec \ref{4.sec tadpole} we could guess a priori that this would have a form proportional to $p^2-m^2$ in order to be consistent with the constraint. Considering the general form of the self-energy \eqref{4.self energy}, this implies that single particle mass $M_1$ has a correction $\delta M_1$ due to the gauge field
\begin{align}
	\delta M_1 = +\frac{g^2}{2}\epsilon^{-1} +\order{N^{-2}}.\label{7. deltam1}
\end{align}
So as the regulator $\epsilon$ goes to zero the energy of the single particle state diverges, and in that sense the non-gauge invariant fundamental representation does indeed leave the spectrum.

\subsection{Spectrum of the squashed sphere sigma model}\label{7.sec squashed}

This regulating quadratic term $\Lagr_\epsilon$ actually has a natural physical interpretation. It modifies the target space of the sigma model to be a \emph{squashed sphere} which interpolates between the $S^{N-1}$ of the $O(N)$ model and $CP^{\bar{N}-1}$ \cite{SquashedSphereSigma}. The $A$ field may be integrated out of the action,
\begin{align}
	\Lagr&= \left(\partial +iA\right)z^\dagger \cdot \left(\partial - iA\right) z + \frac{\epsilon}{2g^2}A^2\non &\rightarrow \partial z^\dagger \cdot \partial z + \frac{2g^2}{1+\epsilon}\left(n^\dagger\cdot \partial n\right)^2.\label{7.squashed sphere}
\end{align}
When $\epsilon=0$ this is the form of the $CP^{\bar{N}-1}$ model in \eqref{7. action}. Any change of $z$ along the fibers of $S^{2\bar{N}-1}$ defined by the gauge orbits $e^{i\phi}z$ is completely canceled. When $\epsilon\neq 0$ this cancelation is incomplete, and the fibers have some finite size. As $\epsilon \rightarrow \infty$ the fibers take precisely the right size so that the $SU(\bar{N})$ symmetry of $CP^{\bar{N}-1}$ and the squashed sphere is extended to the $O(2\bar{N})$ symmetry of $S^{2\bar{N}-1}$ and the action clearly returns to the ordinary $O(2\bar{N})$ model.

Now let us continue to work with the action involving the gauge field $A$, and consider how it affects the spectrum in general. We may define the operators $$O_{r,s}\equiv z^\dagger_{a_1}z^\dagger_{a_2}\dots z^\dagger_{a_r}z_{b_1}\dots z_{b_s},$$
where the particular values of the $a$ and $b$ indices are not so important, but we require that $a_i\neq b_j$ for any $i,j$.\footnote{This requirement on the indices is to ensure that later on there is no mixing of states in the correlation functions, and is not important for the following discussion of gauge fixing.} Now consider the correlation function $\langle O^\dagger_{r,s}(\tau)O_{r,s}(0)\rangle$. If we change variables in the path integral so $z(\tau) = e^{i\int^\tau_{0} A(\sigma)d\sigma}z'(\tau)$, but leave $A$ fixed, the $A$ fields and $z'$ fields completely decouple so the correlation function factorizes,
\begin{align*}
\left\langle O^\dagger_{r,s}(\tau)O_{r,s}(0)\right\rangle=\left\langle e^{i(r-s)\int^\tau_0 A(\sigma)d\sigma}\right\rangle_{A \text{ action}}\left\langle O^\dagger_{r,s}(\tau)O_{r,s}(0)\right\rangle_{O(N)\text{ model} }.
\end{align*}
We see that if $r=s$ so that the operators $O_{r,s}$ are gauge invariant, there is no change in the correlation function compared to the ordinary $O(N)$ model. If $r\neq s$ there is an additional factor which only depends on the squashed sphere term $\Lagr_{\epsilon}$. If this term were absent, the expectation value would be rather ill-defined but it is still plausible that upon integrating over all $A$ this complex exponential would average to zero so that non-gauge invariant correlation functions simply vanish. But given the presence of the regulator $\Lagr_\epsilon$ we can be more precise. Using the propagator $$\left\langle A(\tau)A(0)\right\rangle = g^2\epsilon^{-1}\delta(\tau),$$
and Wick's theorem we can show
\begin{align}
	\left\langle O^\dagger_{r,s}(\tau)O_{r,s}(0)\right\rangle=e^{-(r-s)^2\frac{g^2}{2\epsilon}|\tau|}\left\langle O^\dagger_{r,s}(\tau)O_{r,s}(0)\right\rangle_{O(N)\text{ model} }.
\end{align}
So states in the squashed sphere model with $r\neq s$ get a correction to their mass $\delta M_{r,s}$ which diverges as $\epsilon \rightarrow 0$,
\begin{align}
	\delta M_{r,s}= +(r-s)^2\frac{g^2}{2}\epsilon^{-1}.\label{7.exact correction}
\end{align}
So in particular the $\order{N^{-1}}$ correction to the single particle mass \eqref{7. deltam1} that was calculated in the previous subsection was actually exact. And as we wished to show, gauge-invariant operators have the exact same correlation functions as $O(N)$, and non-gauge invariant states leave the spectrum as we go to the gauge-invariant $CP^{\bar{N}-1}$ limit.

This formula for the spectrum of the squashed sphere model will be derived from another perspective by considering the spectrum of the Laplace-Beltrami operator on the squashed sphere in Appendix \ref{B}. In the remainder of this section we will sketch briefly a third method to find it solely by considering large $N$ diagrams and some combinatorics, much as in Sec \ref{3.sec spectrum} for the ordinary $O(N)$ model.

\begin{figure}
	\centering
	\includegraphics[width=0.8\textwidth]{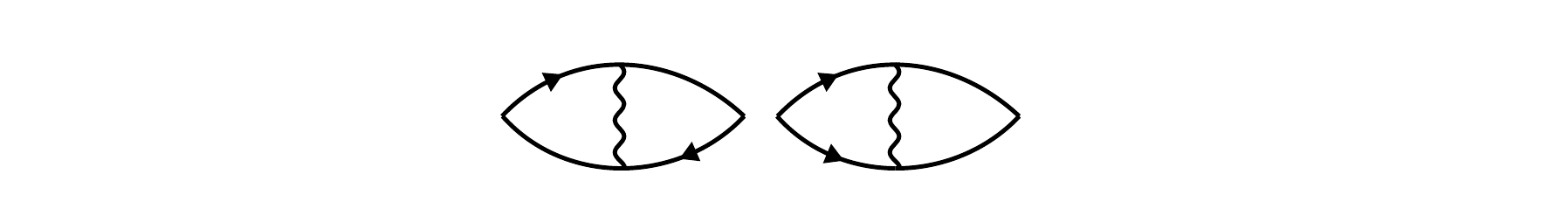}\caption{Two-body interactions due to the gauge field. The sign of the interaction depends on the orientation of the $z$ lines. Opposite charges attract.}\label{7.fig 2body}
\end{figure}

Exactly as in the $O(N)$ model, the corrections to the correlation function $\left\langle O^\dagger_{r,s}(\tau)O_{r,s}(0)\right\rangle$ at $\order{N^{-1}}$ come from corrections to the $r+s$ single particle lines, and also by considering interactions between two distinct lines as in Fig \ref*{7.fig 2body}. Higher $n$-body interactions don't appear at this order.

	Recall that the lowest order two-body correlation function is $m^{-1}\left(p^2+(2m)^2\right)^{-1}$. If a correction to the two-body mass $\delta M_2$ is to be consistent with the lowest order normalization\footnote{Note that the normalization itself has an $\order{N^{-1}}$ correction, but that is entirely accounted for by the $\lambda$ diagrams.} which is fixed by the constraint, it must have the form
	\begin{align*}
\frac{1}{m}\frac{1+\frac{\delta M_2}{2m}}{p^2+(2m+\delta M_2)^2}=\frac{1}{m}\frac{1}{p^2+4m^2}+\frac{p^2-4m^2}{2m^2(p^2+4m^2)^2}\delta M_2+\order{\delta M_2^2}.
	\end{align*}
	Evaluating the diagrams in Fig \ref*{7.fig 2body} leads to a correction precisely of this form, so the two-body interaction energy can be identified as $$\pm 2\frac{g^2}{2\epsilon}$$
	where like charges repel and opposite charges attract. So for the gauge invariant operators $O_{1,1}=z^\dagger_a z_b$ the large positive correction to the two single particle masses is precisely canceled by the large binding energy of the system, and overall the gauge field has no effect.
	
	A simple combinatorial argument allows us to find the correction $\delta M_{r,s}$ for arbitrary states $O_{r,s}|0\rangle$.
	\begin{itemize}
		\item There are $r+s$ single particle lines, each of which has a mass correction \eqref{7. deltam1}:
	$$(r+s)\frac{g^2}{2\epsilon}.$$
	\item There are $rs$ diagrams where two particles with opposite charge interact:
	$$-2rs\frac{g^2}{2\epsilon}.$$
	\item Finally, counting the number of combinations where particles with the same charge interact:
	$$+2\left(\frac{r(r-1)}{2}+\frac{s(s-1)}{2}\right)\frac{g^2}{2\epsilon}.$$
	\end{itemize}
	So in total, there is a correction of $\delta M_{r,s}=(r-s)^2\frac{g^2}{2\epsilon}$ which is exactly what was found above and in Appendix \ref{B} using more abstract arguments.
\section{Supersymmetric $O(N)$ model}\label{8}
Much as the Hamiltonian of an ordinary non-linear sigma model has a very geometric interpretation as the Laplace-Beltrami operator on the target space, the supersymmetric non-linear sigma model Hamiltonian has an interpretation as the Laplace-de Rham operator on the Hilbert space of differential forms on the target space \cite{Witten:1982df}. This system is often considered as it was in the original paper, through canonical quantization (see e.g. \cite{MirrorSymmetry}). But from a Euclidean path integral perspective, this structure is rather less clear to see.

In particular the Laplace-de Rham operator has exactly the same eigenvalues as the Laplace-Beltrami operator when acting upon 0-forms (i.e. ordinary functions). But the path integral in terms of a bosonic $n$ and fermionic $\psi$ is schematically like
$$Z_{\text{SUSY}}=\int \mathcal D n \mathcal D \psi\,\, e^{-S_B[n]}e^{-S_F[n,\psi]},$$
where $S_B$ is the original non-supersymmetric action and the $\psi$ fields are coupled to the $n$ fields in the new SUSY terms. To calculate a correlation function of the $n$ fields alone one may first integrate out $\int\mathcal D\psi e^{-S_F[n,\psi]}$, but it is not at all clear whether the resulting factor is independent of $n$, even though from the wavefunction perspective the poles of the correlation function are expected to be the same regardless of the presence of SUSY.

So we shall consider this question in the case of the SUSY $O(N)$ model, and show how the fermionic corrections to the bosonic correlation functions really do end up vanishing. Then we will consider the equations of motion in the SUSY $O(N)$ model and see that just as in the non-SUSY case there will be non-trivial finite contributions from contact terms that modify the naive equations relating vacuum expectation values in the model.

	\subsection{Zero-dimensional path integral}\label{8.sec 0d}
As was done in Sec \ref{2}, we will introduce the Lagrange multiplier part of the action and the corresponding diagrams in the context of the $d=0$ path integral first. The superfields in $d=0$ work much the same as in $d=1$, and both are quite similar to $\mathcal{N}=1$ SUSY in $d=2$. The two spin components of Majorana spinors in $d=2$ become two independent Grassmann numbers in lower dimensions. So a superfield $\Phi$ in $d=0$ is defined as a function of two Grassmann numbers $\theta, \bar{\theta}$.
\begin{align}
	\Phi(\theta,\bar{\theta})\equiv n+\bar{\theta}\psi+\bar{\psi}\theta +\bar{\theta}\theta F.
\end{align}
	Supersymmetry in $d=0$ simply means invariance under translations $$\theta\rightarrow \theta+\epsilon,\quad \bar\theta\rightarrow \bar\theta+\bar\epsilon.$$
	This induces transformations on the components of $\Phi$,
	\begin{align}
		\delta n = \bar{\epsilon}\psi+\bar{\psi}\epsilon,\qquad \delta \psi = \epsilon F,\qquad \delta \bar{\psi}=\bar{\epsilon} F,\qquad \delta F = 0.\label{8. SUSY trans}
	\end{align}
Note that the $F$ component by itself is completely unchanged under these translations. So a supersymmetric action in $d=0$ may be constructed out of polynomials of \emph{F-terms} which are themselves constructed by integrating polynomials of superfields over $\theta,\bar\theta$. In particular, a `kinetic term' of $-\frac{1}{2}F^2$ is a supersymmetric action.

To create a supersymmetric $O(N)$ model, the Lagrange multiplier terms $(m^2+\lambda)\left(n^2-g^{-2}\right)$ may be promoted to superfields.\footnote{Note that in this section a slightly different convention for $\lambda$ is used where a factor of $i$ is absorbed in the definition.} The superfield version of $\lambda$ is referred to as $\Gamma$.
	\begin{align}
\Gamma(\theta,\bar{\theta})\equiv \sigma+\bar{\theta}u+\bar{u}\theta +\bar{\theta}\theta \lambda_F.
	\end{align}
The $d=0$ $O(N)$ model partition function is defined as
\begin{align}
	Z_{d=0}=\int d\Phi d\Gamma\,\exp\left[\frac{1}{2}F^2+\frac{1}{2}\int d^2\theta \,\left(m+\Gamma\right)\left(\Phi^2-\frac{1}{g^2}\right)\right].
\end{align}
After integrating out $F$ this becomes
\begin{align}
	Z&=\int dn \,d^2\psi\, d\sigma \,d^2u \,d\lambda\,\, e^{-\left(\frac{1}{2}m^2 n^2  +m \bar{\psi}\cdot\psi+\Lagr_\Gamma\right)},\non
\Lagr_\Gamma&\equiv \frac{1}{2}\sigma^2n^2+\sigma\left(\bar{\psi}\cdot\psi +\frac{m}{g^2}\right)-\frac{1}{2}\lambda\left(n^2-\frac{1}{g^2}\right)+n\cdot\left(\bar{\psi}u+\bar{u}\psi\right).\label{8. Lagrange mult action}
\end{align}
Here the original component in the superfield $\lambda_F$ has been redefined to eliminate a quadratic mixing term later on,
\begin{align}
	\lambda\equiv\lambda_F-2m\sigma.\label{8. lambda redef}
\end{align}
Note also that $u, \bar{u}$ may be seen as fermionic Lagrange multiplier fields that enforce the constraints $n\cdot \bar{\psi}=n\cdot \psi =0.$

Now $n$ and $\psi,\bar{\psi}$ may be completely integrated out of the action, leading to logarithms with an $N$ prefactor as before,
\begin{align}
S=\frac{N}{2}\ln\left(1+\frac{1}{m^2}\left(\sigma^2-\lambda-\frac{2\bar{u}u}{\sigma+m}\right)\right)-N\log\left(1+\frac{\sigma}{m}\right)+\frac{\lambda}{2g^2}+\frac{\sigma m}{g^2}.
\end{align}
This is in fact a supersymmetric action for the component fields of $\Gamma$, as can be verified explicitly using the equivalent of \eqref{8. SUSY trans} keeping in mind the redefinition \eqref{8. lambda redef}.

The saddle point condition on $m$ arises from requiring that the linear terms in both $\sigma$ and $\lambda$ vanish upon expanding the logarithms. Both conditions are satisfied by the choice $$m^2= Ng^2,$$ which is the same as the non-SUSY case \eqref{2.saddlepoint}.

The action may be simplified slightly be redefining $\sigma, \lambda,$ and $\bar{u}u$ to absorb factors of $m, m^2, m^3$ respectively. Such factors will instead appear in the vertices connecting the $\Gamma$ fields to $n$ or $\psi$ correlation functions. After rescaling, the action expanded out to fourth order in the fields is
\begin{align}
S&=N\left(\sigma^2-\frac{1}{4}\lambda^2-\bar{u}u-\frac{1}{3}\sigma^3+\frac{1}{2}\sigma^2\lambda-\frac{1}{6}\lambda^3+\bar{u}u\sigma-\bar{u}u\lambda\right.\non
&\qquad\qquad\left.+\frac{1}{2}\sigma^2\lambda^2-\frac{1}{8}\lambda^4-\bar{u}u\lambda^2+\bar{u}u\sigma\lambda\right)+\order{\Gamma^5}.\label{8. 0d action}
\end{align}
Since the SUSY transformations of the field \eqref{8. SUSY trans} are linear, actually each order of homogeneous polynomials in the fields is separately supersymmetric. Practically speaking, this provides a good check that no errors were made in finding the coefficients of the vertices.\footnote{The action may be written explicitly in terms of polynomials of F-terms $\int d^2\theta \, \Gamma^k$ in order to make this order by order SUSY manifest.}

Now we will move on to some calculations in the $d=0$ theory, focusing on behavior that will prefigure what will be seen in the $d=1$ case. Recall that one of the curious things that we wish to explore about the SUSY $O(N)$ model is that apparently in $d=1$ all supersymmetric corrections to correlation functions involving the $n$ field alone should vanish. The same statement will be true for the $d=0$ path integral.

As was seen in the non-SUSY $d=0$ model earlier, the two-point `correlation function' $\langle n^2 \rangle$ will not be modified at all by higher order corrections due to the exact cancelation of each tadpole diagram with a corresponding non-tadpole diagram. So a more interesting test is provided by considering higher order correlation functions, for instance $A_2=\langle n_i n_j n_i n_j\rangle$, which was also considered in the non-SUSY case in Sec \ref{2. Sec two-body}, and found exactly in \eqref{5.Aj}.

\begin{figure}
	\centering
	\includegraphics[width=0.7\textwidth]{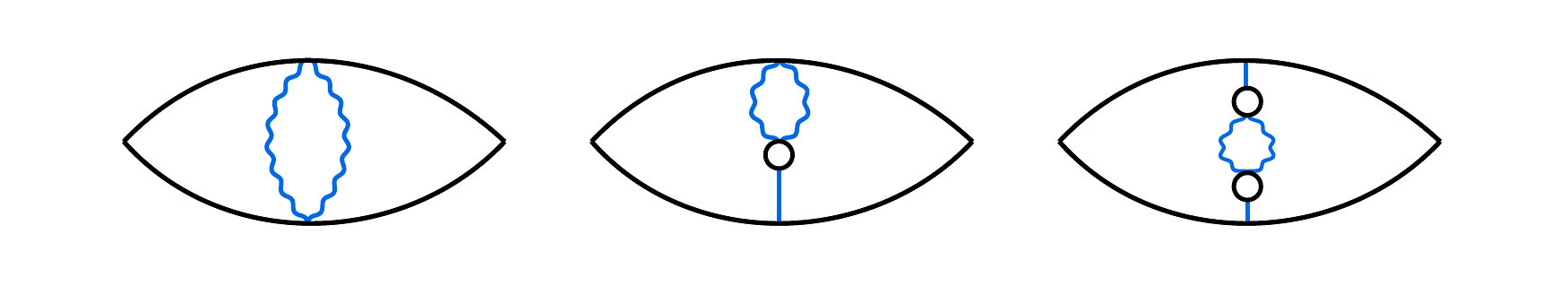}\caption{Supersymmetric corrections to the two-body propagator in $d=0$ at $\order{N^{-2}}$.  Black and blue solid lines indicate $n$ and $\lambda$ fields respectively as before. A wavy blue line indicates a $\sigma$ field. Vertices are indicated by a small black circle to represent that they arise from $n$ fields.}\label{8.fig 0d corrections}
\end{figure}

In fact, there are no new $\order{N^{-1}}$ corrections to this correlation function either. To be clear, after integrating out $\psi, \bar{\psi}$, the terms in the action involving the $n$ field and the rescaled $\Gamma$ fields are
$$-\frac{m^2}{2}\left(1+\sigma^2 -\lambda - \frac{2 \bar{u}u}{1+\sigma}\right)n^2.$$
Since each $\Gamma$ field propagator involves a factor of $1/N$, the only way to connect two $n$ lines at $\order{N^{-1}}$ is through a single $\lambda$ field, which was already taken into consideration in the non-SUSY case. So to find a real non-trivial test of the statement that SUSY corrections to bosonic correlation functions vanish it is necessary to go to $\order{N^{-2}}$. The relevant diagrams are shown in Fig \ref{8.fig 0d corrections}. Several diagrams which appear in the corresponding problem in $d=1$ cancel for more trivial reasons and are not shown here. The complete cancelation of tadpole and non-tadpole pairs was already mentioned, but another simplifying feature of $d=0$ is that since $\bar{u}u$ involves Grassmann numbers and has no spacetime argument or internal indices, it can only appear once in any diagram.

Using the propagators and vertices in the action \eqref{8. 0d action}, it can be shown that the diagrams in Fig \ref{8.fig 0d corrections} do indeed cancel as $\frac{1}{N^2 m^4}\left(\frac{1}{2}-1+\frac{1}{2}\right)$ respectively. Of course it is actually easy enough to integrate out the new fields $\psi, u, \sigma$ from the action directly in order to see that there is no $n$ dependent factor remaining that could alter the bosonic correlation functions. But from a more pedagogical perspective, examples such as this may serve well as exercises to familiarize a student with the combinatorial details of working with Feynman diagrams.

A perhaps more conceptually interesting calculation that can done in a SUSY $O(N)$ model is to consider how supersymmetry restricts various expectation values. In particular if $\delta \Phi$ represents the variation of a superfield as in \eqref{8. SUSY trans}, a simple argument much like the Ward identity in Sec \ref{5} shows that $\langle \delta \Phi \rangle=0$ exactly. Considering the form of $\delta \Phi$, this implies the principle that \emph{the VEV of a F-term of a superfield vanishes}. Since $\lambda_F$ is the F-term of $\Gamma$, this implies by \eqref{8. lambda redef} the exact relation
\begin{align}
	\langle \lambda \rangle = -2m \langle \sigma\rangle. \label{8. SUSY relation lambda}
\end{align}
This was already verified to the lowest order in the sense that we were able to find a saddle point where both $\langle\lambda\rangle^{(0)}=\langle\sigma\rangle^{(0)}=0$.

Furthermore, since the `kinetic term' $\Lagr_K\equiv -\frac{1}{2}F^2$ may be written as the F-term of the superfield $\frac{1}{2}\partial_\theta \Phi \partial_{\bar{\theta} }\Phi$, there is another exact relation which will be useful in $d=1$ as well
	\begin{align}
\langle \Lagr_K \rangle =0.\label{8. SUSY relation kinetic}
	\end{align}
To express $\Lagr_K$ in terms of the $\Gamma$ fields, a $+JF$ term may be added to the exponent of the path integral and differentiated twice. Upon integrating out $F$, this $J$ term is coupled to the $\sigma$ field
$$\frac{1}{2}F^2 + (m+\sigma)n\cdot F + J\cdot F \qimplies -\frac{1}{2}(m+\sigma)^2 n^2 -\frac{1}{2}J^2 - J\cdot n (m+\sigma).  $$
Differentiating by $J$ twice, summing over all components, and setting it to zero, the form of $\Lagr_K$ is found
$$\Lagr_K = -\frac{1}{2}\left(-N+n^2 (m+\sigma)^2\right)= -\frac{N}{2}\left(\frac{\sigma^2}{m^2}+2\frac{\sigma}{m}\right).$$
The $-N$ term in the first equation is due to the presence of the $J^2$ term, and it may be thought of as a contact term for the $F$ equation of motion $F=-(m+\sigma)n$.

Together, the vanishing of the VEV of $\lambda_F$ and $\Lagr_K$ implies the following exact equality for VEVs of the rescaled dimensionless fields
$$\langle \sigma^2\rangle = -2\langle \sigma\rangle = \langle\lambda\rangle.$$

This can be easily explicitly verified at $\order{N^{-1}}$. $\langle\sigma^2\rangle$ is just given by the $\sigma$ propagator $1/(2N)$. $\langle\sigma\rangle$ and $\langle\lambda\rangle$ may be found by considering the three-point vertices in \eqref{8. 0d action} and forming tadpole diagrams, and indeed it is seen that the above equality stemming from the underlying SUSY is satisfied. A similar calculation for $\langle \Lagr_K\rangle$ will be performed in $d=1$, where it will have more relevance to our overall discussion since $\Lagr_K$ will involve terms like $(\partial n)^2$ which lead to non-trivial contact terms.

\subsection{Introduction to the SUSY $O(N)$ model in $d=1$}\label{8.sec d=1}
In $d=1$ the form of the superfields, the mass terms, and the Lagrange multiplier part of the action are essentially exactly the same as the $d=0$ case, but now superfields depend on Euclidean time $\tau$ as well, and so the action will be an integral over $\tau$, and derivatives must appear in the kinetic part of the action.

Superderivatives $D,\bar{D}$ may be defined that anticommute with the generators of SUSY transformations\footnote{Although no use will be made here of the supercharges that anticommute with $D,\bar{D}$, just to be clear, the convention being used is that the SUSY transformation acts as $\tau\rightarrow \tau+\bar{\theta}\epsilon-\bar{\epsilon}\theta.$} 
$$D={\partial_{\bar{\theta}}}+\theta\partial_\tau,\qquad \bar{D}=-\partial_\theta -\bar{\theta}\partial_\tau,$$
and a supersymmetric kinetic term Lagrangian may be constructed from these as
\begin{align}\Lagr_K=-\frac{1}{2}\int d^2\theta \bar{D}\Phi \cdot D\Phi = \frac{1}{2}(\partial{n})^2+\bar{\psi}\cdot\partial{\psi}-\frac{1}{2}F^2.\label{8. Lagr K}\end{align}
After integrating out $F$, the total Lagrangian is
\begin{align}
\Lagr = \frac{1}{2}n\cdot\left(-\partial^2+m^2\right)n+\bar{\psi}\cdot\left(\partial+m\right)\psi + \Lagr_\Gamma,\label{8. Lagr total}
\end{align}
with $\Lagr_\Gamma$ defined in \eqref{8. Lagrange mult action}.

From this, the bare fermion propagator (with no contraction of indices) may be read off as
\begin{align}
	\langle \psi(\tau)\bar{\psi}(0)\rangle^{(0)}=\int \frac{dp}{2\pi}\left(\frac{1}{-ip+m}\right)e^{-ip\tau}.
\end{align}
This just evaluates to $e^{-m\tau}$ for $\tau>0$ and vanishes for $\tau<0$, so it is consistent with the quantum mechanical interpretation of $\bar{\psi}, \psi$ as creation and annihilation operators of a state of mass $m$.

To evaluate $m$ using the saddle point condition $\langle \lambda\rangle^{(0)}=\langle \sigma\rangle^{(0)}=0$, consider the relevant terms in $\Lagr_\Gamma$,
$$\sigma\left(\bar{\psi}\cdot\psi +\frac{m}{g^2}\right)-\frac{1}{2}\lambda\left(n^2-\frac{1}{g^2}\right).$$
As in the non-SUSY case, the condition that the linear terms in $\lambda$ vanish leads to the same condition as before \eqref{3.saddle point}
$$m=\frac{Ng^2}{2}.$$

Perhaps unsurprisingly, this is the same condition that is required for the linear terms in $\sigma$ to vanish. This just follows from the same supersymmetric relation \eqref{8. SUSY relation lambda} which was considered in the $d=0$ case. To see the vanishing of the terms linear in $\sigma$ explicitly, consider the fermion loop from the $\sigma \bar{\psi}\cdot\psi$ term,
\begin{align*}
\langle\bar{\psi}\cdot\psi\rangle^{(0)} = -N \int \frac{dp}{2\pi}\left(\frac{1}{-ip+m}\right).
\end{align*}
This integral is logarithmically divergent, which is ultimately a result of the operator ordering ambiguity for $\bar{\psi}\cdot \psi$ taken at the same time. If this is understood as $\bar{\psi}(\tau)\cdot \psi(\tau-\epsilon)$ that would introduce a regulating $e^{ip\epsilon}$ factor into the propagator that makes this integral well-defined in the one-sided $\epsilon\rightarrow 0$ limit. But actually this point-splitting prescription would not be consistent with SUSY for either sign of $\epsilon$ since the fermion loop would not cancel with the $m/g^2$ term and $\sigma$ may pick up a VEV even when $\lambda$ does not.

A consistent regulation is actually given by the very same regularization we have being using throughout,
\begin{align*}
	\langle\bar{\psi}\cdot\psi\rangle^{(0)} &= -N \int \frac{dp}{2\pi}\left(\frac{{ip}+m}{p^2+m^2}\right)=-\frac{N}{2}.
\end{align*}
The logarithmically divergent $ip$ term is simply taken to vanish due to symmetric integration, and the $m$ term gives a finite contribution which is effectively the average of the point-splitting prescription for the two signs of $\epsilon$.\footnote{The same ordering prescription $\bar{\psi}\psi \sim \frac{1}{2}[\bar{\psi},\psi]$ is used to derive SUSY QM from a path integral action in \cite{Gildener:1977hm, Cooper:1994eh}.} In this regularization the fermion loop does cancel with $m/g^2$ for a consistent $\langle \lambda\rangle^{(0)}=\langle \sigma\rangle^{(0)}=0$ saddle point.

		\begin{figure}
	\centering
	\includegraphics[width=0.8\textwidth]{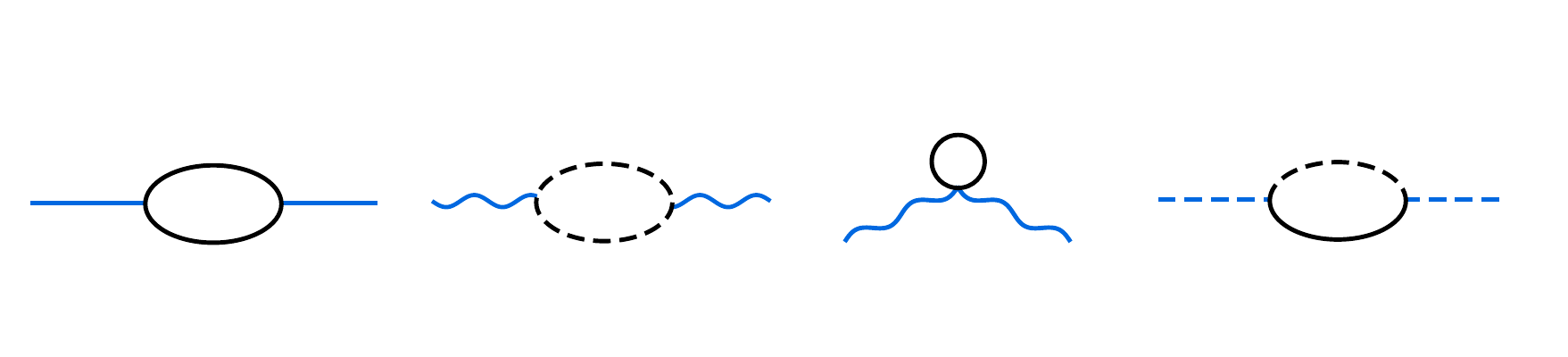}\caption{The diagrams leading to propagators for the Lagrange multiplier fields. A black dashed line indicates $\psi$ and a blue dashed line indicates $u$. When relevant, an arrow may indicate the direction from $\psi$ to $\bar{\psi}$ or $u$ to $\bar{u}$.}\label{8.fig two point diagrams}
\end{figure}

Now that the linear terms of the Lagrange multiplier fields have been considered and shown to cancel, the next step is to consider their propagators which arise from the two-point diagrams in Fig \ref{8.fig two point diagrams}. The exact same diagrams will arise in any dimension, but a peculiar feature of $d=1$ is that the fermion bubble involved in the $\sigma$ propagator vanishes. Since the $\psi$ propagator has the form $(-ip+m)^{-1}$ the poles in loops involving only fermions all fall in the negative imaginary half plane and the integral vanishes. But it may be instructive to find the propagators in any dimension by considering the $\psi$ propagator to be $(-\gamma^\mu p_\mu+m)^{-1}$, where $\gamma_\mu$ satisfies $\{\gamma^\mu,\gamma^\nu\}=-2\delta^{\mu\nu}$. In $d=1$ we just have $\gamma=i$.

The results for the propagators are as follows (recall the definition of $J$ \eqref{3.J def}),
\begin{align}
	\frac{2}{N}\frac{1}{(p^2+4m^2)J(p)}&=\qquad\frac{2m}{N}\qquad &\text{($\sigma$ propagator)}\non
	\frac{2}{N}\frac{\gamma p-2m}{(p^2+4m^2)J(p)}&=\qquad\frac{2m}{N}(ip-2m)\qquad &\text{($u$ propagator)}\non
		\frac{2}{N}\frac{-1}{J(p)}&=\qquad-\frac{2m}{N}(k^2+4m^2)\qquad &\text{($\lambda$ propagator)}.\label{8.propagators}
\end{align}

As in the non-SUSY case in Sec \ref{4.sec lambda and contact terms}, the fact that the propagators of the Lagrange multiplier fields are the inverse of bubbles of the ordinary $n, \psi$ fields is connected with the appearance of contact terms associated with the constraints implied by the Lagrange multiplier fields. The following relations hold to all orders
\begin{align}
	\left\langle\lambda(\tau)\,n^2(0)\right\rangle_{\text{connected}}=-2\delta(\tau),\qquad \left\langle u(\tau)\,\left(\bar{\psi}\cdot n\right)\!(0) \right\rangle= \left\langle\left( n\cdot{\psi}\right)\!(\tau) \,\bar{u}(0)\right\rangle=\delta(\tau).\label{8.contact term constraint}
\end{align}

The $\sigma$ field is not associated to an equation of constraint in the same way. But there is still a similar relation that may be derived from the equation of motion,
\begin{align}
\left\langle\sigma(\tau)\,\left(\bar{\psi}\cdot\psi\right)\!(0)\right\rangle_{\text{connected}}=\delta(\tau)-\frac{1}{g^2}\left\langle\sigma(\tau)\,\sigma(0)\right\rangle_{\text{connected}}.\label{8.contact term sigma}
\end{align}
In $d=1$ at lowest order this is clearly satisfied, since the left side involves a vanishing fermion loop, and the right side involves the $\sigma$ propagator \eqref{8.propagators}, which is $g^2\delta(\tau)$ in real space.

\subsection{Supersymmetric corrections to propagators}
\begin{figure}
	\centering
	\includegraphics[width=0.5\textwidth]{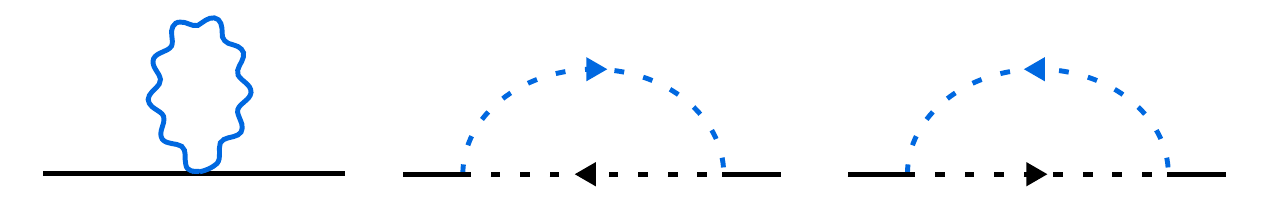}\hfil
		\caption{$\order{N^{-1}}$ SUSY corrections to the $n$ propagator. The $u\bar{u}$ arc diagrams depend on orientation. Only non-tadpole diagrams are shown.}\label{8.fig n prop}
\end{figure}

The corrections to the $n$ propagator due to the $\lambda$ field alone are exactly the same as the non-SUSY case. Recall that the self-energy correction due to the $\lambda$ arc in Fig \ref{2.Fig n and n2} was \eqref{3.propArc}
$$\Pi_{\lambda}=\frac{2m}{N}\left(\Lambda+\frac{p^2+3m^2}{2m}\right).$$

The new SUSY corrections are shown in Fig \ref{8.fig n prop}. The $\sigma^2$ loop involves an integral over a constant propagator leading to 
$$\Pi_{\sigma}=\frac{2m}{N}\Lambda.$$

There are two corrections with different orientations involving $u$ lines. Each one individually involves a term linear in $p$, but this $p$ dependence cancels summing over both orientations.
$$\Pi_{u}=-\frac{2m}{N}\left(2\Lambda+m\right).$$

Now note in passing that the sum of all these non-tadpole diagrams leads to
$$\Pi_\lambda+\Pi_{\sigma}+\Pi_{u}=\frac{1}{N}\left(p^2+m^2\right).$$
Exactly the same formula is found in $d=2$, for instance in \cite{GraceyEtAl}. This is because it follows from rather general cancelations between the diagrams when expressed in terms of $J(p)$ and the $\gamma$ matrices.\footnote{the two different orientations of the $u$ arc diagrams in $d=1$ corresponds to the trace over spin indices in $d=2$.}

Recall from Sec \ref{4.sec tadpole} that if a non-tadpole diagram evaluates to $Ap^2+B$ its corresponding $\lambda$ tadpole evaluates to $-Am^2-B$. So from the sum of the non-tadpole diagrams above the VEV of $\lambda$ may be calculated (recalling the minus sign in the definition of $\Pi$). And then from the supersymmetric relation \eqref{8. SUSY relation lambda} the VEV of $\sigma$ may be calculated
\begin{align}
	\langle \lambda\rangle=\frac{2m^2}{N},\qquad\langle \sigma \rangle = -\frac{m}{N}. \label{8. vevs}
\end{align}
The VEV of $\sigma$ will be checked independently momentarily when we consider the $\psi$ propagator.

But first note that since the corrections $\Pi_{\sigma}$ and $\Pi_{u}$ both have no $p$ dependence, they completely cancel with their corresponding $\lambda$ tadpole. \emph{So all SUSY corrections to the bosonic correlation function vanish in $d=1$}. As discussed in the introduction to this section, this is what we would expect from the idea that the Hamiltonian is the Laplace-de Rham operator, but it is not so obvious a priori.

However there is an all-orders argument that may be made, and it also clarifies why this statement no longer holds true in $d=2$ and above. All the Lagrange multiplier fields may be completely integrated out of the action \eqref{8. Lagrange mult action}, leading to the fermionic sector
\begin{align*}
	\Lagr_\psi = \bar{\psi}\cdot\left(\partial+m\right)\psi -\frac{g^2}{2}\left(\bar{\psi}\cdot\psi\right)^2,\qquad n\cdot \psi=n\cdot\bar{\psi}=0.
\end{align*}
Despite appearances, $\Lagr_\psi$ is coupled to $n$ through the constraint. To make this more explicit, at each time $\tau$ we choose $N-1$ orthonormal vectors $e^a(\tau)$, all of which are orthogonal to $n(\tau)$, then $\psi$ and $\bar{\psi}$ may be written in terms of $N-1$ unconstrained fermions $q_a,\bar{q}_a$,
$$\psi= q_ae^a,\quad \bar{\psi}= \bar{q}_ae^a,\qquad e^a\cdot e^b=\delta^{ab},\quad e^a\cdot n=0.$$ 

Now the fermionic sector action may be written as
\begin{align*}
\Lagr_\psi = \bar{q}_a\left(\partial+m\right)q_a+ \bar{q}_a \left(e^a \cdot\partial e^b\right)q_b -\frac{g^2}{2}\left(\bar{q}_a q_a\right)^2.
\end{align*}
There is clearly a $O(N-1)$ gauge symmetry in the definition of $q$, and the term $e\cdot\partial e$ is acting as an $O(N-1)$ gauge field to ensure that the derivative of $q$ is covariant. This gauge field implicitly couples $q$ to the bosonic sector. Its field strength may be written in terms of the $n$ field alone.

But here we see what is special about $d=1$. There is no field strength, and the gauge field  $e\cdot \partial e$ may be chosen to vanish.\footnote{Explicitly choose a new basis $e_b^\prime = e_a O_{ab}$, with $O_{ab}\equiv \left(e^{-\mathcal{P}\int e\cdot \partial e}\right)_{ab}$.} So the fermionic sector really is decoupled from $n$, and thus the $n$ correlation functions are the same as their non-SUSY counterparts. In other words, the same feature of gauge fields in $d=1$ that led to a relation between the spectra of the $O(N-1)$ model and the $CP^{\bar{N}-1}$ model in Sec \ref{7} leads to a relation between the spectra of the SUSY and non-SUSY models.

\begin{figure}
	\centering
	\includegraphics[width=0.5\textwidth]{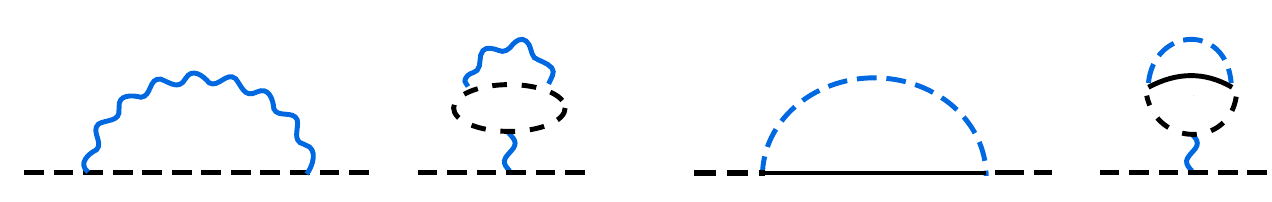}\hfil
	\caption{$\order{N^{-1}}$ corrections to the $\psi$ propagator.}\label{8.fig psi prop}
\end{figure}

But of course besides the bosonic states there are also fermionic states to consider in the SUSY model. We expect on general grounds that the mass of the fermionic state excited by $\bar{\psi}$ is corrected to be exactly the same as the bosonic $M_1=m(1-N^{-1})$. This can be shown to be the case by considering the corrections to the $\psi$ propagator in Fig \ref{8.fig psi prop}.

The self energy $\Sigma$ arising from the $\sigma$ arc is straightforward given the discussion above on the proper regularization of the fermion propagator loop, and likewise for the $u$ arc. The sum of the arc diagrams is
$$\Sigma_{\text{arc}}=-\frac{1}{N}\left(ip-m\right).$$

As suggested in Fig \ref{8.fig psi prop}, each non-tadpole correction to the $\psi$ propagator has a corresponding $\sigma$ tadpole. This won't be quite as powerful as the case for the $n$ propagator and $\lambda$ tadpoles, since $\sigma$ is not associated to a constraint. But it is still useful to pair tadpole and non-tadpole in this manner, and note that if the non-tadpole diagram evaluates to $Ap+B$, the tadpole diagram evaluates to $-iAm$. This implies
$$\Sigma_{\text{tadpole}} = \langle\sigma\rangle = -\frac{m}{N},$$
which agrees with the result \eqref{8. vevs} found from the general supersymmetric relation \eqref{8. SUSY relation lambda}.

So the fermion propagator is corrected to
\begin{align}
	\langle \psi(p)\bar{\psi}(-p)\rangle=\frac{1}{-ip+m+\Sigma_{\text{arc}}+\Sigma_{\text{tadpole}}}=\frac{1-N^{-1}}{-ip+m(1-N^{-1})}+\order{N^{-2}}.
\end{align}
And the mass is indeed equal to the boson mass $M_1$.

To interpret the amplitude correction $(1-N^{-1})$, consider the VEV of the equation of motion associated to the $\sigma$ field,
\begin{align}
\left\langle\bar{\psi}\cdot \psi\right\rangle=-\frac{m+\langle \sigma\rangle}{g^2}.\label{8. sigma tadpole equation}
\end{align}
This relation may also be seen diagrammatically by considering the analogue of Fig \ref{4.fig constraint} for the $\sigma$ tadpoles.

As seen in Sec \ref{8.sec d=1}, the result of calculating $\left\langle\bar{\psi}\cdot \psi\right\rangle$, does not depend on the mass of the fermion propagator at all, only the amplitude. So \eqref{8. sigma tadpole equation} gives us a general expression for the fermion propagator
\begin{align}
\langle \psi(p)\bar{\psi}(-p)\rangle=\frac{1+{\langle \sigma\rangle}m^{-1}}{-ip+M_1}.\label{8.psi propagator exact}
\end{align}
In the next subsection much use will be made of this relation between $\left\langle\bar{\psi}\cdot \psi\right\rangle$ and $\langle \sigma\rangle$ in order to discuss the supersymmetric identity \eqref{8. Lagr K} $\langle\Lagr_K\rangle=0$. 

\subsection{Equations of motion, VEVs, and contact terms}\label{8.sec contact terms}
Now finally let us consider more carefully the operator $\Lagr_K$ \eqref{8. Lagr K}. This is the SUSY extension of the (negative) Hamiltonian $\frac{1}{2}\left(\partial n\right)^2$ which led to some interesting considerations in Sec \ref{4.secEqM}. As before, much can be done by considering equations of motion. From \eqref{8. Lagr total}, the equations for $n$ and $\bar{\psi}$ are
\begin{align}
\left(-\partial^2+m^2\right)n+\left(\sigma^2-\lambda\right)n+\bar{\psi}u+\bar{u}\psi=0\label{8.eq mot n}\\
\left(\partial+m\right)\psi+\sigma \psi + nu=0.\label{8.eq mot psi}
\end{align}
Once again, we may dot by $n$ and $\bar{\psi}$ respectively to find `naive' equations for the VEVs, which ignore contact terms
\begin{align}
	\frac{1}{2}\left\langle n\cdot \left(-\partial^2+m^2\right)n\right\rangle\cong\frac{1}{2g^2}\left(\langle\lambda\rangle-\langle\sigma^2\rangle\right)\label{8.naive n}\\
	\left\langle \bar{\psi}\cdot\left(\partial+m\right)\psi\right\rangle\cong \frac{1}{g^2}\left(m\langle\sigma\rangle+\langle\sigma^2\rangle\right).\label{8.naive psi}
\end{align}
In the second relation the $\sigma$ equation of motion \eqref{8. sigma tadpole equation} was used to naively set $-\langle \sigma \bar{\psi}\psi\rangle \cong g^{-2}\langle \sigma \left(m+\sigma\right)\rangle$.

Surprisingly, the naive relations for the VEVs actually do satisfy the supersymmetric relation $\langle\Lagr_K\rangle=0$. To see this, first rewrite the $-\frac{1}{2}F^2$ term in $\langle\Lagr_K\rangle$ in terms of $\sigma$ fields. There will be a `trivial' contact term when this is done, exactly as in the discussion at the end of the $d=0$ case in Sec \ref{8.sec 0d}. But for now we simply ignore all contact terms,
\begin{align}
\Lagr_K &\cong \frac{1}{2}(\partial{n})^2+\bar{\psi}\cdot\partial{\psi}-\frac{1}{2g^2}\left(m+\sigma\right)^2\\
\langle\Lagr_K\rangle&\cong \frac{1}{2g^2}\left(-2m\langle\sigma\rangle-\langle\sigma^2\rangle\right)-\frac{m^2}{2g^2}\non
&\qquad+\frac{1}{g^2}\left(m\langle\sigma\rangle+\langle\sigma^2\rangle\right)+\frac{m}{g^2}\left(m+{\langle\sigma\rangle}\right)-\frac{1}{2g^2}\left\langle\left(m+\sigma\right)^2\right\rangle=0.\label{8.susy kinetic identity}
\end{align}
What is going on? In the non-SUSY case $\frac{1}{2}\!\left(\partial n\right)^2$ didn't agree at all with the naive expression beyond lowest order. But now both finite and divergent $\order{N^{-1}}$ corrections ($\langle\sigma\rangle$ and $\langle\sigma^2\rangle$ respectively) in the naive expressions are canceling among themselves, suggesting that the naive expression might actually be correct up to trivial contact terms.

Considering first the $\bar{\psi}$ equations of motion \eqref{8.eq mot psi} by dotting by $\bar{\psi}$ at a different point and including the trivial contact term we find
\begin{align}
	\left\langle \bar{\psi}\cdot\left(\partial+m\right)\psi\right\rangle &= -N\Lambda+\frac{1}{g^2}\langle\sigma\rangle\left(m+\langle \sigma\rangle\right)\non
	&\qquad+\lim_{\tau\rightarrow 0}\left\langle\left(\sigma\psi\right)\!(\tau)\cdot\bar\psi(0)\right\rangle_\text{connected}+\lim_{\tau\rightarrow 0}\left\langle\left(nu\right)\!(\tau)\cdot\bar\psi(0)\right\rangle.\label{8.psi propagator subleading}
\end{align}
The first term is the trivial contact term, and the second term is similar to the naive expression \eqref{8.naive psi}, but apparently a fortuitous mistake was made there since rather than $g^{-2}\langle \sigma^2\rangle$, which is divergent and $\order{N^{0}}$, the proper term should be $g^{-2}\langle \sigma\rangle^2$ which is finite and $\order{N^{-1}}$, so it can be effectively disregarded at the order of analysis we are considering here. The mistake was fortuitous because actually the same $\order{N^{0}}$ divergence arises from the last two terms on the second line, which are new examples of non-trivial contact terms in the same mold as the term considered in Sec \ref{4.secEqM}.

\begin{figure}
	\centering
	\includegraphics[width=0.8\textwidth]{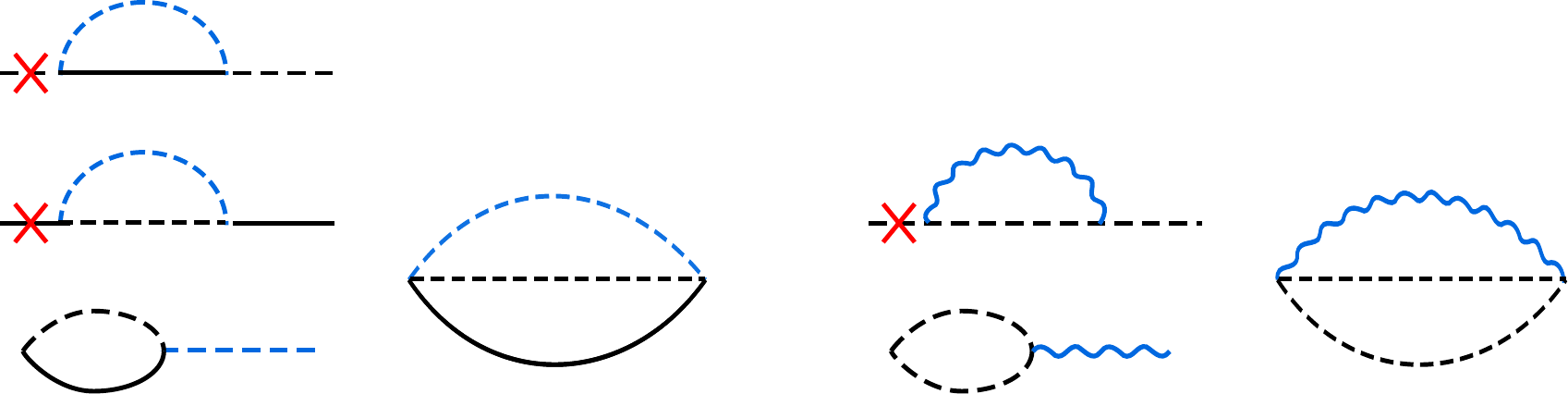}\hfil
	\caption{New non-trivial contact terms in the SUSY $O(N)$ model, compare with Fig \ref{4.fig nontrivial contact}. On the left, going from top to bottom, the integrals \eqref{8.nontrivial u}, \eqref{8.nontrivial u n prop}, and \eqref{8.contact term constraint} are represented. Upon integration over external momentum, all of these diagrams formally become the two-loop diagram to their right. On the right side of the figure, the integrals \eqref{8.nontrivial sigma} and \eqref{8.contact term sigma} are represented, as well as their formal two-loop diagram limit.}\label{8.fig contact term}
\end{figure}

These non-trivial contact terms are related to the contact term expressions associated to the $\sigma$ and $u$ fields in \eqref{8.contact term sigma} and \eqref{8.contact term constraint}, respectively. The situation is shown in Fig \ref{8.fig contact term}, compare to Fig \ref{4.fig nontrivial contact}. The integral for the term associated to $\sigma$ is
\begin{align}
\lim_{\tau\rightarrow 0}\left\langle\left(\sigma\psi\right)\!(\tau)\cdot\bar\psi(0)\right\rangle_\text{connected}=-N\int \frac{dp}{2\pi}\frac{1}{-ip+m}\int\frac{dk}{2\pi}\frac{2m}{N}\frac{1}{-i\left(p-k\right)+m}=-\frac{m}{2}.\label{8.nontrivial sigma}
\end{align}
In the case of the limit associated to the expression \eqref{8.contact term sigma}, $p$ is integrated over first. In this case the inner integral is a fermion loop, and as discussed earlier, the result simply vanishes. As before, the two-loop diagram in Fig \ref{8.fig contact term} is ambiguous, since it depends on how the $\tau=0$ limit is taken.

The integral for the term associated to the $u$ field is
\begin{align}
\lim_{\tau\rightarrow 0}\left\langle\left(nu\right)\!(\tau)\cdot\bar\psi(0)\right\rangle=-N\int \frac{dp}{2\pi}\frac{1}{-ip+m}\int\frac{dk}{2\pi}\frac{2m}{N}\frac{ik-2m}{(p-k)^2+m^2}=\Lambda+\frac{m}{2}.\label{8.nontrivial u}
\end{align}
If $p$ is integrated first this evaluates to $\Lambda$ and represents the contact term associated to the constraint $n\cdot \bar{\psi}=0$ in \eqref{8.contact term constraint}. Note that just as was seen for the non-trivial contact term associated to $\lambda$ in Fig \ref{4.fig nontrivial contact}, even though the finite part depends on order of integration, the divergent part does not.

In this case the finite parts of \eqref{8.nontrivial sigma} and \eqref{8.nontrivial u} cancel, and the divergent part happens to be equal to the conceptually mistaken term $g^{-2}\langle \sigma^2\rangle$ which explains why the naive expression actually seems to work. As one last check, note that the exact $\psi$ propagator \eqref{8.psi propagator exact} may be differentiated (taking into account the discontinuity at $\tau=0$) to produce the exact result
\begin{align}
\langle \bar{\psi}\cdot\partial\psi\rangle=\frac{N}{2}M_1\left(1+{\langle \sigma\rangle}m^{-1}\right)-N\left(1+{\langle \sigma\rangle}m^{-1}\right)\Lambda.
\end{align}
Using the result $\langle \sigma\rangle =-m/N +\order{N^{-2}}$ calculated in the last subsection, and the results for the non-trivial contact terms above, it is seen that this agrees to the calculated order with the expression \eqref{8.psi propagator subleading}, so the results are consistent.

Now let us turn our attention briefly to the bosonic equation of motion \eqref{8.eq mot n},
\begin{align}
		\frac{1}{2}\left\langle n\cdot \left(-\partial^2+m^2\right)n\right\rangle&=\frac{N}{2}\Lambda+\frac{1}{2g^2}\left(\langle\lambda\rangle-\langle\sigma^2\rangle\right)\non
		&\quad\lim_{\tau\rightarrow 0}\left[\,\,\left\langle\left(\lambda n\right)\!(\tau)\cdot n(0)\right\rangle+\left\langle\left(\bar{\psi}u+\bar{u}\psi\right)\!(\tau)\cdot n(0)\right\rangle\,\,\right]+\order{N^{-1}}.\label{8.n propagator subleading}
\end{align}
The first term is the trivial contact term and the next terms are exactly the same as the naive expression \eqref{8.naive n}. The next terms on the second line are the non-trivial contact terms.\footnote{Note that to the order considered, there is no non-trivial contact term associated to the $\sigma^2$ operator.} The first involving $\lambda$ is exactly the same as the non-SUSY case in \eqref{4.contact}, leading to $-2\Lambda-m$. The next two terms involve $u$ and $\bar{u}$, and the only difference is the sign of the linear term in $p$, which ends up vanishing due to the integration over $p$ in the $\tau\rightarrow 0$ limit. So both of these terms are like the $u$ non-trivial contact term in Fig \ref{8.fig contact term}. The integral is
\begin{align}
\lim_{\tau\rightarrow 0}\left\langle\left(\bar{u}\psi\right)\!(\tau)\cdot n(0)\right\rangle=-N\int \frac{dp}{2\pi}\frac{1}{p^2+m^2}\int\frac{dk}{2\pi}\frac{2m}{N}\frac{ik-2m}{-i\left(p+k\right)+m}=\Lambda+\frac{m}{2}.\label{8.nontrivial u n prop}
\end{align}
The result is the same as the $\psi$ propagator case. And since the $\bar{\psi}u$ term takes the same value, these two non-trivial contact terms from $u$ and $\bar{u}$ combine to cancel with the the non-trivial contact term arising from $\lambda$ \eqref{4.contact}. So the exact expression \eqref{4.nddn} is just equal to the naive expression \eqref{8.naive n} up to trivial contact terms.

Finally, to put to rest the issue of trivial contact terms, the contact term in $\Lagr_K$ \eqref{8.susy kinetic identity} arising from the $-\frac{1}{2}F^2$ term is $+\frac{N}{2}\Lambda$ (in the same way as in Sec \ref{8.sec 0d}). So in total, the trivial contact terms arising form  $-\frac{1}{2}F^2$ and from \eqref{8.psi propagator subleading} and \eqref{8.n propagator subleading} do indeed cancel. Considering that the exact expressions involving non-trivial contact terms were shown to be equal to the naive expressions used in \eqref{8.susy kinetic identity} we have just shown that the supersymmetric relation $\langle\Lagr_K\rangle=0$ is consistent to the order considered.

\section{Summary}

A more detailed summary of the results presented in this paper is presented in Sec \ref{1.Sec outline}, but considering that a number of disparate topics have been presented here, it is worthwhile to make a brief statement on what has been shown and how the various topics relate to each other.

We have just considered the supersymmetric $O(N)$ model in $d=1$ and shown that much as in the ordinary $O(N)$ model there are subtle corrections to the VEVs of composite operators which look similar to contact terms arising from the constraint, but in fact involve a different limiting procedure. In the lattice regularization in the ordinary $O(N)$ model, it was shown in Sec \ref{4.sec lattice} that these `non-trivial contact terms' lack some of the subtlety of the continuum case. Considering SUSY on a lattice is not quite so straightforward, but it might be interesting to consider how different regularizations in the supersymmetric case change the situation as well. Furthermore it may be interesting to explore the situation taking into consideration the regularization dependent counterterms discussed for more general sigma models in $d=1$ path integrals \cite{Bastianelli:1991be}. Also note that these VEVs are very important to the study of IR renormalons and the resurgence program in $d=2$ \cite{Dunne:2015ywa}. Given that the issue seems to be related to the proper UV regularization of operators, and given that the ambiguities in the operator product expansion were shown to cancel using the naive form of the equations of motion \cite{Schubring:2021hrw}, one might at first expect that this will not affect the study of IR renormalons significantly, but certainly the issue is worthy of further study.

Beyond this technical point involving VEVs and contact terms, we have investigated a broader statement that in the $d=1$ SUSY $O(N)$ model correlation functions involving the bosonic $n$ fields alone are exactly the same as the non-SUSY case. This is clearly not true in higher dimension, and we have demonstrated that the peculiarity of $d\leq 1$ that allows for this is due to the triviality of gauge theories in low dimensions. A similar situation occurs for the $CP^{\bar{N}-1}$ sigma model which is equivalent to a gauged $O(2\bar{N})$ sigma model, as discussed in Sec \ref{7}. Those correlation functions of $O(2\bar{N})$ which are gauge invariant are exactly the same in the $CP^{\bar{N}-1}$ and $O(2\bar{N})$ models, but such a fact is not true in higher dimension.

Nevertheless, we have been emphasizing results which are valid in any dimension, and the relations between diagrams arising from the equation of constraint discussed in Sec \ref{4} are an example of this kind of result. By considering how the constraint is manifested in terms of correlation functions and diagrams, in Sec \ref{6} we have been led to a form of SCSA which is valid in the limit of an exact $O(N)$ model with a hard constraint. But as emphasized there, in $d=1$ for the $O(N)$ limit, the SCSA is not so useful. It would be interesting to see how the $O(N)$ limit of the SCSA compares with a truncated large $N$ expansion in higher dimensions as well.

Finally, the core of this paper has been about calculating the spectrum of quantum mechanical models in a down-to-earth pedagogical way using Feynman diagrams arising from the large $N$ expansion. This was considered first in terms of $d=0$ ordinary integrals in Sec \ref{2} and Sec \ref{8.sec 0d}, where the evaluation of Stirling's series, expectation values on the sphere, and some SUSY identities suggestive of the situation in higher dimension were discussed. In \eqref{7.squashed sphere} we have found the correction to the energy levels in a sigma model with a  squashed sphere target space in a physically suggestive way arising from repulsion and attraction of particles charged under the $U(1)$ group associated with the fibers. Perhaps the most important case, at least from a pedagogical perspective, is simply the ordinary $O(N)$ model considered in Sec \ref{3}. The entire exact spectrum is given by two-body interactions alone, and the possible three-body interactions at next order are shown explicitly to cancel in Appendix \ref{A} below. Although the spectrum was known long ago from exact results, it is curious to what extent this physical picture of particles interacting solely through two-body interactions extends to higher dimensions.

		\section*{Acknowledgments}

I am grateful to Mikhail Katsnelson and Vitaly Vanchurin for discussions on the self-consistent screening approximation and an invitation to give a talk on an early form of this work. 

\appendix
	\section{$\mathcal{O}(N^{-2})$ corrections}\label{A}
	
	The calculation of corrections to the spectrum at $\order{N^{-2}}$ follows much the same general idea as the calculation at $\order{N^{-1}}$. The correlation function $\langle O_j(\tau)O_j(0)\rangle$ involves $j$ single particle lines, and to this order these lines may only interact in 2-body and now also 3-body interactions. So the idea will be to first show that the corrections $M_1^{(2)}, M_2^{(2)}, M_3^{(2)}$ all vanish. And then given the fact that $M_1^{(2)}, M_2^{(2)}, M_3^{(2)}$ are uncorrected, the vanishing of $M_j^{(2)}$ for general $j$ will be shown as well.
		
			\begin{figure}[t]
			\centering
		\includegraphics[width=0.9\textwidth]{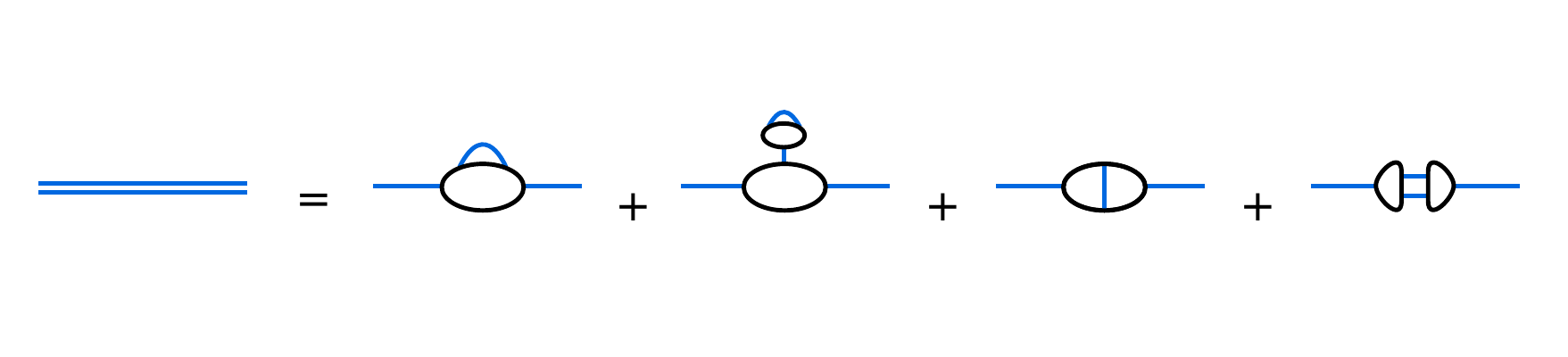}\caption{The $\order{N^{-2}}$ corrections to the $\lambda$ propagator. The self-energy corrections are just the 1PI corrected bubbles.}\label{4.fig lambda ON2}
		\end{figure}
		
	First it is useful to consider the $\order{N^{-2}}$ corrections to the $\lambda$ propagator itself shown in Fig \ref*{4.fig lambda ON2}, since this will appear as a sub-diagram in several cases. As discussed in Sec \ref{4.sec lambda and contact terms}, the self-energy corrections to the $\lambda$ propagator are just the 1PI corrections to the $\langle n^2(\tau)n^2(0)\rangle$ bubble. Except for the rightmost diagram in Fig \ref*{4.fig lambda ON2}, these diagrams were all calculated in the context of finding the $\order{N^{-1}}$ correction to the two-body mass $M_2^{(1)}$. The correction to the $\lambda$ propagator ends up being divergent and independent of external momentum.
	\begin{align}
		\langle \lambda(-p)\lambda(p)\rangle^{(2)}=-\frac{8m^2}{N^2}\left[\Lambda+4m\right].
	\end{align}
	
	\subsection{Single particle diagrams}

\begin{figure}[t]
		\centering
		\includegraphics[width=0.8\textwidth]{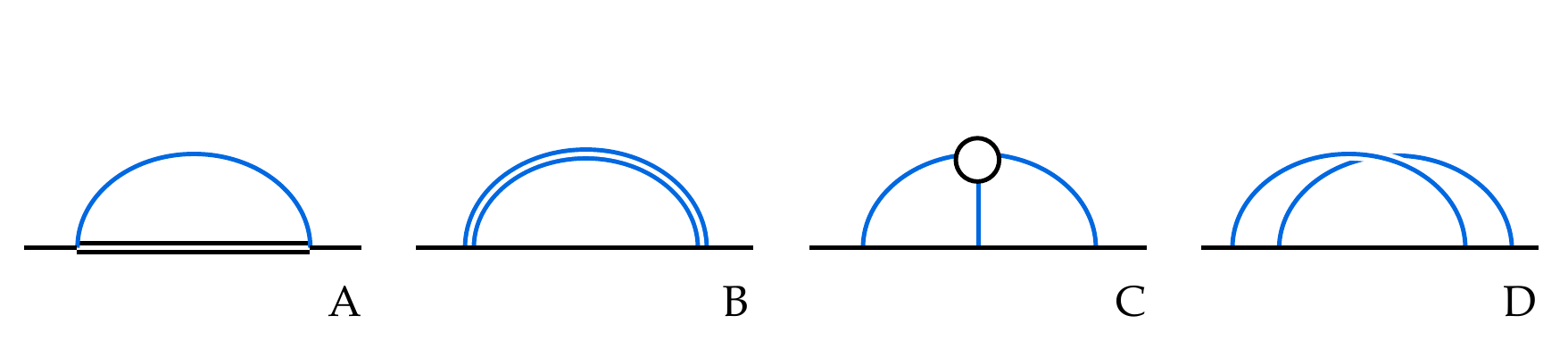}\caption{The $\order{N^{-2}}$ non-tadpole corrections to $\langle n(-p)n(p)\rangle$. A double line indicates a corrected $n$ or $\lambda$ propagator.}\label{A.fig n}
	\end{figure}

	As discussed in Sec \ref{4.sec tadpole}, there is a correspondence between tadpole and non-tadpole diagrams, and the contribution from the tadpole diagrams may be calculated simply from the result of the non-tadpole diagrams. So we will focus only on the non-tadpole diagrams in Fig \ref{A.fig n}. The results in terms of the self-energy will be simply summarized:
	\begin{align}
		\Pi_A&=-2\left[\Lambda-m\right]\frac{m}{N^2},
		&\Pi_B=-4\left[\Lambda+4m\right]\frac{m}{N^2},\non
		\Pi_C&=+8\left[\Lambda+\frac{p^2+10m^2}{4m}\right]\frac{m}{N^2},
		&\Pi_D=-2\left[\Lambda+\frac{p^2+9m^2}{2m}\right]\frac{m}{N^2}.
	\end{align}
	The divergences cancel and the total result is
	\begin{align}
		\Pi_\text{non-tadpole}(p^2)=\frac{1}{N^2}\left[p^2-3m^2\right].
	\end{align}
We can already see that $M_1^{(2)}=0$, from the coefficient of $p^2$ in \eqref{4.self energy}, and the earlier result $M_1^{(1)}=-m/N$. Furthermore, $M_1^{(2)}=0$ implies that the coefficient of $m^2$ must vanish, so the sum of all tadpoles must be
\begin{align}
\Pi_\text{tadpole}(p^2)=+\frac{3m^2}{N^2}.
\end{align}
	This is an honest calculation of the tadpole diagrams since the form of \eqref{4.self energy} was fixed by the diagrammatic relation between tadpoles and non-tadpoles. Note that in our convention for the self-energy $\Pi_\text{tadpole}=-i\langle \lambda\rangle$ and the calculated value of $\langle \lambda\rangle ^{(2)}$ agrees with the value found in lattice regularization \eqref{4.lambda vev lattice}.
	\subsection{Two- and three-body diagrams}
\begin{figure}
		\centering
		\includegraphics[width=0.6\textwidth]{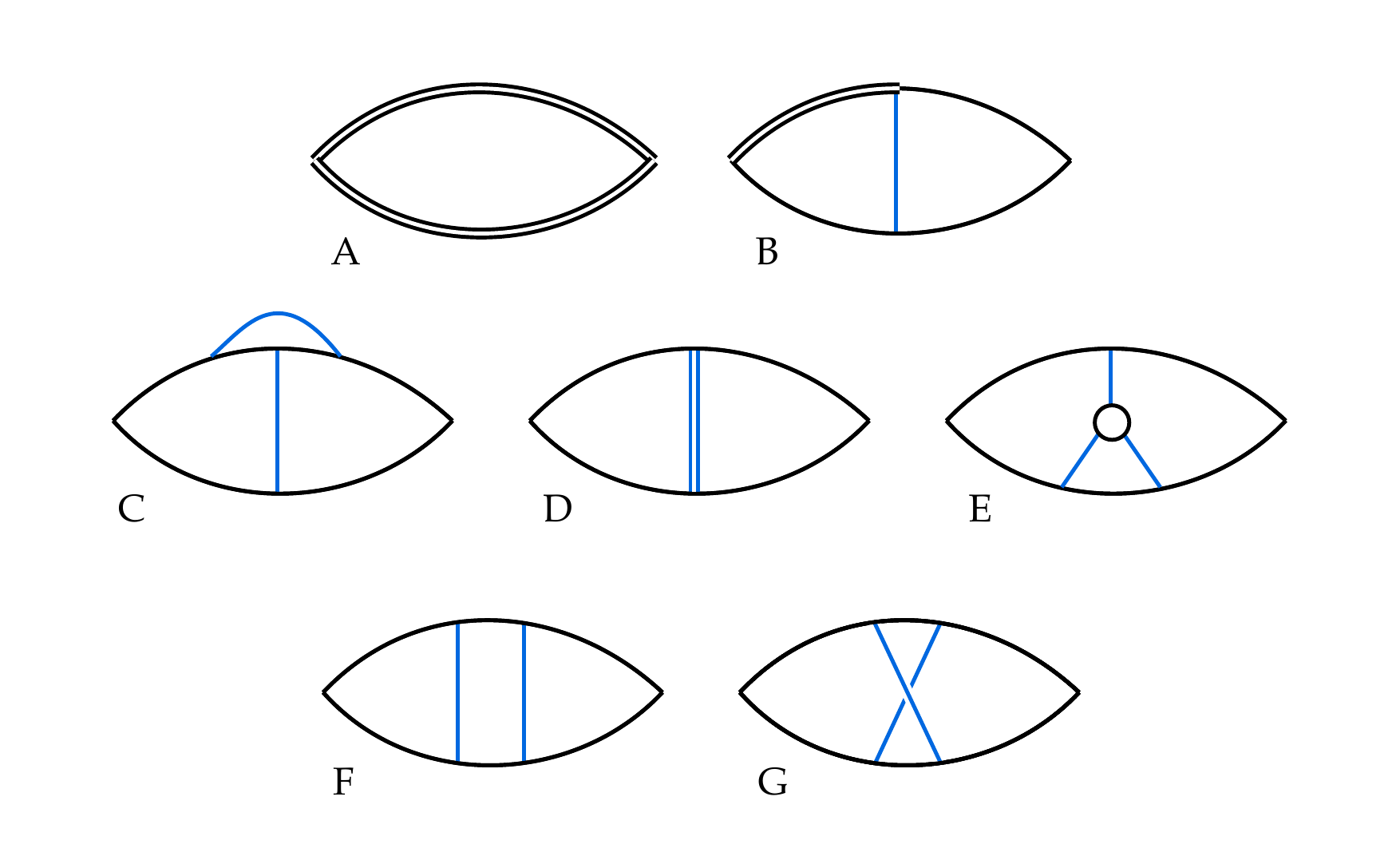}\caption{$\order{N^{-2}}$ corrections to $\langle O_2(-p)O_2(p)\rangle$.}\label{A.fig n2}
	\end{figure}

	The corrections to the 2-body correlation function $g_2(p)\equiv\langle O_2(-p)O_2(p)\rangle$ are shown in Fig \ref*{A.fig n2}. To simplify both reporting and summing the individual diagrams, the dimensionless quantity $a$ is defined,
	$$a\equiv \frac{m^2}{p^2+4m^2},$$
	and the dimensionless amplitudes $G\equiv m^3 N^2g_2^{(2)}$ are reported:
\begin{align}
	G_A=-12a^2+64a^3, \quad G_B=2a+8a^2-128a^3,\quad G_C=4a+32a^2,\non \quad G_D=32a^2,\quad G_E=-6a-80a^2,\quad G_F=a+12a^2+64a^3,\quad G_G=3a+8a^2.
\end{align}	
Diagrams $D, E, F, G$ each individually involves a divergence in the inner loop which was not reported here, but they mutually cancel. The sum of all diagrams adds up to $4a$. The absence of an $a^2$ term implies the correction $M_2^{(2)}=0$. The coefficient of $a$ agrees with the $\order{N^{-2}}$ correction to the exact amplitude $Z_2 = m^{-1}\left(1+2N^{-1}\right)^{-1}$.

\begin{figure}
		\centering
		\includegraphics[width=0.6\textwidth]{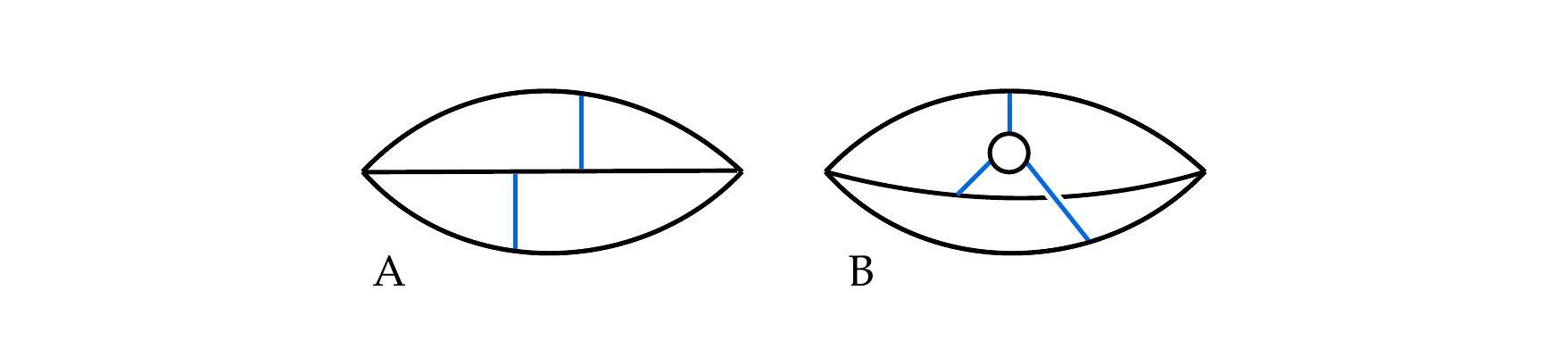}\caption{New $\order{N^{-2}}$ corrections to $\langle O_3(-p)O_3(p)\rangle$.}\label{A.fig n3}
	\end{figure}

	The 3-body correlation function $g_3(p)$ is now quite a bit easier than the one- and two-body cases since there are only two new diagrams, both of which are convergent. Once again we define a dimensionless parameter $b$,
	$$b\equiv \frac{m^2}{p^2+9m^2},$$
	and dimensionless amplitudes $H\equiv m^4 N^2g_3^{(2)}$. The two diagrams in Fig. \ref*{A.fig n3} evaluate to
	\begin{align}
		H_A=9b+108b^2+648b^3,\quad H_B=-3b-54b^2.
	\end{align}
In addition to these new diagrams, there are the diagrams involving corrections to single particle propagators $g_1$, and one pair of lines forming a $g_2$ propagator, exactly as in the $\order{N^{-1}}$ case \eqref{3. gj convolution}, but taken to the next order,
\begin{align}
	H_A+H_B+m^4N^2\left(3 g_1\ast g_2 - 2 g_1^{\ast 3}\right)=\frac{33}{2}b+\frac{243}{4}b^2+243b^3
\end{align}
This is slightly more difficult to interpret than $g_2$ since $M_3^{(1)}\neq 0$, but by considering the general expansion \eqref{A. expansion of gj} below it is seen that indeed $M_3^{(2)}=0$, and the next order term in the amplitude $ Z_3^{(2)}=\frac{33}{2m^2 N^2}$ is also consistent with \eqref{5.Aj}.
	\subsection{Full spectrum to $\order{N^{-2}}$}
	Now to find $g_j(p)$ for arbitrary $j$, an extension of the combinatorial argument in Sec \ref{3.sec spectrum} is used. The diagrams are grouped into four cases: four legs acting as two copies of $g_2$, three legs as $g_3$, two legs as a single factor of $g_2$, and all legs as non-interacting $g_1$. Taking care to avoid overcounting, the total is
	\begin{align}
g_j&=\frac{j(j-1)(j-2)(j-3)}{8}\left(g_2^{\ast 2}\ast g_1^{\ast\left(j-4\right)}-2g_2\ast g_1^{\ast\left(j-2\right)}+g_1^{\ast j}\right)\non
&\qquad+\frac{j(j-1)(j-2)}{6}\left(g_3\ast g_1^{\ast\left(j-3\right)}-3g_2\ast g_1^{\ast\left(j-2\right)}+2g_1^{\ast j}\right)\non
&\qquad+\frac{j(j-1)}{2}\left(g_2\ast g_1^{\ast\left(j-2\right)}-g_1^{\ast j}\right)+g_1^{\ast j}+\order{N^{-3}}.\label{A. convolution}
	\end{align}

	To interpret the result, the general form of $g_j$ is expanded to second order,
	\begin{align}
g_j^{(2)}(p)&=\frac{Z_j^{(2)}}{p^2+(jm)^2}+\frac{Z_j^{(0)}\left(2jm M_j^{(1)}\right)^2}{\left(p^2+(jm)^2\right)^3}\non &\quad-\left[Z_j^{(1)}2jmM_j^{(1)}+Z_j^{(0)}\left(\left(M_j^{(1)}\right)^2+2jm M_j^{(2)}\right)\right]\frac{1}{\left(p^2+(jm)^2\right)^2}.\label{A. expansion of gj}
	\end{align}
	The last term contains the only appearance of $M_j^{(2)}$, so some small effort may be saved by only considering the terms proportional to $\left(p^2+(jm)^2\right)^{-2}$ in \eqref{A. convolution}. The result is
	$$g^{(2)}_j(p)=-\frac{m^2Z_j^{(0)}}{N^2(p^2+(jm)^2)^2}\left[(j-2-j(j-1))2j^2(j-2)+j^2(j-2)^2\right]\quad+\left(\text{other powers of $p^2+(jm)^2$}\right),$$
	and given the first order results \eqref{3. gj 1st order} this finally implies $M_j^{(2)}=0$.
		\section{Quantum mechanics on homogeneous spaces}\label{B}
		Here we will give a very brief sketch of the wavefunction approach to finding the spectrum of a sigma model in $d=1$ with a homogeneous target space such as the sphere $S^{N-1}$ or complex projective space $CP^{\bar{N}-1}$. A more general and rigorous account can found for instance in the appendices of Camporesi's review on harmonic analysis on homogeneous spaces \cite{Camporesi:1990wm}. Here the focus will be on deriving exact results for the sigma model on a squashed sphere which will agree with the two path integral methods considered in Sec \ref{7}.
		
		Given some Lie group $G$ and a subgroup $H\subset G$, consider the coset space $G/H=\{gH:g\in G\}$. There is a natural $G$ action on $G/H$ given by left multiplication, and we can consider a metric $g$ on $G/H$ which is invariant under this action.
		
		For example, the sphere
		 $S^{N-1}$ may be considered to be the homogeneous space
		\linebreak$SO(N)/ SO(N-1)$. If $N=2\bar{N}$ is even, $S^{N-1}$ may instead be considered in terms of the coset space $SU(\bar{N})/SU(\bar{N}-1)$ but the metric will of course (being the same space) still have the full $SO(N)$ symmetry. A more general metric on $SU(\bar{N})/SU(\bar{N}-1)$ that only has the required $SU(\bar{N})$ symmetry\footnote{Actually the squashed sphere metric will also have an extra Killing vector associated with the $U(1)$ subgroup which is modded out in the $CP^{\bar{N}-1}$ case.}  will instead describe a squashed sphere. If we mod out a larger subgroup $H=SU(\bar{N}-1)\times U(1)$, then the homogeneous space $SU(\bar{N})/H$ will describe $CP^{\bar{N}-1}$. 
		\subsection{Irreducible representations}
		We can define a Hilbert space $\mathcal{H}$ consisting of complex valued wavefunctions $\psi: G/H\rightarrow \mathbb{C}$, with the obvious inner product given by integration over $G/H$ using the measure induced by the $G$ invariant metric $g$. $G$ has a natural representation $\rho$ on the wavefunctions induced by its left action on $G/H$. Given $g\in G$, define the linear map $\rho(g):\mathcal{H}\rightarrow \mathcal{H}$ by
		$$\left[\rho(g)\psi\right](x)= \psi(g^{-1}x).$$
		Given the $G$ invariant measure involved in the inner product it is easy to see this is in fact a unitary representation of $G$. However, it is a very reducible representation. It will be useful to decompose the representation of $G$ into irreducible representations, much in the same way as rotations of general wavefunctions on $S^2$ may be decomposed into representations on spherical harmonics.
		
		Suppose we consider some arbitrary unitary irreducible representation $\lambda$ of $G$ acting on some vector space $V$. Given a $g\in G$ the action of the representation on $v\in V$  may be written as $U^{(\lambda)}(g)v$. Given a basis on $V$ we can write this in components as $U^{(\lambda)}(g)^i_j v^j$.
		
		Now suppose there is some unit vector $\xi$, that is invariant under the representation for all $h\in H$,
		$$U^{(\lambda)}(h)^i_j \xi^j= \xi^i.$$
		For convenience we may choose $\xi$ to be one of the basis vectors, say at the index $i=I$. Now consider the following functions defined on $G/H$,
		$$\psi^{(\xi)}_j(gH) \equiv U^{(\lambda)}(g^{-1})^I_j.$$
		Due to the invariance of $\xi$ under the representation restricted to $H$ this is in fact a well-defined function on $G/H$ that does not depend on choice of coset representative. Furthermore under action by $\rho$ these wavefunctions transform as
		$$\rho(g)\psi^{(\xi)}_j = U^{(\lambda)}(g)^k_j\psi^{(\xi)}_k,$$ 
		so this is in fact an irreducible subrepresentation of $\rho$ unitarily equivalent to $\lambda$. In fact conversely the number of times that the irreducible representation $\lambda$ appears in the decomposition of $\rho$ is equal to the number of linearly independent vectors $\xi$ which are invariant under $H$ in the irrep $\lambda$. This is a special case of the idea of Frobenius reciprocity \cite{MackeyInducedRep}.
		
		In particular this means that if some irreducible representation of $G$ has no vector invariant under $H$, then that irreducible representation can not appear in the Hilbert space. To be more concrete, in the case of $CP^{\bar{N}-1}$, the fundamental representation of $SU(\bar{N})$ given by complex vectors $z^a$ does have an invariant vector under the $SU(\bar{N}-1)$ factor of $H$. Indeed every vector has a stabilizer isomorphic to $SU(\bar{N}-1)$, and the factor of $SU(\bar{N}-1)$ in $H$ is just the stabilizer of some reference vector. But that reference vector is not invariant under the $U(1)$ factor of $H$, which is specifically chosen to change the phase of that reference vector. The statement that the only representations that exist in the Hilbert space are those which have a vector invariant under the full subgroup $H$ amounts to being the same as the notion that only gauge invariant combinations of operators acting on the vacuum are true states in the Hilbert space.

		\subsection{Hamiltonian and eigenvalues}

		The Hamiltonian $\hat{H}$ will be given by the Laplace-Beltrami operator on $G/H$
		$$\hat{H} = -\frac{1}{2}g^{ab}\nabla_a\nabla_b,$$
		where $\nabla$ is the Levi-Civita connection associated to $g$. $\hat{H}$ commutes with the unitary representation $\rho$ of $G$, so by Schur's lemma the irreducible representations $\psi^{(\xi)}$ will be eigenvectors of $\hat{H}$. To find the eigenvalues, it will be helpful to rewrite $\hat{H}$ in a form that takes advantage of the properties of $G$ as a Lie group.
		
		$G$ may be considered as a fiber bundle over the base space $G/H$ with the natural projection map $\pi(g)=gH.$ $G$ has a family of right-invariant vector fields $R_a$, where $a$ is an index of the Lie algebra, and because they are right-invariant the pushforward $$K_a\equiv\pi_\star R_a$$ does not depend on the choice of coset representative. Since the metric on $G/H$ was chosen to be invariant under left action by $G$,  $K_a$ are in fact Killing vectors on $G/H$.
		
		$G$ also has left-invariant vectors $L_a$, but the pushforward of the left-invariant vectors does depend on the choice of coset representative. Even so, a section $\sigma:G/H\rightarrow G$ may be chosen and basis vectors $e_a$ on $G/H$ may be defined through the pushforward of $L_a$ at points of $G$ specified by the section
		$$e_a\equiv \pi_\star|_\sigma L_a.$$ 
		The metrics that are being considered here have the nice property that they take the same form in the basis $e_a$ regardless of choice of $\sigma$. In particular, the metric of the squashed sphere target space is
		$$g_{ab}=\frac{1}{g^2}C_a \delta_{ab},$$
		where $C_a$ is equal to $1$ for every index except that associated to the left-invariant vector $L_\phi$ that moves in the direction of overall change of phase of the $z$ fields. This distinct value $C_\phi$ is determined by the coefficient of the $\left(z^\dagger\cdot \partial z\right)^2$ term in the Lagrangian \eqref{7.squashed sphere}. Comparing this coefficient to the parameters in previous works \cite{SquashedSphereSigma},
		\begin{align}
			C_\phi = \frac{\epsilon}{1+\epsilon}\frac{2(\bar{N}-1)}{\bar{N}}.
		\end{align}
	
		To proceed with the evaluation of the Laplace-Beltrami operator for this metric, it is first useful to state a rather well-known result\footnote{This is stated without proof in \cite{Camporesi:1990wm}. To prove it, it is helpful to consider the properties of the adjoint map relating $L$ and $R$, which is directly related to the components $K_a = K^b_a e_b$. A useful lemma is to first show $K^b_a\nabla_b K^c_a = 0.$} for the slightly simpler metric of $g_{ab}=\delta_{ab}$, which is the metric of $CP^{\bar{N}-1}$, or $S^{N-1}$ considered as the homogeneous space $SO(N)/SO(N-1)$. In this special case, the Laplace-Beltrami operator may be written in terms of the Killing vectors considered as differential operators,
		\begin{align}
			\delta^{ab}\nabla_a\left(\nabla_b \psi\right)= K_a\left( K_a \psi\right).
		\end{align}
		Then the more general Laplace-Beltrami operator with $C_\phi\neq 1$ may be written in these terms as well 
		\begin{align}
			\hat{H}=-\frac{g^2}{2}\left(\left(K_a\right)^2 + \left(C_\phi^{-1}-1\right)\left(e_\phi\right)^2\right).
		\end{align}
		
		Now consider how the differential operators in $\bar{H}$ act on the eigenfunctions $\psi_j(gH)= U^{(\lambda)}(g^{-1})^I_j$ above,
		\begin{align*}
K_a \psi_j&=-i\psi_k\,{\tau^{(\lambda)}_{a}}^k_j \qimplies \left(K_a \right)^2\psi_j = -C_2^{(\lambda)} \psi_j\\
e_\phi \psi_j &= -i{\tau^{(\lambda)}_{\phi}}^I_k{U^{(\lambda)}}^k_j\qimplies \left(e_\phi\right)^2 \psi_j = -\left(C_\xi\right)^{2}\psi_j.
		\end{align*}
	Here $C_2$ is the quadratic Casimir $\left(\tau_a\right)^2=C_2I$. $C_\xi$ is the diagonal coefficient $C_\xi\equiv {\tau_{\phi}}^I_{\,\,I}$. Note that in the squashed sphere case $\tau_\phi$ commutes with all the generators of $H$, and takes a diagonal form, so $\psi_j$ is indeed an eigenvector of $e_\phi$. To summarize,
	\begin{align}
		\hat{H}\psi_j=\frac{g^2}{2}\left(C_2^{(\lambda)}+\left(C_\phi^{-1}-1\right)\left(C_\xi\right)^2\right)\psi_j.
	\end{align}

This formula then may be used to find the exact eigenvalue for any representation $\lambda$ with a singlet vector $\xi$ under $H$. As an important special case, consider the fundamental representation of $SU(\bar{N})$, which has eigenvalue $M_1$. The singlet vector $\xi$ may be taken as the unit vector with $1$ in the $\bar{N}$th component, and the phase generator that commutes with the stabilizer $H$ of $\xi$ is
$$\tau_\phi = \sqrt{\frac{2}{(\bar{N}-1)\bar{N}}}\text{diag}\left(1,1,\dots,1,-\bar{N}+1\right).$$
The quadratic Casimir for the fundamental representation, given our normalization convention $\text{Tr}\left(\tau_a\tau_b\right)=2\delta_{ab}$ is $$C_2=\frac{2}{\bar{N}}\left(\bar{N}^2-1\right).$$
So now plugging in it can be shown
\begin{align}
	M_1 = \frac{g^2}{2}\left(2\bar{N}-1\right)+\frac{g^2}{2\epsilon},
\end{align}
which (recalling $N=2\bar{N},\,m=\frac{Ng^2}{2}$) agrees with the renormalized mass found in \eqref{3.n propagator} and the squashed sphere correction found in \eqref{7. deltam1} and \eqref{7.exact correction}.

\end{document}